\documentclass[aps,prb,eqsecnum, amsmath,floatfix, twocolumn, showpacs,amssymb,superscriptaddress,10pt]{revtex4-2}
\usepackage[english, english]{babel}[2018/11/13]
\usepackage{graphicx}
\usepackage{float,rotating}
\usepackage[caption=false]{subfig}
\usepackage{subeqnarray}
\usepackage{bbold}
\usepackage{amsmath, amscd}
\usepackage{verbatim}
\usepackage{amssymb}
\usepackage{mathrsfs}
\usepackage{bm} 

\usepackage{float}
\usepackage{color}
\usepackage{lettrine}
\usepackage{txfonts}
\usepackage[classicReIm]{kpfonts}
\usepackage [autostyle, english = american]{csquotes}\MakeOuterQuote{"}
\usepackage[breaklinks=true,colorlinks=true,linkcolor=blue,urlcolor=blue,citecolor=magenta]{hyperref}
\usepackage{mathtools}
\DeclarePairedDelimiter\bra{\langle}{\rvert}
\DeclarePairedDelimiter\ket{\lvert}{\rangle}
\DeclarePairedDelimiterX\braket[2]{\langle}{\rangle}{#1 \delimsize\vert #2}
\newcounter{multifig}

\usepackage[ruled]{algorithm2e}
\SetKwComment{Comment}{/* }{ */}
\SetKw{KwBy}{by}

\begin{document}
	\title{Dirac Representation for Lattice Spin Operators: Spin-$1/2$ and  Spin-$1$ cases}
	\author{Maseim B. Kenmoe}
	\affiliation{Mesoscopic and Multilayer Structures Laboratory, Faculty of Science, Department of Physics, University of Dschang, Cameroon}
	\date{\today}
	
\begin{abstract}
We introduce a quantum representation of lattice spin operators (LSOs) by mapping quantum spins to their classical analogues for spin values $S=1/2$ and $S=1$. The resulting Dirac (bra--ket) forms are derived from the structure of binary and ternary digit distributions in non-negative integers.  The construction requires only an enumeration of the lattice sites and is therefore independent on the spatial dimension and lattice geometry.  Analytical expressions for extracting the $j$th  bit or trit of an integer without performing an explicit binary or ternary decomposition  are obtained. As an application of the representation,  we analyze the one-dimensional  spin-$1/2$ {XXZ} Heisenberg model with open boundaries in a magnetic field in both absence (uniform magnetic field) and presence of disorder (random magnetic field). In the absence of disorder (clean system), we demonstrate that the corresponding eigenvalue problem can be reduced to a tight-binding problem on a graph. For the isotropic case, this is solved analytically without resorting to any spinless transformation or the Bethe Ansatz. In the presence of disorder, a convergent perturbation theory is elaborated. Our analytical results are benchmarked against exact diagonalization for systems up to $K\leq 18$ spins. These computations exploit the global $U(1)$ symmetry to block-diagonalize the Hamiltonian and the spin-inversion symmetry to further decompose the zero-magnetization sector. Quantitative agreement between analytical and numerical results is observed.
\end{abstract}
\maketitle
	
\section{Introduction}\label{Section1}
\lettrine{D}irac once wrote\cite{Dirac}: {\it ``A good notation can be of great value in helping the development of a theory ''}. In quantum Many-Body (MB) theory for spin systems and statistical physics, depending on the problem, it is highly desired to achieve appropriate representation of lattice spin operators (LSOs). These fields rapidly grew during the early periods of quantum mechanics thanks to the efforts due in historical order to Jordan and Wigner (JW)\cite{Jordan}, Holstein and Primakoff (HP)\cite{Holstein}, J. Schwinger (JS) \cite{Schwinger} and Dyson and Maleev(DM)\cite{Dyson1956, Maleev1957}, among others. LSOs at the same site span the SU(2) group exhibiting unique features: they commute at different sites and satisfy non-trivial commutation relations at the same site and for these reasons do not admit neither Fermi nor Boson statistics\cite{Kiselev2000, Kiselev2001}. Despite this difficulty, the above authors succeeded in  mapping LSOs onto fermions (in the case of JW) and bosons (in the case of HP, JS and DM). These representations still serve as fundamental tools in the construction of quantum theories for spin systems, including spin-wave theories\cite{Herring1951, Auerbach1994, Plihal}. They have led to hallmark step-forward in understanding spin systems with various applications. 

Despite the immense list of successes attributed to the aforementioned pioneering works, each of the representations comes with certain drawbacks that severely limits its range of applications. Among the limitations one may list challenges related to dimensionality, non-locality, spin magnitude, population constraints, non-linearity, non-Hermiticity and temperature dependence. For instance, JW transformations are well-suited for $1D$ systems as in higher dimension they induce non-locality\cite{Comment5} in the form of string operators. HP transformations are advantageous for large spin-size systems and are adapted for higher dimensions as they preserve locality. However, they require truncation/approximation of square root leading to non-linearity and are consequently not convenient for spin-$1/2$ due to poor convergence of the approximations. The DM transformation exhibits non-Hermiticity, yet it remains preferable to the HP transformation --not only due to its simplicity in spin-wave theory but primarily because non-Hermitian effects are negligible. Several alternative representations were developed to cure these limitations among which include, the  pseudofermions Abrikosov\cite{Abrikosov2012}, Popov-Fedotov semionic\cite{Popov1987, Popov1994} and Majorana representations\cite{Tsvelik1992, Tsvelik2003, Coleman1993}. Yet, like the previously mentioned transformations these ones also come with severe constraints. To the best of our knowledge, no universal representation that copes with the entire spectrum of unsolved problems in Condensed Matter Physics and related fields has so far been invented. 

We propose viable Dirac (bra-ket) representations for spin-$1/2$ and spin-$1$ LSOs free from any non-locality, non-linearity and non-Hermiticity. The construction is formulated by enumerating lattice sites rather than their spatial coordinates and as such, is not intrinsically restricted to $1D$ systems. It applies to arbitrary lattice geometries and spatial dimensions once a labeling of lattice sites is specified.  We realize a quantum-to-classical correspondence by writing LSOs in a classical basis (binary and ternary bases) while preserving their quantum nature. This became possible by mapping qubits and qutrits onto their classical analogue, the classical bits and trits. Similar tasks are accomplished in Refs.\cite{Ivanov2018, Ivanov2021, Ivanov2024} for numerical treatment of coordinates and momentum on one hand, energy and time on the other hand. LSOs representation remaining of valuable importance. Dirac notations are typically desired because of their inherent abstract algebra that alleviates manipulation of the associated observables. They could hold the promise to be useful in the MB theory to decipher the non-trivial mechanism behind quantum phenomena such as quantum entanglement\cite{EPR1935, Aspect1982}, topological phase of matter\cite{Moore2021}, spin liquids and MB localization (MBL)\cite{Schliemann2021, Siegl2023} just to list a few. They pave the way for further insight into the spin-wave theory. For example, in Ref.\onlinecite{Gluza} such a notation is achieved for the LSO in the quantization direction in terms of the floor function and used to identify the complete set of integrals/constants of motion claimed as potential candidates for describing MBL. 

 Although the derived Dirac representation  is independent of dimensionality and lattice geometry, in the present work we restrict its physical application to a $1D$ model. In the following,   it is employed to revisit the eigenspectrum of the $1D$ spin-$1/2$ Heisenberg model incorporating anisotropy in an open boundary configuration without resorting to the Bethe Ansatz. An additional  Ising term is included, involving local magnetic fields $h_j$ with $j$ indexing lattice sites. The model has a global $U(1)$ symmetry allowing treatments in sectors corresponding to fixed magnetization quantum numbers. In the sector with zero magnetization, spin-inversion symmetry is invoked allowing to break down the Hamiltonian into yet two disconnected sectors corresponding to the eigenvalues of the spin-inversion operator. Two complementary limits of the magnetic field are considered: (i) the uniform and non-uniform magnetic fields. For uniform magnetic fields and in the absence of anisotropy ($J_z=0$), the eigenvalue equations reduce to tight-binding-type recurrence relations that can be solved exactly in each magnetization sector. In the case of non-uniform local magnetic fields (all $h_j$ different e.g. disordered or contaminated models), the eigenspectra are derived using a perturbation theory with respect to the spin-spin exchange $J$. Our results are compared with data from exact diagonalization obtained by invoking the aforementioned symmetries for relatively large-size spin systems and validated in both the antiferromagnetic ($J>0$) and ferromagnetic ($J<0$) exchange regimes. An excellent agreement between both data is observed when the perturbation parameter satisfies $|J|/h \lesssim 1$.

The remaining part of the paper consists of $5+1$ Sections structured as follows: Section \ref{Section2} aims at clarifying the position of the problem addressed in this piece of paper. We demonstrate that the state-of-the-art of our current knowledge on LSOs permits Dirac notations  that only  hold for few-body problems and become increasingly cumbersome for many-body systems. Hence, the need to invent a more convenient representation. The novel representation is presented for spin-$1/2$.  Section \ref{Section3} is devoted to demonstrating that a well-defined problem can easily be solved. The impact of our representations in the Physics of integrable models including the $1D$ Heisenberg model is highlighted. A non-integrable version is discussed in Section \ref{Section4}. The representation for spin-$1$ is elaborated in Section \ref{Section5} but applications are deferred to future projects. Section \ref{Section6} concludes the paper and a few Appendices are provided as additional materials for deeper understanding of our main results.

\section{Quantum-to-classical mapping}\label{Section2}
\subsection{Quantum Representation}\label{Section2A}
Consider a  lattice   of $K$ identical spins of size $S$. Each spin lives in a local Hilbert space $\boldsymbol{\mathcal{H}}^{(j)}$ (with $j=1,\cdots, K$) spanned by the basis vectors $\ket{m_S^{(j)}}$  (with $-S \le m_S^{(j)}\le S,$ the eigenvalues of spin vectors along the quantization direction). The composite Hilbert space necessary to capture the dynamics of the ensemble is given by the tensor product of local sub-spaces,
\begin{eqnarray}\label{equ1}
	\boldsymbol{\mathcal{H}}=\bigotimes_{j=1}^{K}\boldsymbol{\mathcal{H}}^{(j)}.
\end{eqnarray}
 
Here, $j$ labels the sites and should not be interpreted as a $1D$ spatial coordinate. For a given lattice, including higher-dimensional lattices, one may introduce a correspondence $j\leftrightarrow\mathbf R_j$ between $j$ and the lattice sites. The tensor-product \eqref{equ1} and the Dirac representation developed below are invariant under this choice of enumeration.  The lattice geometry only enters when interactions between specified sites (neighboring sites for example) are introduced through the Hamiltonian.

This space generated by the construction \eqref{equ1} is of dimension $D=d^K$ with $d=2S+1$ and spanned by the basis vectors,
\begin{eqnarray}\label{equ2}
	\ket{m_S^{(1)}\cdots m_S^{(j)}\cdots m_S^{(K)}}\equiv \otimes_{j=1}^{K}\ket{m_S^{(j)}}.
\end{eqnarray}
These vectors are mutually orthogonal with a unique completeness relation thus forming a complete set of vectors.  Denoting as $\hat{\vec{S}}_j=[\hat{S}_j^x, \hat{S}_j^y, \hat{S}_j^z]$ the spin vector acting on the local subspace $\boldsymbol{\mathcal{H}}^{(j)}$  and introducing the corresponding ladder  operators $\hat{S}^{\pm}_j=\hat{S}_j^x\pm i\hat{S}_j^y$ (which are non-Hermitian, $(\hat{S}_j^{+})^{\dagger}=\hat{S}_j^{-}
$), then the spin action at a specified site is evaluated through ($\hbar=1$)
\begin{eqnarray}\label{equ2a}
\hat{S}^{\pm}_j\ket{m_S}_j=\lambda^{\pm}_{m_S^{(j)}}\ket{m_S^{(j)}\pm1}_j,
\end{eqnarray}
\begin{eqnarray}\label{equ2b}
\hat{S}^{z}_j\ket{m_S}_j=m_S^{(j)}\ket{m_S^{(j)}}_j,
\end{eqnarray}
with $\lambda^{\pm}_{m_S^{(j)}}=[S(S+1)-m_S^{(j)}(m_S^{(j)}\pm 1)]^{1/2}$ as found in many textbooks of quantum mechanics (see for e.g. Ref.\cite{Farnell}). In the above relations $\ket{m_S}_j$ is a shorthand notation for the basis vector in Eq.\eqref{equ2}. The extra subscript emphasizes the lattice site on which the operator acts. One easily observes that the action of the lattice ladder spin operators $\hat{S}^{\pm}_j$  onto a given configuration increases/decreases the $j$th quantum number by a unit. We shall see in our representation that this remark is rather counter-intuitive (see Section \ref{Section2B}). 

Although the above relations at first glance look simple and elegant as one might expect the action of quantum operators to be, the intrinsic reality is rather complex. Indeed, a concrete  Dirac ("bra" and "ket") representation inferred from Eqs.\eqref{equ2a} and \eqref{equ2b} reads 
\begin{eqnarray}\label{equ2c}
	\nonumber\hat{S}^{\pm}_j=\sum_{m_S^{(1)}=-S}^S\cdots \sum_{m_S^{(j)}=-S}^S\cdots \sum_{m_S^{(K)}=-S}^S\lambda^{\pm}_{m_S^{(j)}}\\\times\ket{m_S^{(1)}\cdots m_S^{(j)}\pm 1\cdots m_S^{(K)}}\bra{m_S^{(1)}\cdots m_S^{(j)}\cdots m_S^{(K)}},
\end{eqnarray}
\begin{eqnarray}\label{equ2d}
	\nonumber\hat{S}^{z}_j=\sum_{m_S^{(1)}=-S}^S\cdots \sum_{m_S^{(j)}=-S}^S\cdots \sum_{m_S^{(K)}=-S}^S m_S^{(j)}\\\times\ket{m_S^{(1)}\cdots m_S^{(j)}\cdots m_S^{(K)}}\bra{m_S^{(1)}\cdots m_S^{(j)}\cdots m_S^{(K)}}.
\end{eqnarray}
We adopt the standard convention that $\lambda_{S}^{+}=0$ and $\lambda_{-S}^{-}=0$, so that the above relations remain valid at the boundaries of the spectrum. And we see that Eqs.\eqref{equ2a} and \eqref{equ2b} concretely involve as many summations as there are spins in the system. They are suitable for analytical treatments of few-body systems as one can manually write down all possible configurations and handle a few summations. This suggests that the full Hilbert space can be constructed. For example, consider a spin-$1/2$ three-body system. The Hilbert space is spanned by $2^3=8$ basis vectors. They can all be manually constructed. Two possible such vectors are $\ket{\frac{1}{2}\frac{1}{2}\frac{1}{2}}$ and $\ket{-\frac{1}{2}\frac{1}{2}\frac{1}{2}}$. A matrix element of an observable $\hat{A}$ is $\bra{-\frac{1}{2}\frac{1}{2}\frac{1}{2}}\hat{A}\ket{\frac{1}{2}\frac{1}{2}\frac{1}{2}}$. We already observe from here that for large spin systems similar notations will be cumbersome involving a large number of summands. This is one of the manifestations of the finite-size effect, a drastic drawback occurring as a consequence of the exponential growth of the Hilbert space  with the total number of spins $K$. For numerical analyses,  this effect extensively restricts the number of implementable spins in actual computers ($K\le 24$ to the best of our knowledge).  While the Hilbert space dimension remains exponential, the Dirac representation significantly simplifies the structure of matrix elements and allows more efficient numerical implementations. Relatively large spin systems can be numerically integrated on small computers ($K\leq 18$, RAM $32$Gb with Mathematica configured with the highly optimized Intel Math Kernel Library (MKL)). Our objective is to show that there exists better alternatives to Eqs.\eqref{equ2c} and \eqref{equ2d} easier to handle both analytically and numerically.

\subsection{Binary theory}\label{Section2B}

Before diving into the classical representation, let us elaborate on a binary theory for spin-$1/2$ systems. To this end, we consider the above example and relabel the basis vectors as  $\ket{\frac{1}{2}\frac{1}{2}\frac{1}{2}}\equiv \ket{0}$ and $\ket{-\frac{1}{2}\frac{1}{2}\frac{1}{2}}\equiv \ket{4}$ such that $\bra{-\frac{1}{2}\frac{1}{2}\frac{1}{2}}\hat{A}\ket{\frac{1}{2}\frac{1}{2}\frac{1}{2}}\equiv \bra{4}\hat{A}\ket{0}$. This is the binary representation\cite{Sandvik}. It is already simpler compared to the spin-projection notation in Eq.~\eqref{equ2}. Indeed, qubits are mapped onto their classical counterparts, the classical bits $0$ and $1$ as $\ket{\frac{1}{2}}\equiv \ket{\uparrow}\to\ket{0}$ and $\ket{-\frac{1}{2}}\equiv \ket{\downarrow}\to\ket{1}$ with $\ket{0}=[1,0]^T$ and $\ket{1}=[0,1]^T$. Thus, 
$\ket{\frac{1}{2}\frac{1}{2}\frac{1}{2}}=\ket{000}=\ket{0}$, $\ket{-\frac{1}{2}\frac{1}{2}\frac{1}{2}}=\ket{100}=\ket{4}$. 

The binary representation is convenient for numerical analysis, exact diagonalization for example. The Hilbert space can easily be constructed by performing binary decomposition of the basis vectors, which in turn facilitates the construction of the Hamiltonian (see \href{https://github.com/Kenmax15}{GitHub repository}). We shall show that this representation is not only convenient for numerical simulations but also advantageous for analytical manipulations \cite{Kenmoe}. It improves on the readability of observables with consequences on their physical interpretations. 
 
It is very important to emphasize at this point that throughout this work we use the local basis
$\ket{\uparrow}=[1,0]^{T}$ and $\ket{\downarrow}=[0,1]^{T}$.
In the binary encoding of a configuration $\ket{n}$ we assign
\begin{eqnarray}\label{equ3a}
\sigma_n^{(j)}=0 \;\Longleftrightarrow\; \ket{\uparrow}_j,
\qquad
\sigma_n^{(j)}=1 \;\Longleftrightarrow\; \ket{\downarrow}_j,
\end{eqnarray}
so that $s_n^{(j)z}=\frac{1-2\sigma_n^{(j)}}{2}$. With this convention the ladder operators act as $\hat{S}_j^+\ket{\downarrow}_j=\ket{\uparrow}_j$ and $\hat{S}_j^+\ket{\uparrow}_j=0$, hence $\hat{S}_j^+$ flips the local bit $1\to 0$ (and annihilates for $\sigma_n^{(j)}=0$), while $\hat{S}_j^-$ flips $0\to 1$ (and annihilates for $\sigma_n^{(j)}=1$).

We denote as $\boldsymbol{\mathcal{H}}_{\mathcal{B}}$  the binary space analogue to \eqref{equ1} and spanned by $\{\ket{n_S}\}_{n_S=0}^{D-1}$ (basis vectors sorted in lexicographical order). This space is isomorphic to $\boldsymbol{\mathcal{H}}$. Indeed, there exists a linear bijective  map (i.e. both injective and surjective) between them. This map is injective i.e. every element in $\boldsymbol{\mathcal{H}}$ has a unique correspondence in $\boldsymbol{\mathcal{H}}_{\mathcal{B}}$ (one-to-one). The map is surjective (onto) i.e. every element in $\boldsymbol{\mathcal{H}}_{\mathcal{B}}$ has a unique image in $\boldsymbol{\mathcal{H}}$. If we denote as $\sigma_{n_S}^{(j)}=\left\{0,1\right\}$ the $j$th digit in the binary decomposition of $n_S$ counting from $j=1$ in the reading/coding direction--- from the left to the right then,
\begin{eqnarray}\label{equ3}
	n_S=\left(\sigma_{n_S}^{(1)}\cdots \sigma_{n_S}^{(j)}\cdots \sigma_{n_S}^{(K)}\right)_{2}, \quad  n_S=\sum_{j=1}^{K}\left(\sigma_{n_S}^{(j)}\right)q_j,
\end{eqnarray}

The weight $q_j=2^{K-j}$ embebded the ordering of the basis vectors and does not encode a physical distance between lattice sites. As a consequence, the binary mapping \eqref{equ3} does not require the sites $j$ and $j+1$ to be spatial neighbors. Spatial adjacency and distances are specified independently by the interaction part of the Hamiltonian.

The  weight $q_j$ is also defined such that the largest number is $\sum_{j=1}^{K}q_j=(2^K-1)$ (number with all digits equals to $1$). For reading/coding direction--- from the right to the left $n_S=\sum_{j=1}^{K}\left(\sigma_{n_S}^{(K+1-j)}\right)r_j$ where $r_j=2^j$. 

 In the first relation in \eqref{equ3}, subscript $2$ refers to the base $b_2=\left\{0,1\right\}$. Representations \eqref{equ3} are easier to interpret compared to \eqref{equ2} which involves alternating $\ket{1/2}$ and $\ket{-1/2}$ in a non-trivial manner especially for large systems. Indeed, geometrically, if we define the bits-vector $\boldsymbol{\sigma}_{n_{S}}=[\sigma_{n_S}^{(1)}, \sigma_{n_S}^{(2)},\cdots, \sigma_{n_S}^{(K)}]$ and the weight-vector $\bm{q}=[q_1,q_2,\cdots,q_K]$ (arithmetic basis) we see that $n_S=\boldsymbol{\sigma}_{n_{S}}\cdot \bm{q}$ is a weighted sum of $\sigma_{n_S}^{(j)}$ i.e. a scalar product (not in the Euclidean sense). The $\sigma_{n_S}^{(j)}$ being the coefficient of the binary expansion of $n_S$ determine the presence or absence of $q_j$ in the weighted sum. 
 
 As indicated in Refs.\onlinecite{Sandvik, Jung} the binary representation of $n_S$ encapsulates the exact configuration of spins in the state \eqref{equ2} provided that one starts from $n_S=0$, thence the quantum-to-classical mapping $m_S^{(j)}\to\sigma_{n_S}^{(j)}$. Let us now describe in the frame of this mapping how LSOs act onto the basis vectors of $\boldsymbol{\mathcal{H}}_{\mathcal{B}}$. With our encoding Eq.\eqref{equ3a}, the action of the
 ladder operators acquires a simple binary interpretation. 
 
 When $\hat{S}_j^{+}$ acts on a configuration $\ket{n}$, it acts non-trivially only if the $j$th spin is in the state $\ket{\downarrow}_j$, i.e. when $\sigma_n^{(j)}=1$.
 In that case it flips the local bit $1\to 0$ and produces the configuration $\ket{\mu_j^{[n]}}$, where $\mu_j^{[n]}={\rm \bf flip}(n,j)$ denotes the integer obtained after flipping the $j$th digit in the binary representation of $n$ (see Refs.\onlinecite{Sandvik, Jung}).  If instead $\sigma_n^{(j)}=0$ (spin up), then $\hat{S}_j^{+}\ket{n}=0$.
 
 Similarly, $\hat{S}_j^{-}$ acts non-trivially only when $\sigma_n^{(j)}=0$ (spin up) in which case it flips the bit $0\to 1$; otherwise it annihilates the state. The diagonal operator $\hat{S}_j^{z}$ leaves the configuration unchanged and simply multiplies it by the eigenvalue 
 $s_n^{(j)z}=(1-2\sigma_n^{(j)})/2$. These observations lead to the compact matrix elements\cite{Comment3}
 \begin{subeqnarray}\label{equ3b}
 \slabel{equ3b1}	\bra{m}\hat{S}_j^{\pm}\ket{n} &=& s_n^{(j)\pm}\,\delta_{m,\mu_j^{[n]}},\\
 \slabel{equ3b2}	\bra{m}\hat{S}_j^{z}\ket{n}   &=& s_n^{(j)z}\,\delta_{m,n},
 \end{subeqnarray}
 where
 \[
 s_n^{(j)+}=\frac{1}{2}\bigl(1-2s_n^{(j)z}\bigr)=\sigma_n^{(j)},
 \qquad
 s_n^{(j)-}=\frac{1}{2}\bigl(1+2s_n^{(j)z}\bigr)=1-\sigma_n^{(j)}.
 \]
A few of their properties are displayed in Appendix \ref{AppB}. 

\begin{figure}[]
	\centering
	\begin{center} 
		\includegraphics[width=4.5cm, height=5cm]{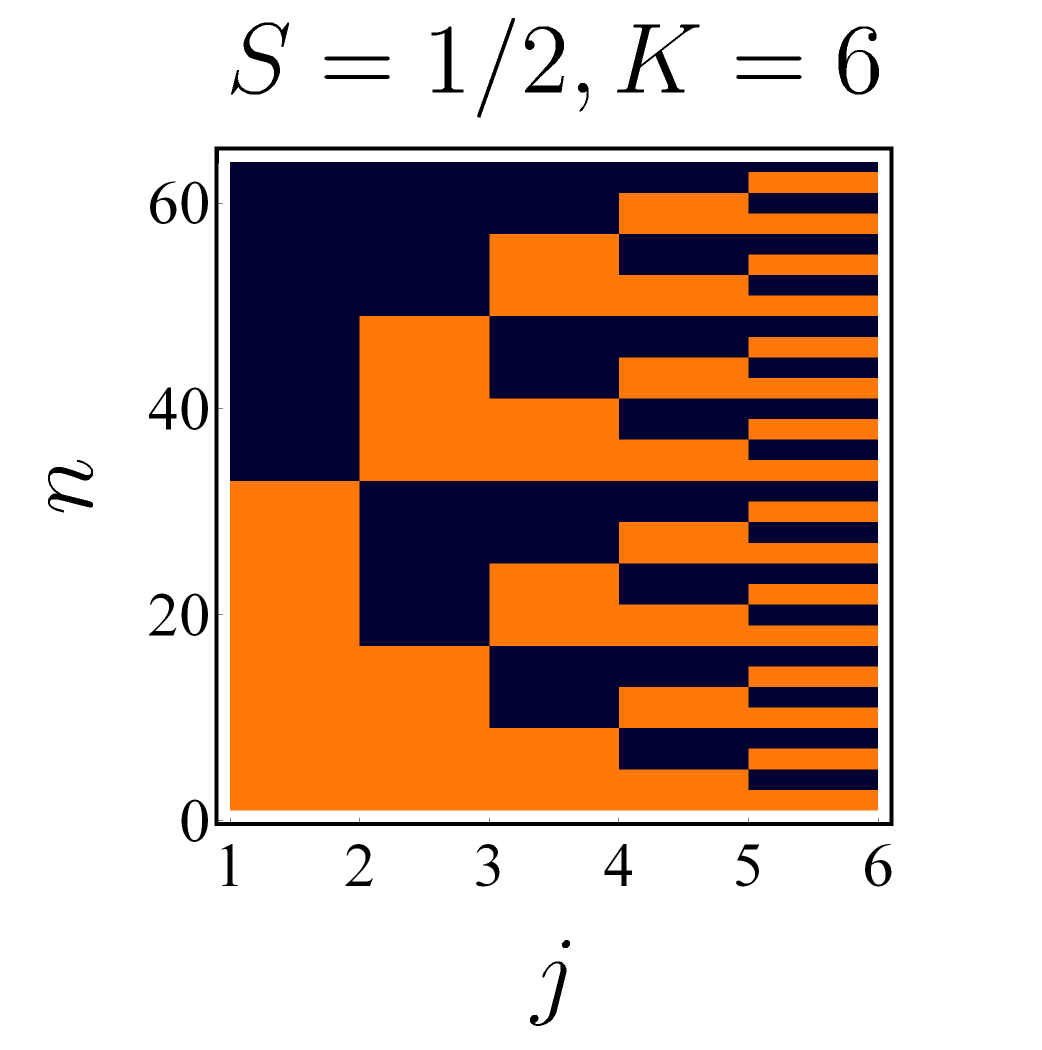}\hspace{-0.7cm}
		\includegraphics[width=4.7cm, height=5cm]{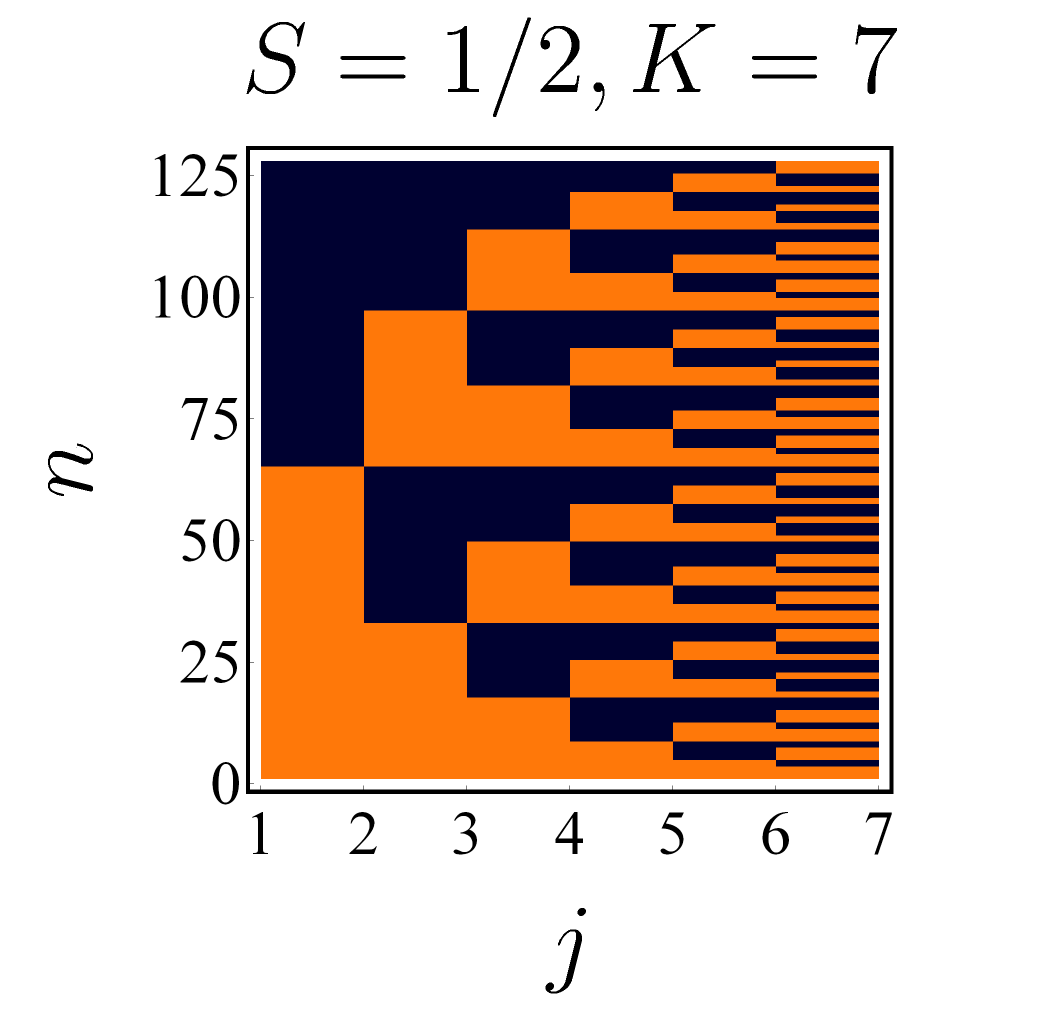} 
		\vspace{-0.65cm}
		\caption{Spin-$1/2$ distribution for two configurations of bit-field lengths $K=6$ and $K=7$. For generating data, we considered integers in $0\leq n\leq 2^K-1$ and $1\le j\le K$ and perform binary decomposition.  We see that spins along the quantization direction follow a {\it defined} pattern and are {\it not} randomly distributed as one may think. As yet another observation, consecutive even and odd numbers only differ by the least significant digits (LSDs). For example $0$ (even) and $1$ (odd) only differ by LSDs same for  $2$ (even) and $3$ (odd) etc (See Tables \ref{Table2} and \ref{Table3}). The color palette contains only two colors. Yellow color $\to 1/2$ and dark-blue $\to -1/2$ and this explains why it is not depicted.}
		\label{Figure1} 
	\end{center}
\end{figure}

 We gain further insight into the ${\bf flip}()$ function by noting from Eq.\eqref{equ3} that as $\bar{\sigma}_{n}^{\left(j\right)}$ denotes the reversed digit of $\sigma_{n}^{\left(j\right)}$ then,
\begin{eqnarray}\label{equ3d}
	\mu_j^{\left[n\right]} = n + \left(\bar{\sigma}_{n}^{\left(j\right)}-\sigma_{n}^{\left(j\right)}\right)q_j.
\end{eqnarray}
We learn from here that flipping an integer $n$ of bit-field length $K$ at the $j$th site is equivalent to adding $q_j$ (if $\sigma^{(j)}_{n}=0$) or subtracting $q_j$ (if $\sigma^{(j)}_{n}=1$) to the said number. For instance, consider the set of non-negative integers of bit-field length $K=6$ i.e. $0\le n\le 2^{6}-1$. Consider again $n=5$ in the binary basis, $5=\left(000101\right)_2$. Let us reverse the second digit counting from the left to right i.e. $j=2$. Because $\sigma_{5}^{\left(2\right)}=0$ reversing this digit yields $\left(010101\right)_2=21$ which is nothing but $5+2^{6-2}=21$ ascertaining our assertion (see complete proof in Appendix \ref{AppB}). In the same vein, it can be proven that $\mu_j^{[0]}=q_j $ and $\mu_j^{[D-1]}=D-1-q_j $.  Similarly, the integer $\bar{n}$ obtained after reversing all digits in an integer $n$ reads,
$
	\bar{n} = n + \sum_{j=1}^{K}\left(\bar{\sigma}_{n}^{\left(j\right)}-\sigma_{n}^{\left(j\right)}\right)q_j=\sum_{j=1}^{K}\bar{\sigma}_{n}^{\left(j\right)}q_j=D-1-n
$
whereby, $\sigma_{\bar{n}}^{\left(j\right)}=\bar{\sigma}_{n}^{\left(j\right)}$.
 For even integers, the least significant digit (LSD) $\sigma^{(K)}_n=0$, $\mu_K^{[n]} = n+1$ while for odd integers the LSD is $\sigma^{(K)}_n=1$ and $\mu_K^{[n]} = n-1$. Note that the LSD is the digit at $j=K$ as we read from the left to the right. Eq.\eqref{equ3d} can now be regarded as
\begin{eqnarray}\label{equation3e}
	\mu_j^{\left[n\right]}=
	\left\{\begin{array}{c}
		n+q_j, \quad {\rm if} \quad \sigma^{(j)}_n=0\\ 
		n-q_j, \quad {\rm if} \quad \sigma^{(j)}_n=1 
	\end{array}
	\right. , \quad \mu_j^{[\mu_j^{\left[n\right]}]}=n.
\end{eqnarray}
Then,  $n$ and $\mu_j^{\left[n\right]}$ only differ at the $j$th site where a bit is reversed. Remarkably, as for $\sigma^{(j)}_n=0$ the integer obtained after flipping this bit is $n+q_j$   we infer that $\sigma^{(j)}_{n+q_j}=1$. Similarly, as $\sigma^{(j)}_n=1$ and the bit resulting from inversion is $n-q_j$ then $\sigma^{(j)}_{n-q_j}=0$. Note also that given an integer $q_j$ (i.e. a number consisting of $0$ everywhere excepted the $j$th position) it can be verified that 
$\sigma_{q_j}^{(i)}=\delta_{i,j}$.

\subsection{Classical representation}
Now, based on Eqs.\eqref{equ3b1}-\eqref{equ3b2} and their trivial relations with ladder spin  operators and thanks to Eq.\eqref{equation3e}, we can now make manifest the promised Dirac representation for spin-$1/2$ (See also Ref.\onlinecite{Gluza}),
\begin{subeqnarray}\label{equ3f}
	\slabel{equ3f1}\hat{S}_{j}^{+} &=& \sum_{n=0}^{D-1}s_n^{\left(j\right)-}\ket{n}\bra{n+q_j}, \\ 
	\slabel{equ3f2}\hat{S}_{j}^{-} &=& \sum_{n=0}^{D-1}s_n^{\left(j\right)-}\ket{n+q_j}\bra{n},\\
	\slabel{equ3f3}\hat{S}_{j}^{z} &=& \sum_{n=0}^{D-1}s_n^{\left(j\right)z}\ket{n}\bra{n}.
\end{subeqnarray}
The appearance of $s_n^{(j)-}$ in both $\hat{S}_j^{+}$ and $\hat{S}_j^{-}$ reflects our encoding
$\sigma_n^{(j)}=0 \leftrightarrow \ket{\uparrow}_j$ and $\sigma_n^{(j)}=1 \leftrightarrow \ket{\downarrow}_j$. This ensures that $(\hat{S}_{j}^{\pm})^{\dagger}=\hat{S}_{j}^{\mp}$ given that the embedded classical objects are real, $(s_{n}^{\left(j\right)\pm})^{\dagger}=s_{n}^{\left(j\right)\pm}$. The summation index $n$ is chosen such that the pair $(n,n+q_j)$ represents two configurations that only differ by the $j$th bit. Hence $s_n^{(j)-}$ projects onto the subset with $\sigma_n^{(j)}=0$ (spin up), so that $\hat{S}_j^{-}$ maps $\ket{n}$ to $\ket{n+q_j}$, while $\hat{S}_j^{+}$ is the Hermitian conjugate maps from $\ket{n+q_j}$ back to $\ket{n}$.

 A second representation in terms of $s_n^{\left(j\right)+}$ can be constructed. Indeed, using $\mu_j^{[n]}=n\pm q_j$ together with the conjugation relations
$s_{n-q_j}^{(j)-}=s_n^{(j)+}$ and $s_{n+q_j}^{(j)+}=s_n^{(j)-}$  (see Appendix~\ref{AppB}),
one obtains an equivalent representation,
\begin{subeqnarray}\label{equ3g}
\slabel{equ3g1} \hat S_j^+ &=& \sum_{n=0}^{D-1} s_n^{(j)+}\ket{n-q_j}\bra{n},\\
\slabel{equ3g2}\hat S_j^- &=& \sum_{n=0}^{D-1} s_{n}^{(j)+}\ket{n}\bra{n-q_j}.
\end{subeqnarray}
Relabeling $n\mapsto n+q_j$ in these expressions immediately recovers Eqs.~\eqref{equ3f1} and \eqref{equ3f2}.

As promised, the above representations are simpler than their quantum counterparts. Representation of LSOs in terms of classical variables involves a single summation over the entire Hilbert space $\boldsymbol{\mathcal{H}}_{\mathcal{B}}$ whereas their equivalent quantum counterparts require as many nested summations as the number of spins in the chain and this is clearly a major advantage. While the Hilbert-space dimension remains exponential, the resulting Dirac forms reduce the number of explicit nested sums and yield sparse, structured matrix elements that are advantageous both analytically and numerically.

We emphasize that Eqs.\eqref{equ3f1}-\eqref{equ3f3} contain no assumption regarding the spatial dimension and lattice geometry. As already pinpointed, $j$ labels lattice sites. Therefore, our representations apply for one-, two-, and higher-dimensional lattices as well as to arbitrary interaction graphs. Dimensionality and geometry enter only through the choice of interacting site pairs in the Hamiltonian.

It can also be verified that our results do not suffer from any non-locality nor non-Hermiticity and are extendable to higher spin-size (See Section \ref{Section5} for spin-$1$). Importantly, they preserve the $su(2)$ algebra. 

For a single spin system, $K=1$ i.e. $j=1$, we retrieve the traditional representation for Pauli matrices $\sigma^{\alpha}$ (with $\alpha=x,y,z$)\cite{sakurai}, $\hat{S}_{1}^{+}=\ket{0}\bra{1}\equiv\frac{1}{2}\sigma^+$, $\hat{S}_{1}^{-}=\ket{1}\bra{0}\equiv \frac{1}{2}\sigma^-$ and $\hat{S}_{1}^{z}=\frac{1}{2}\left(\ket{0}\bra{0}-\ket{1}\bra{1}\right)\equiv\frac{1}{2}\sigma^z$ where $\sigma^{\pm}=\sigma^{x}\pm i\sigma^{y}$. The above representations can also be regarded as generalization of "braket" representation of Pauli's matrices for LSOs. 

At this stage, our construction is not yet complete. As already mentioned,  $s_n^{\left(j\right)+}$ is the coefficient of $q_j$ in the binary decomposition of $n$. This remark suggests from common perception that $s_n^{\left(j\right)+}$ can only be numerically evaluated (i.e.  for large numbers with the aid of a computer and/or manually for small numbers). We have successfully disproved this claim by writing $s_n^{\left(j\right)+}$ in analytical form such that one can obtain the $j$th digit of a number without explicit binary decomposition. Indeed, after a profound inspection into the binary structure of a few numbers (see Appendix.\ref{AppA}  for complete proof), we realized that binary digits enter non-negative integers through a well-defined pattern (see Fig.\ref{Figure1}). We identified this pattern and came up with
\begin{subeqnarray}\label{equ6}
	\hspace{-0.75cm}\slabel{equ6a}	s_n^{\left(j\right)+} &=& \sum_{m=0}^{r_{j-1}-1}\left[u\left(n-(2m+1)q_j\right)-u\left(n-(2m+2)q_j\right)\right],\quad\\
	\hspace{-0.75cm}\slabel{equ6b}	s_n^{\left(j\right)-} &=& \sum_{m=0}^{r_{j-1}-1}\left[u\left(n-2mq_j\right)-u\left(n-(2m+1)q_j\right)\right], \\
	\hspace{-0.75cm}\slabel{equ6c}	s_n^{\left(j\right)z} &=& \frac{1}{2}\sum_{m=0}^{r_j-1}\left(-1\right)^m\left[u\left(n-mq_j\right)-u\left(n-(m+1)q_j\right)\right],  
\end{subeqnarray}
where $r_j=2^j$ and
\begin{eqnarray}\label{eq:unit-step_def}
u\left(x\right)=\left\{\begin{array}{c}
	1, \quad x\ge0\\ 
	0, \quad x< 0 
\end{array}\right.,
\end{eqnarray}
denoting the unit-step function\cite{UnitStep}. This completes our goal. It is worth emphasizing that $u\left(x\right)$ is {\it\bf not} the Heaviside step function $\Theta\left(x\right)$, in other words $u\left(0\right)\neq 1/2$ but as stated $u\left(0\right)=1$ (further clarifications are provided shortly). A simple Mathematica program allows one to verify the correctness of these results. This is done and added to the following \href{https://github.com/Kenmax15}{GitHub repository}. It was shown in Ref.\onlinecite{Venetis2014} that
\begin{eqnarray}\label{equ8b}
	u\left(x\right)=\frac{1}{\pi}\left(\arctan(x-1)+\arctan\left(\frac{2-x}{x}\right)+\frac{3\pi}{4}\right).
\end{eqnarray}
Although derived in the context of spin systems, Eqs.~\eqref{equ6a}--\eqref{equ6c} provide a closed analytical expression for binary digits and may find applications in other areas
where digit extraction plays a role. As a consequence in physics this offers the possibility to infer the direction of a spin along the quantization direction in a given configuration. Cryptography aims at manipulating binary digits for securing communication and ensuring that only the intended recipients read the information even if it falls into the hands of unauthorized users. The possibility offered by Eqs.\eqref{equ6a}-\eqref{equ6c} to obtain the $j$th digit of a number $n$ without binary decomposition could lead to the development of new cryptography techniques that are more robust, secure and harder to break. Number theory is devoted to the properties of numbers. This could lead to new insight into the properties of numbers and their relationships with one another. In computer science, these could allow the development of faster algorithms for performing binary operations especially for large numbers. 

It can be seen that our transformations preserve the $su(2)$ algebra at the same site. They do not induce any non-locality as in the case of JW transformations nor non-linearity  as for HP. They remain local and linear. There is also no constraint related to population occupancy. Let us now clarify what we mean by non-locality and non-linearity. LSOs in the spin representation are local i.e. each operator acts solely on a single site and no coupling to other sites is required. In the case of JW representation, a ladder operator acting at site $j$ reads $\hat{S}_j^+=\hat{c}^{\dagger}_j\prod_{k=1}^{j-1}\exp[i\pi \hat{c}^{\dagger}_k \hat{c}_k]$ where $\hat{c}^{\dagger}_k$ and $\hat{c}_k$ are respectively fermion creation and annihilation operators. The factor $\prod_{k=1}^{j-1}\exp[i\pi \hat{c}^{\dagger}_k \hat{c}_k]$  is a non-local string operator as it involves all fermionic sites from $1$ to $j-1$. Non-locality arises as consequence of the fact that the string operator must fulfill the fermionic anti-commutation relations. Non-locality makes it hard to extend JW to higher dimensions. For HP, $\hat{S}_j^+=\sqrt{2S-\hat{a}^{\dagger}_j \hat{a}_j}\hat{a}_j$ where $S$ is the spin size, $\hat{a}^{\dagger}_j$ and $\hat{a}_j$ are boson creation and annihilation operators  respectively. This representation is local but it usage requires approximating the square root. This is done by assuming $S\gg 1$ leading to non-linearity in the boson representation. For spin $S=1/2$, this leads to poor convergence of the truncation.

A few general remarks are in order. Firstly, because of the constraint $n\ge 0$ our treatments hold for
\begin{eqnarray}\label{equ3g}
	s_n^{\left(j\right)\pm} = 0, \quad n< 0.
\end{eqnarray}
Secondly, $s_0^{(j)+} = 0 (s_0^{(j)-}=1, s_0^{(j)z}=\frac{1}{2})$ and  $s_{D-1}^{(j)+} = 1 (s_{D-1}^{(j)-} = 0, s_{D-1}^{(j)z}=-\frac{1}{2})$ for arbitrary $j$ and $K$. In what follows, the spectrum of $q_j$ is assumed to fall within the window $1\leq q_j\le 2^{K-1}$. Thus, every value of $q_j$ that lies outside this region is conventionally set to $0$ i.e. $q_{-j}=0$ and,
\begin{eqnarray}\label{equ3h}
	q_0 = q_{K+1} = q_{K+2} = \cdots = 0.
\end{eqnarray}
While in the conventional spin representation the action of $\hat{S}_{j}^{\pm}$ changes the local magnetic quantum number  $m_S^{(j)}$ by $\pm 1$, in the binary ordering of  $\boldsymbol{\mathcal{H}}_{\mathcal{B}}$ the same operation shifts the 
global basis index by $\mp q_j$. Indeed, it can easily be verified that $\hat{S}_{j}^{\pm}\ket{n}=s_n^{\left(j\right)\pm}\ket{n\mp q_j}$. Therefore, the action of $\hat{S}_{j}^{+}$ onto the basis vector $\ket{n}$ returns the $j$th bit in the binary representation of $n$ and shifts it to the $(n-q_j)$th position in $\boldsymbol{\mathcal{H}}_{\mathcal{B}}$ while $\hat{S}_{j}^{-}$ returns the complementary bit $s_n^{\left(j\right)-}$ and shifts the basis vector to the $(n+q_j)$th position. 

We have so far achieved a Dirac representation in terms of unit-step function. This is a discrete notation. It would be interesting to move from the discrete to a continuous representation. We use the Heaviside function $\Theta(x)$. It is worth noting that $u(x)$ coincides with $\Theta(x)$ everywhere on the real axis except at the origin (turning point) where $u(0)=1$ while $\Theta(0)=1/2$. We can rewrite Eqs.\eqref{equ6a}-\eqref{equ6c} in terms of $\Theta(x)$ by excluding every point where the argument of $u(x)$ vanishes. This is achieved by discriminating between even and odd positive integers. Thus, for even $n$,
\begin{eqnarray}\label{equ9g}
	\hspace{-0.25cm}\nonumber s_{n}^{\left(j\right)z} =\frac{1}{2}\sum_{m=0}^{r_j-1}\left(-1\right)^m\left( \Theta\left(\mathcal{Q}_{n+1,m}^{(j)}\right)+\Theta\left(\mathcal{Q}_{n+1,m+1}^{(j)}\right)\right)+\frac{1}{2}\delta_{j,K},\\  
\end{eqnarray}
while for odd $n$,
\begin{eqnarray}\label{equ9qg}
	\hspace{-1cm} s_{n}^{\left(j\right)z} =\frac{1}{2}\sum_{m=0}^{r_j-1}\left(-1\right)^m\left( \Theta\left(\mathcal{Q}_{n,m}^{(j)}\right)+\Theta\left(\mathcal{Q}_{n,m+1}^{(j)}\right)\right)-\frac{1}{2}\delta_{j,K},  
\end{eqnarray}
and where we have defined $\mathcal{Q}_{n,m}^{(j)}=n-mq_j$ and a smooth regularization of the Heaviside distribution\cite{sakurai}
\begin{eqnarray}\label{equ9c}
	\Theta\left(x\right)=\lim_{\lambda\to\infty}\frac{1}{1+e^{-\lambda x}}. 
\end{eqnarray}
$s_n^{\left(j\right)+}$ and $s_n^{\left(j\right)-}$ are calculated from  \eqref{equA3a}. Numerical proofs are available in the \href{https://github.com/Kenmax15}{GitHub repository}. We have verified symbolically (using exact arithmetic) that the discrete Dirac representation expressed in terms of the unit-step function $u(x)$, with the convention $u(0)=1$, is exactly equivalent, for all $n\in\{0,\dots,2^K-1\}$ and all $j\in\{1,\dots,K\}$, to the continuous (distributional) representation written in terms of the symmetric Heaviside distribution $\Theta(x)$ satisfying $\Theta(0)=\tfrac12$, provided that the turning points $n-m q_j=0$ are treated with the symmetric prescription. In particular, the identities $s_n^{(j)+}+s_n^{(j)-}=1$, $n=\sum_{j=1}^{K} s_n^{(j)+}q_j$ hold exactly in both representations. We also confirm that the LSD for even integers $s_{n}^{\left(K\right)z}=1/2$ whereas $s_{n}^{\left(K\right)z}=-1/2$  for odd ones. The proof is computer-assisted and performed symbolically without any floating-point approximation.	

\subsection{States representation}
As already indicated $\ket{n+q_j}$ is the state obtained by flipping the $j$th bit in the binary representation of $\ket{n}$ provided $s_n^{\left(j\right)+}=0$. Therefore $\ket{n+mq_j}$ could be viewed as the state achieved by flipping the $j$th bit $m$ times. If $m$ is even, then $s_{n+mq_j}^{\left(j\right)+}=s_{n}^{\left(j\right)+}$ i.e. the action of flipping a single digit an even number of times takes it back to its initial position preserving the original state. If in opposition $m$ is odd, thus $s_{n+mq_j}^{\left(j\right)+}=s_{n+q_j}^{\left(j\right)+}=s_{n}^{\left(j\right)-}$.  As a consequence of these observations, every basis vector can be constructed from $\ket{0}\equiv \ket{000\cdots 000}$ (vacuum state, all spins up in our encoding) by inverting local  spin directions only once as it is done in quantum mechanics by operational application. Thus,
\begin{eqnarray}\label{equ8aa}
	\ket{q_j} = \ket{000\cdots 1_{j}\cdots 000},
\end{eqnarray}
describes all states with one spin in down state $(K_{\downarrow}=1)$.
For spin-$1/2$, the total magnetization is
$\mathcal{J}^z=(K_{\uparrow}-K_{\downarrow})/2=K/2-K_{\downarrow}$,
so this sector corresponds to $\mathcal{J}^z=K/2-1$ (assuming the Hamiltonian possesses a global $U(1)$ symmetry and thus conserves $\mathcal{J}^z$). The sub-Hilbert space consists of the complete set of basis vectors $\{\ket{q_j}\}_{1\leq j\leq K}$ equipped with  $\braket{q_i}{q_j}=\delta_{i,j}$. 

Similarly, if from the fully symmetric state one inverts two spins respectively located at positions $j_1$ and $j_2$ the resulting state reads,
\begin{eqnarray}\label{equ8ab}
	\ket{q_{j_1}+q_{j_2}} = \ket{000\cdots 1_{j_1}\cdots 1_{j_2}\cdots 000},
\end{eqnarray}
with $1\leq j_1< j_2\leq K$. The sub-Hilbert space in this sector contains $\begin{pmatrix} K \\ 2\end{pmatrix}=K(K-1)/2$ basis vectors. If we equally assume the canonical ordering $i_1<i_2$ such that the basis vectors are uniquely labeled then
$\braket{q_{i_1}+q_{i_2}}{q_{j_1}+q_{j_2}}=\delta_{i_1,j_1}\delta_{i_2,j_2}$. The basis vectors \eqref{equ8ab} exactly match the set obtained by addition of two distinct elements in the previous subset. This process can be repeated until the entire set of sub-Hilbert spaces is generated the last but one of which contains $\binom{K}{K-1}=K$ states with $K-1$ spins in the down state,
\begin{eqnarray}\label{equ8ac}
	\ket{q_{j_1}+q_{j_2}+\cdots +q_{j_{K-1}}} = \ket{1_{j_1}1_{j_2}\cdots 0\cdots 1_{j_{K-1}}},
\end{eqnarray}
with $1\leq j_1< j_2<\cdots< j_{K-1}\leq K$. It should be noted that $\ket{q_{j_1}+q_{j_2}+\cdots +q_{j_K}} = \ket{1_{j_1}1_{j_2}\cdots 1_{j_K}}\equiv \ket{2^K-1}$. In each sector, the basis vectors normalizes as
\begin{eqnarray}\label{equ8acc}
	\braket{q_{i_1}+q_{i_2}+\cdots +q_{i_{K_{\downarrow}}} }{q_{j_1}+q_{j_2}+\cdots +q_{j_{K_{\downarrow}}}} = \prod_{k=1}^{K_{\downarrow}}\delta_{i_k,j_k}.\qquad
\end{eqnarray}
with $i_1< i_2<\cdots< i_{K_{\downarrow}}$ and $j_1< j_2<\cdots< j_{K_{\downarrow}}$. 
For two distinct sectors, this identity is identically zero.

These observations are useful to evaluate the eigenspectrum of spin Hamiltonians on arbitrary lattices or interaction graphs. In the following Sections, we illustrate this general construction for a $1D$ chain.

In the same logic, except for $s_{q_{j_1}+q_{j_2}+\cdots + q_{j_{K}}}^{\left(i\right)+} = 1$, it can easily be seen that
\begin{subeqnarray}\label{equ3gh}
&	s_{q_j}^{\left(i\right)+} = \delta_{i,j},\\
&	s_{q_{j_1}+q_{j_2}}^{\left(i\right)+} = \delta_{i,j_1} + \delta_{i,j_2},\\
&   \nonumber\vdots \\
&	s^{\left(i\right)+}_{q_{j_1}+q_{j_2}+\cdots + q_{j_{K}}} = \sum_{\ell=1}^{K}\delta_{i,j_{\ell}}.
\end{subeqnarray}
Hereby, we verified for further purposes that $\sum_{i=1}^{K}s_{q_j}^{\left(i\right)+}=1$, $\sum_{i=1}^{K}s_{q_{j_1}+q_{j_2}}^{\left(i\right)+}=2$ gradually until $\sum_{i=1}^{K}s^{\left(i\right)+}_{q_{j_1}+q_{j_2}+\cdots + q_{j_{K-1}}}=K-1$. Similarly, using Eq.~\eqref{equA3}, one obtains
\begin{subeqnarray}\label{equ3gi}
	&	s_{q_j}^{\left(i\right)z} = \frac{1}{2}\bar{\delta}_{i,j},\\
	&	s_{q_{j_1}+q_{j_2}}^{\left(i\right)z} = \frac{1}{2}\left(\bar{\delta}_{i,j_1} + \bar{\delta}_{i,j_2}\right),\\
	&   \nonumber\vdots \\
	&	s^{\left(i\right)z}_{q_{j_1}+q_{j_2}+\cdots + q_{j_{K}}} = \frac{1}{2}\sum_{\ell=1}^{K}\bar{\delta}_{i,j_{\ell}},
\end{subeqnarray}
where we have defined
\begin{eqnarray}\label{eq:bardelta_def}
	\bar{\delta}_{i,j}=\left\{\begin{array}{c}
		-1, \quad i=j\\ 
		+1, \quad i\neq j 
	\end{array}
	\right. ,
\end{eqnarray}
Importantly,
\begin{eqnarray}\label{eq:bardelta_prod}
	\bar{\delta}_{i,j}\bar{\delta}_{i,j+1}=
	\left\{\begin{array}{cc}
		-1, &\quad i=j\quad \text{or}\quad i=j+1,\\[2pt]
		+1, &\quad i\neq j\quad \text{and}\quad i\neq j+1.
	\end{array}\right.
\end{eqnarray}
in opposition to $\delta_{i,j}\delta_{i,j+1}=0$ as proven below. Thus, because of the constraint $j_1<j_2$ we always have $\delta_{j_1,j_2}=\delta_{j_1,j_2+1}=0$ i.e. in general $\delta_{j_{k},j_{\ell}}=\delta_{j_{k},j_{\ell}+1}=0$ for $k < \ell$; the only contribution coming from $k > \ell$ ($k=\ell$ being excluded). It is also important for further development to keep in mind that $\left(1+\delta_{i,j}\right)\left(1-\delta_{i,j}\right)=1-\delta_{i,j}$ and $\left(1-\delta_{i,j}\right)\left(1-\delta_{i,j}\right)=1-\delta_{i,j}$.

\begin{figure}[]
	\centering
	\begin{center} 
		\includegraphics[width=6cm, height=6cm]{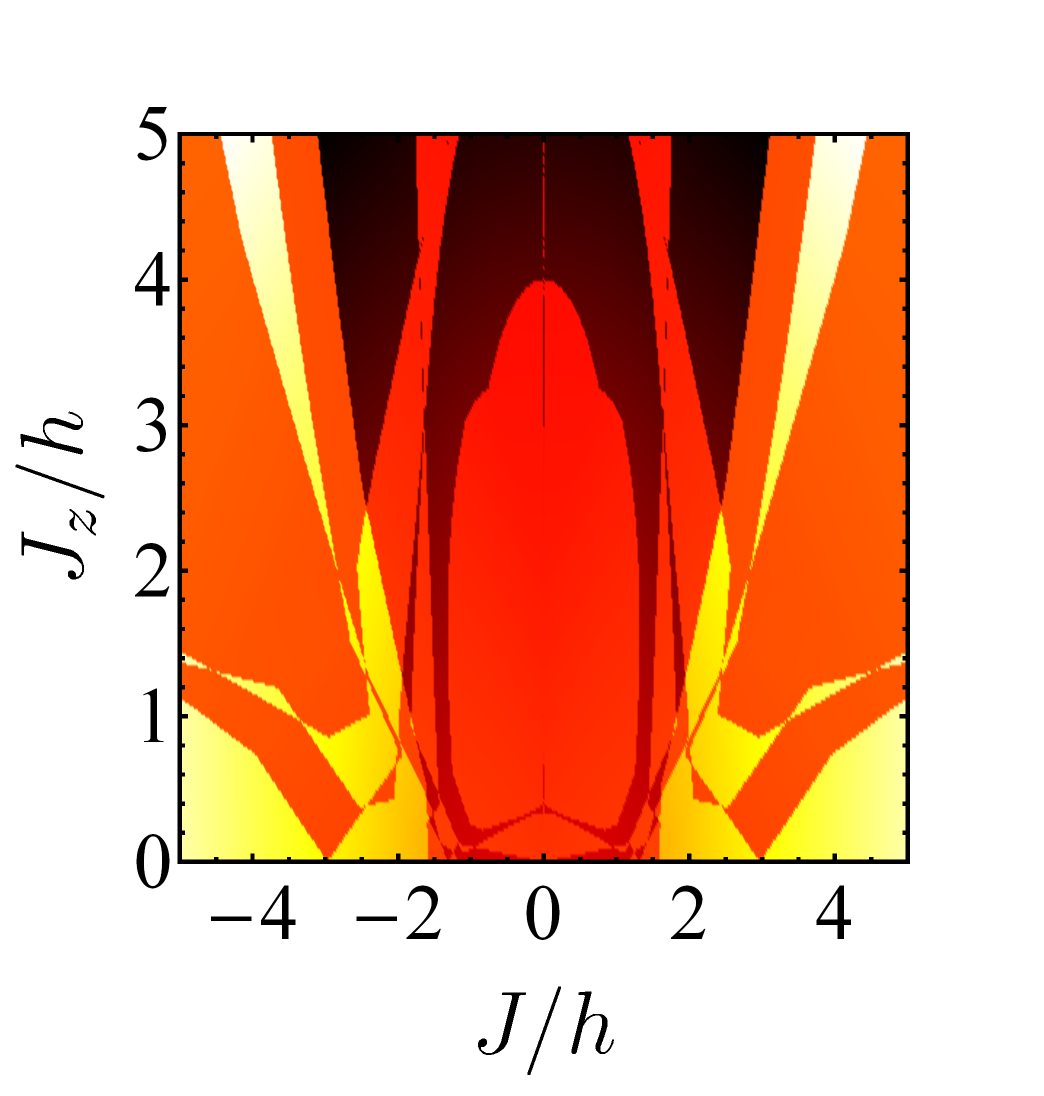}\hspace{-1.5cm}
		\includegraphics[width=2.8cm, height=5.5cm]{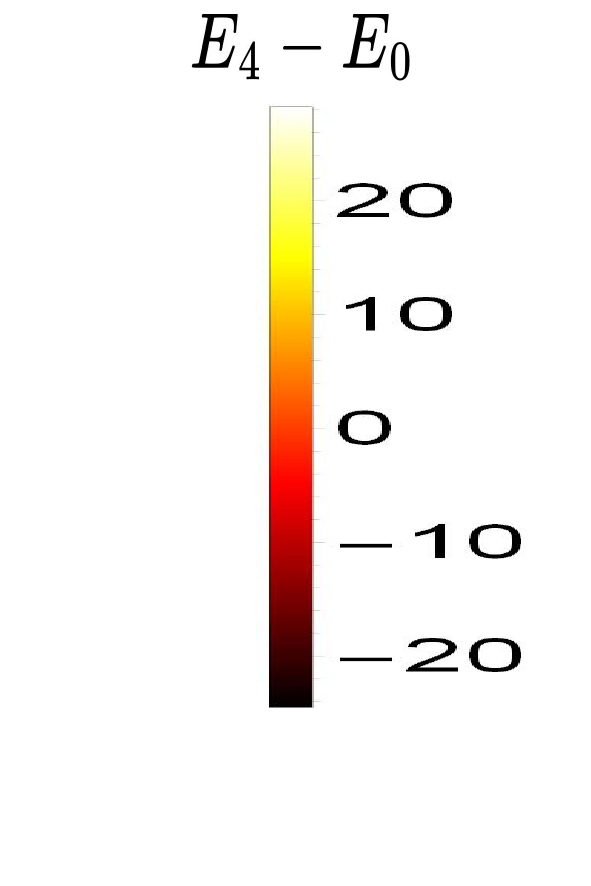}
		\vspace{-0.5cm}
		\caption{Numerical calculation of the gap $E_4-E_0$ as a function of the dimensionless spin-spin exchange $J/h$ and the dimensionless anisotropy parameter $J_z/h$ in the spin-$1/2$ XXZ Heisenberg model for $K=8$ spins. These two parameters interplay in a non-trivial manner leading to the color palette running from $-20$ to $20$. Three main colors dominate the portrait. From the bottom to the top we go from the minimum to the maximum gap through the red area corresponding to the region where the gap closes ($E_4-E_0=0$). The yellow region corresponds to the region where the gap tends to be maximal while the black region corresponds to the minimum. The presence of fringes is an indication of intrinsic interference pattern i.e. level crossings between $E_4$ and $E_0$ by linear variation of $J/h$. This portrait is not unique, any combination of $E_n$ and $E_0$ may give a different pattern.}
		\label{figure2} 
	\end{center}
\end{figure}

\section{Applications: 1$D$ Spin-1/2 Integrable $\mathrm{XXZ}$ Heisenberg model}\label{Section3}
 After establishing a dimension-independent representation for LSOs, free from any non-locality and non-Hermicity we now consider a $1D$ system in order to illustrate its usefulness in an analytically tractable many-body problem.  We consider the $1D$ spin-$1/2$ $\mathrm{XXZ}$ Heisenberg model with open boundaries in a magnetic field (MF) and compute its eigenspectrum. Two distinct cases are of interest: the case of uniform and random MFs. For uniform MF (integrable model) the eigenspectrum is in general obtained analytically by invoking the Bethe Ansatz\cite{Bethe1931, Zhang1999} which amounts to complicated calculations or by resorting to fermionic or bosonic transformations. We demonstrate that our representation transforms the corresponding eigenvalue problem into recurrence equations that can be solved in closed form.  In the case of random MF (non-integrable model) a comprehensive perturbation theory is constructed. The prototype Hamiltonian of the study reads\cite{Hetterich, Schliemann2021}
\begin{eqnarray}\label{equ18}
	\hat{H} =\sum_{\alpha=x,y,z}\sum^{K-1}_{j=1}J_{\alpha}\hat{S}^{\alpha}_{j}\hat{S}^{\alpha}_{j+1}+\sum^K_{j=1}h_{j}\hat{S}^{z}_{j},
\end{eqnarray}
to which we further impose the constraint $J_{x}=J_{y}=J$ (anisotropic XXZ spin-$1/2$) with $J$ measuring spin-spin interactions; $J_z$ is the spin anisotropy parameter,  $h_{j}$ are local magnetic fields applied along the $z$-direction onto individual spins. Our goal is to analyze this model both numerically in the Dirac (binary) representation with the aid of exact diagonalization and analytically without invoking the Bethe Ansatz or any fermionic or bosonic transformation.

\subsection{Numerical Treatment}\label{Section NT}
Exact diagonalization (ED) in the model \eqref{equ18} is achieved by exploiting two main symmetries: the global $U(1)$ and spin-inversion symmetries.  These allow us to considerably reduce the size of the Hilbert space.  Our choice of spin-spin interaction ensures that $\hat{H}$ has a global $U(1)$ symmetry thus preserving the total magnetization. In the binary basis, this reads
\begin{eqnarray}\label{equ18a}
	\hat{\mathcal{J}}^z=\sum_{n=0}^{D-1}m_n^{z}\ket{n}\bra{n}, \quad \left[\hat{H},\hat{\mathcal{J}}^z\right]=0.
\end{eqnarray}
Here, $m_n^{z}=\sum^K_{j=1}s_n^{\left(j\right)z}$ is the $n$th magnetic quantum number written in explicit form in \eqref{equA9e}. The conservation law \eqref{equ18a} decomposes the Hilbert space into disconnected magnetization sectors\cite{Yahui2023, Ganguli2025}. There are various ways for distinguishing between the sectors. One way consists of labeling them using the total magnetization $\hat{\mathcal{J}}^z$ in which case $-K/2\leq \hat{\mathcal{J}}^z\leq K/2$ for even number of spins. An alternative way consists of using the number of spins in either the state $\ket{\uparrow}$ or $\ket{\downarrow}$. We label the sectors with the number $K_{\downarrow}$ of spins in down state starting from the smallest values; $0\leq K_{\downarrow}\leq K$. 
The tensor product of subspaces in \eqref{equ1} decomposes into direct sum of subspaces as
\begin{eqnarray}\label{equ18b}
	\boldsymbol{\mathcal{H}}=\bigoplus_{K_{\downarrow}=0}^{K}\boldsymbol{\mathcal{H}}^{\left(K_{\downarrow}\right)}.
\end{eqnarray}
Each subspace is of dimension ${\rm\bf dim}\boldsymbol{\mathcal{H}}^{\left(K_{\downarrow}\right)}=\begin{pmatrix} K \\ K_{\downarrow}\end{pmatrix}$. This is smaller than the $D=2^K$ dimension of the original Hilbert space. The Hamiltonian decomposes in a similar fashion. Thanks to this symmetry, states with different value of $\hat{\mathcal{J}}^z$ are not mixed and the Hamiltonian can be diagonalized independently in each subspace. The gaps between the energy of the fourth excited state and the ground-state energies ($E_4-E_0$) is calculated as a function of $J/h$ and $J_z/h$ and the results are depicted in Fig.\ref{figure2}. A similar task is accomplished for $E_8-E_0$ and displayed in Fig.\ref{Figure5}. We observe interesting interference patterns revealing the existence of level-crossings between $E_4$ and $E_0$ by linear variation of $J/h$ and where the wave functions split and recombine. These calculations can unfortunately only be done for a few number of spins as although exploiting the $U(1)$ global symmetry, ED continues suffering from finite-size effects. We invoke an additional symmetry.

In the absence of local magnetic fields ($h_j=0$ for all $j$), the Hamiltonian acquires an additional global spin-inversion symmetry\cite{Sandvik}. It commutes with the spin-inversion operator,
\begin{eqnarray}\label{equ18c}
	\hat{Z} = \sum_{n=0}^{D-1}\ket{D-1-n}\bra{n}, \quad \left[\hat{H},\hat{Z} \right]=0.
\end{eqnarray}
Because of the binary encoding basis states,  $D-1-n$ corresponds to the bitwise complement. Essentially, $\hat{Z}^{\dagger}\hat{S}_{j}^{\pm}\hat{Z}=\hat{S}_{j}^{\mp}$ and $\hat{Z} ^{\dagger}\hat{S}_{j}^{z}\hat{Z} =-\hat{S}_{j}^{z}$ given in addition that $\hat{Z}  =\hat{Z} ^{-1}=\hat{Z} ^{\dagger}$ (unitary operator). The operator $\hat{Z} $ is an anti-diagonal matrix that has zeros everywhere excepted on the anti-diagonal where it possesses ones. Therefore, given that $\hat{Z} ^2=\hat{I}$, its eigenvalues are $z=\pm 1$ and its normalized eigenvectors are 
\begin{eqnarray}\label{equ18d}
\ket{n_z}=\frac{1}{\sqrt{2}}(\ket{n}+z\ket{D-1-n}).
\end{eqnarray}
They are mutually orthogonal and are equipped with a completeness relation that allows them to form a complete set of vectors,
\begin{eqnarray}\label{equ18e}
\braket{m_{z'}}{n_z}=\delta_{n,m}\delta_{z,z'},\quad \frac{1}{2}\sum_{z=\pm 1}\sum_{n=0}^{D-1}\ket{n_z}\bra{n_z}=\hat{I}.
\end{eqnarray}
When the global $U(1)$ symmetry is implemented, the spin-inversion symmetry is applicable only in the sector  $\hat{\mathcal{J}}^z=0$ (largest sector) where $\hat{Z}\ket{s_{n}^{\left(1\right)z}, s_{n}^{\left(2\right)z},\cdots, s_{n}^{\left(K\right)z}}=\ket{-s_{n}^{\left(1\right)z}, -s_{n}^{\left(2\right)z},\cdots, -s_{n}^{\left(K\right)z}}$.  This permits further fragmentation of this subspace into yet two disconnected sectors associated with the eigenvalues $z$,
\begin{eqnarray}\label{equ18f}
	\boldsymbol{\mathcal{H}}^{\left(\hat{\mathcal{J}}^z=0\right)}=\boldsymbol{\mathcal{H}}_+\bigoplus\boldsymbol{\mathcal{H}}_-.
\end{eqnarray}
The projection out of an operator (not necessary a matrix) onto each subspaces can be accomplished with the aid of the projection operator (see also Ref.\cite{Barentzen})
\begin{eqnarray}\label{equ3g-clean}
	\hat{\mathcal{P}}_z = \frac{1}{2}\left(\hat{I}+z\hat{Z}\right).
\end{eqnarray}
It is Hermitian with properties $\hat{\mathcal{P}}_{z}\hat{\mathcal{P}}_{z'}=\delta_{z,z'}\hat{\mathcal{P}}_{z}$, $\sum_{z=\pm}\hat{\mathcal{P}}_z=\hat{I}$ and $\hat{\mathcal{P}}_z\hat{Z}=z\hat{\mathcal{P}}_z$. The projection of a (basis) vector initially in the eigenbasis of $\hat{\mathcal{J}}^z$ onto each new subspace in \eqref{equ18f} can be realized as 
\begin{eqnarray}\label{equ3h}
	\ket{n}=\sum_{z=\pm 1}\sum_{k=0}^{D-1}C_{k_z,n}\ket{k_z}.
\end{eqnarray}
Here, the $C_{k_z,n}$ denoting the projection of $\ket{n}$ onto the direction of $\ket{k_z}$ in the subspace $\boldsymbol{\mathcal{H}}_z$ are determined with the aid of the relations in \eqref{equ18e} as
\begin{eqnarray}\label{equ3j}
	C_{k_z,n}=\frac{1}{\sqrt{2}}\left(\delta_{k,n}+z\delta_{k,D-1-n}\right).
\end{eqnarray}
We see that  the basis vectors mix the two subspaces as indicated by $\ket{n}=\frac{1}{\sqrt{2}}\sum_{z=\pm 1}\left(\ket{n_z}+z\ket{\left(D-1-n\right)_z}\right)$. Every matrix operator $\hat{\mathcal{O}}=\sum_{n,m=0}^{D-1}\hat{\mathcal{O}}_{n,m}\ket{n}\bra{m}$ in the eigenspace \eqref{equ18b} is projected out onto the eigenspaces $\boldsymbol{\mathcal{H}}_z$ as
\begin{eqnarray}\label{equ3k}
	\hat{\mathcal{O}}=\sum_{z',z}\sum_{n,m=0}^{D-1}\left(\bra{m_{z'}}\hat{\mathcal{O}}\ket{n_z}\right)\ket{n_z}\bra{m_{z'}},
\end{eqnarray}
where
\begin{eqnarray}\label{equ3i}
\hspace{-0.5cm}	\nonumber\bra{m_{z'}}\hat{\mathcal{O}}\ket{n_z}=\frac{1}{2}\Big(\hat{\mathcal{O}}_{n,m}+z'\hat{\mathcal{O}}_{n,D-1-m}+z\hat{\mathcal{O}}_{D-1-n,m}\\+zz'\hat{\mathcal{O}}_{D-1-n,D-1-m}\Big),
\end{eqnarray}
are the matrix elements of $\hat{\mathcal O}$ in the $\{\ket{n_z}\}$ basis. The $\hat{\mathcal{O}}_{\beta,\alpha}$ in the left hand side of \eqref{equ3i} are components of $\hat{\mathcal{O}}$ in the eigenbasis \eqref{equ18b}. In general, $\hat{\mathcal{O}}$ mixes states with different $z$. When it preserves $\hat{Z}$ it does not mix them anymore also falling apart into two disconnected blocks corresponding each to a value of $z$.

\begin{figure}[]
	\centering
	\begin{center} 
		\includegraphics[width=7.5cm, height=6cm]{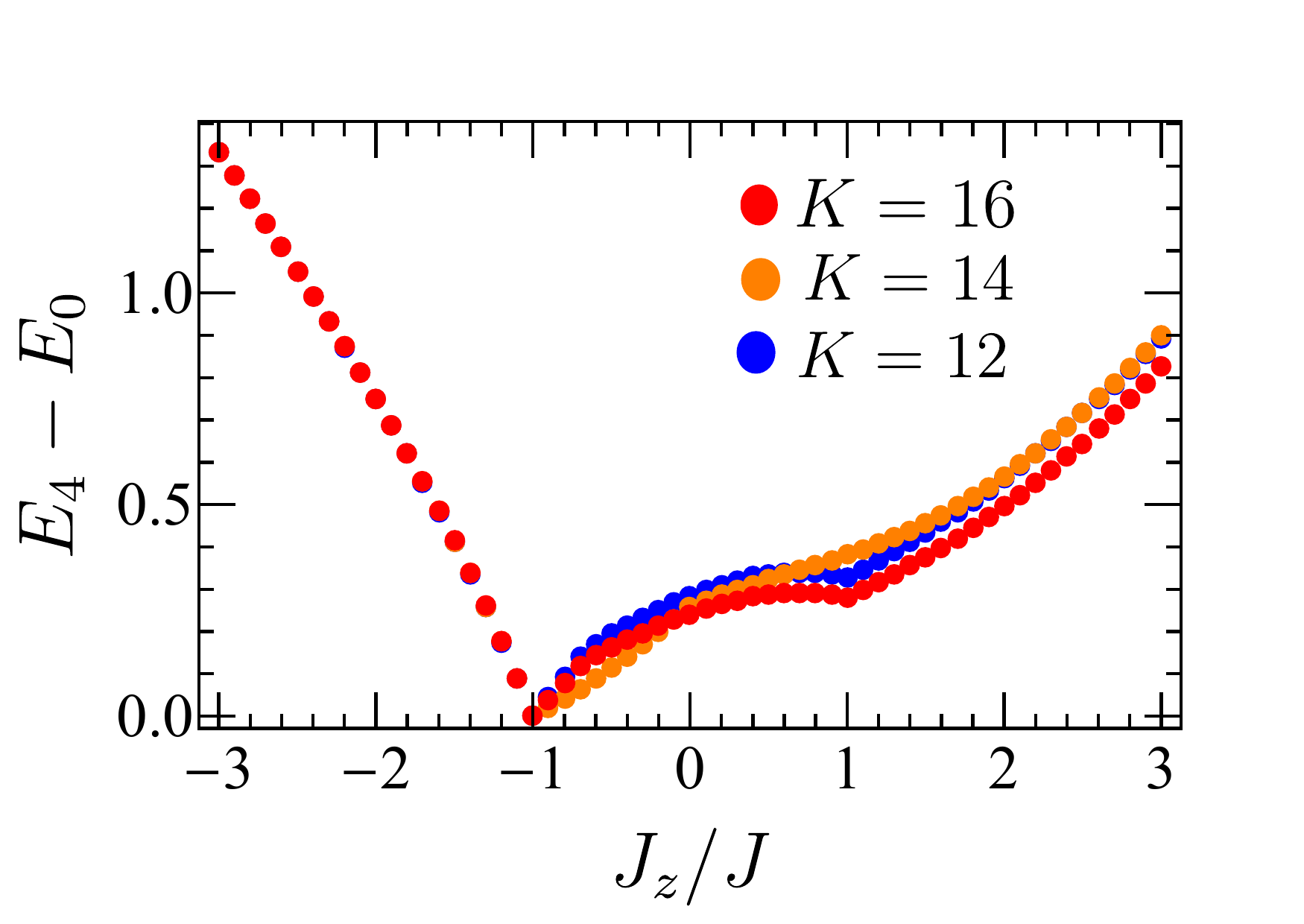}
		\vspace{-0.5cm}
		\caption{Energy difference $E_4-E_0$ in the XXZ Heisenberg model with open boundaries versus the anisotropy parameter $J_z/J$ for $h=0.J$ and various values of the total number of spins. This figure depicts three phases: (i) the regime $J_z/J< -1$ (ferromagnetic gapped phase) where we observe that the gap is essentially independent on the system size in this regime indicating a {\it bulk gap} that survives in the thermodynamic limit $K\to\infty$. We also note that as the anisotropy parameter increases the gap closes and vanishes near $J_z/J=-1$ (critical point) indicating a phase transition.  (ii) The regime $-1<J_z/J\le 1$ (Luttinger liquid) and (iii) the regime $J_z/J> 1$ (anti-ferromagnetic gapped phase).}
		\label{figure3} 
	\end{center}
\end{figure}

\begin{figure}[]
	\centering
	\begin{center} 
		\includegraphics[width=7.5cm, height=6cm]{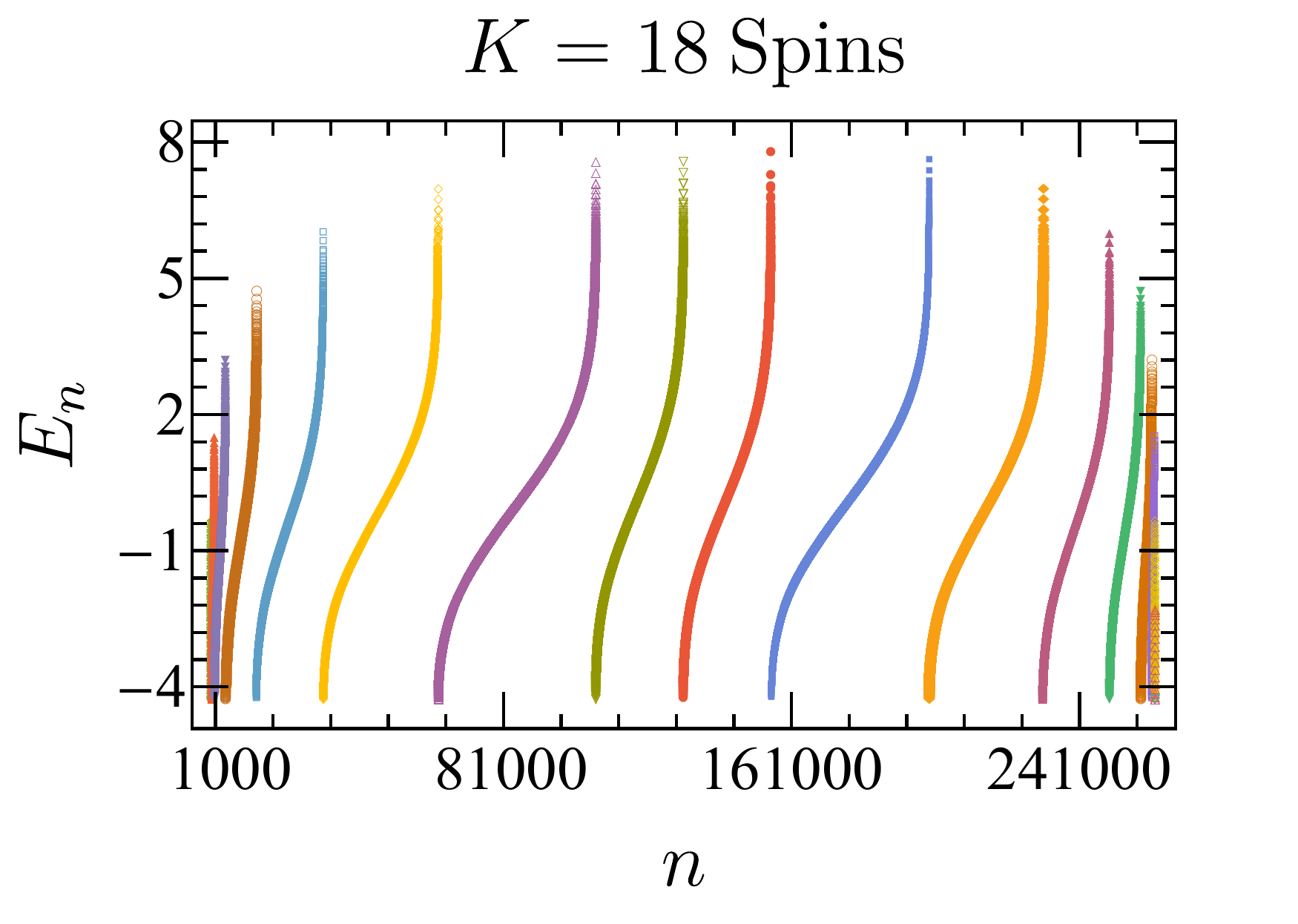}\vspace{-0.2cm}\\\hspace{0.5cm}
		\vspace{-0.5cm}
		\caption{Exact diagonalization results of a spin-$1/2$ XXZ Heisenberg model of length $K=18$ for $J_z=-J$ in the absence of local magnetic field,  $h=0.0J$.  Calculations are numerically done by block-diagonalization of the Hamiltonian and implementation of the spin-inversion symmetry in the sector $\mathcal{J}^z=0$ i.e. each sector is exactly diagonalized independently. Each colored-solid object corresponds to an eigenvalue of the total magnetization (block). Vertical curves tend to become horizontal as $h\neq 0$ (see Fig.\ref{Figure7}).}
		\label{figure4} 
	\end{center}
\end{figure}

These symmetries considerably mitigate finite-size limitations. Indeed, on a  small computer with 32Gb of RAM (6-cores Ryzen 5 processor) we have been able on the Mathematica software configured with the highly optimized Intel Math Kernel Library (MKL) to numerically diagonalize (eigenvalues only) $\hat{H}$ for systems up to $K\le18$ spins without resorting to any parallelization. Thanks to the binary representation it only takes a few seconds to build the Hamiltonian as no matrix multiplication is required. Block Hamiltonians are diagonalized one after the other. We expect that further optimization in compiled languages 
will allow treatment of even larger systems. The algorithm is available in the following \href{https://github.com/Kenmax15}{GitHub repository}.
Figures \ref{figure3}  and \ref{figure4} depict our numerical results.

\subsection{Analytical Treatment}\label{Section AT}
After gaining  insight into the numerical behavior of the generic model $\hat{H}$, we are now in position to obtain its eigenspectrum analytically. In the representation \eqref{equ3f1}-\eqref{equ3f3} it casts the form $\hat{H} = \sum_{n=0}^{D-1}\hat{\mathcal{H}}_n$ where
\begin{eqnarray}\label{equ19}
	\nonumber\hat{\mathcal{H}}_n &=& \sum_{j=1}^{K-1}\left(\Delta^{(j)+-}_{n}\ket{n-Q_j}\bra{n}+\Delta^{(j)-+}_{n}\ket{n+Q_j}\bra{n}\right)+\Delta^{zz}_n\ket{n}\bra{n},\\
	&=&  \sum_{j=1}^{K-1}\Delta^{(j)-+}_{n}\left(\ket{n}\bra{n+Q_j}+\ket{n+Q_j}\bra{n}\right)+\Delta^{zz}_n\ket{n}\bra{n},
\end{eqnarray}
and where we have defined
\begin{eqnarray}\label{equ20}
	\Delta^{(j)+-}_{n} = \frac{J}{2}s_n^{\left(j\right)+}s_{n}^{\left(j+1\right)-},
	\quad
	\Delta^{(j)-+}_{n} = \frac{J}{2}s_n^{\left(j\right)-}s_{n}^{\left(j+1\right)+},
\end{eqnarray}
and  the classical Hamiltonian
\begin{eqnarray}\label{equ21}
	\Delta^{zz}_n = J_{z}\sum_{j = 1}^{K-1}s_n^{\left(j\right)z}s_n^{\left(j+1\right)z}  + \sum_{j = 1}^{K}{h_{j}s_n^{\left(j\right)z}},
\end{eqnarray}
with $Q_{j}=q_{j}-q_{j+1}$. The quantities $\Delta^{(j)+-}_{n}$ and $\Delta^{(j)-+}_{n}$ encapsulate the microscopical structure of neighboring spins in the configuration  $\ket{n}$. For parallel neighboring spins one has $\Delta^{(j)+-}_{n}=\Delta^{(j)-+}_{n}=0$.
For anti-parallel ones, exactly one of the two coefficients is nonzero:
$\Delta^{(j)+-}_{n}=J/2$ and $\Delta^{(j)-+}_{n}=0$ if the local pattern is $(+,-)$,
whereas $\Delta^{(j)-+}_{n}=J/2$ and $\Delta^{(j)+-}_{n}=0$ if it is $(-,+)$.
Equivalently, $\Delta^{(j)+-}_{n}\,\Delta^{(j)-+}_{n}=0$ for all $n$.
Thus, when neighboring spins are simultaneously inverted, this action interchanges the quantities in \eqref{equ20}, $\Delta^{(j)+-}_{n+Q_j}=\Delta^{(j)-+}_{n}$ and $\Delta^{(j)-+}_{n-Q_j}=\Delta^{(j)+-}_{n}$. These identities are used together with a relabeling of the summation index $n \to n-Q_j$ in the full sum over $n$, to rewrite the exchange part of $\hat{H}$ in the symmetric form in the second line in Eq.\eqref{equ19}.

Also, $\Delta^{(j)+-}_{0}=\Delta^{(j)+-}_{D-1}=0$ and $\Delta^{(j)-+}_{0}=\Delta^{(j)-+}_{D-1}=0$ as all adjacent spins in the states $\ket{0}$ and $\ket{D-1}$ with extremal values of the total magnetization are parallel. We also observe that $\Delta_{\bar{n}}^{(j)+-} = \Delta_{n}^{(j)-+}$ and $\Delta_{\bar{n}}^{(j)-+} = \Delta_{n}^{(j)+-}$ for all lattice site. $(\Delta^{(j)+-}_{n})^2=\frac{J}{2}\Delta^{(j)+-}_{n}$,  $(\Delta^{(j)-+}_{n})^2=\frac{J}{2}\Delta^{(j)-+}_{n}$ by virtue of  Eq.\eqref{equ9a}.  Thereby, we infer that $(\Delta^{(j)+-}_{n})^{m}=\left(\frac{J}{2}\right)^{m-1}\Delta^{(j)+-}_{n}$ and $(\Delta^{(j)-+}_{n})^{m}=\left(\frac{J}{2}\right)^{m-1}\Delta^{(j)-+}_{n}$.

$\Delta^{zz}_n$ in Eq.\eqref{equ21} is nothing but the classical Ising model\cite{Ising1925, Yessen2014}. It corresponds to the non-interacting limit $J=0$ of \eqref{equ19}. When local magnetic fields are randomly distributed as in the MBL  case\cite{Schliemann2021, Siegl2023}, it describes a single-body localization\cite{Anderson, AndersonLoc1}.  Therefore, in this limit a wide class of problems ranging from magnetism\cite{Jiles1998}, phase transitions\cite{Subir2011}, spin glasses\cite{Nishimori2001} through critical phenomena\cite{stanley1987} can be handled more easily as there is no need anymore to manipulate quantum but classical objects. We have considerably reduced the level of complexity associated with the quantum nature of spin Hamiltonians. This is clearly an advantage of \eqref{equ21} over \eqref{equ18}. As yet another advantage, we will see later in the work that $\Delta^{(j)+-}_{n}$ and $\Delta^{(j)-+}_{n}$ acquire simple forms when we exploit the global $U(1)$ symmetry of $\hat{H}$ by sitting in a given sector of the fragmented Hilbert space.
It should be noted for further purposes that $\Delta^{zz}_0 = J_{z}(K-1)/4+{\bf h}/2$ and $\Delta^{zz}_{D-1} = -J_{z}(K-1)/4 -{\bf h}/2$ with the sample sum ${\bf h}=\sum_{j = 1}^{K}h_{j}$. 

A special case of paramount interest to be discussed before diving into the fragmented Hilbert space is the case of {\it perfect} Neel's states. They correspond to configuration where any  adjacent pair of spins are in opposite direction $\ket{\uparrow\downarrow\uparrow\downarrow\cdots}$ and $\ket{\downarrow\uparrow\downarrow\uparrow\cdots}$ accounting for the state of the first spin. In other words, $s^{\left(j\right)z}_{\rm Neel}=-s^{\left(j+1\right)z}_{\rm Neel}$. As a consequence, the total magnetization $\hat{\mathcal{J}}^z=0$; for every pair $s^{\left(j\right)z}_{\rm Neel}s^{\left(j+1\right)z}_{\rm Neel}=-1/4$, $\Delta^{(j)+-}_{\rm Neel}=J/2,\ \Delta^{(j)-+}_{\rm Neel}=0$ for $(+,-)$ bonds,
while $\Delta^{(j)-+}_{\rm Neel}=J/2,\ \Delta^{(j)+-}_{\rm Neel}=0$ for $(-,+)$ bonds independently on whether the first component of the chain is found in the state $\ket{\uparrow}$ or $\ket{\downarrow}$. For homogeneous local magnetic fields $\Delta^{zz}_{\rm Neel}=-J_z\left(K-1\right)/4$. This is an indication that effects of $h$ in the sector accommodating $\ket{{\rm Neel}}$  are suppressed by mutual compensation of flipped spins i.e. $h\sum_{j=1}^{K}s_n^{(j)z}=h\hat{\mathcal{J}}^z=0$ for all $\ket{n}$. These remarks confirm that Neel's states (two-fold degenerate) are ground states of the $1D$ antiferromagnetic spin-$1/2$ Heisenberg model in the classical limit with energy $\Delta^{zz}_{\rm Neel}$\cite{Pires2021}.  For an even number of spins, $\ket{{\rm Neel}}=\ket{\uparrow\downarrow\uparrow\downarrow\cdots}=\ket{\frac{2}{3}\left(2^K-1\right)}$ and $\ket{{\rm Neel}}=\ket{\downarrow\uparrow\downarrow\uparrow\cdots}=\ket{\frac{1}{3}\left(2^K-1\right)}$.

The central pole of interest in this Section, is the eigenvalue equation $\hat{H}\ket{\psi}=E\ket{\psi}$ with eigenvalue $E$ and the associated eigenstates $\ket{\psi}$. Our goal is to solve this equation. To this end, we recall that in the absence of spin-spin exchange ($J=0$) spins are in the quantization direction; the eigenvalues of the Hamiltonian are $E_n=\Delta_n^{zz}$ whilst the eigenstates are $\ket{n}$. When interactions are turned on ($J\neq 0$) spins are in various superpositions of  $\ket{n}$ weighted by certain probability amplitudes. The eigenstates of the Hamiltonian can be constructed as, 
\begin{eqnarray}\label{equ22}
\ket{\psi}=\sum_{n=0}^{D-1}C_{n}\ket{n},
\end{eqnarray}
where $C_{n}$ are projections of  $\ket{\psi}$ onto the direction of  $\ket{n}$ in the Hilbert space or the probability amplitude of observing spins in the configuration $\ket{n}$. They are subjected to the constraint $\sum_{n=0}^{D-1}|C_{n}|^2=1$. Plugging \eqref{equ22} into the eigenvalue equation results into the system of recurrence equations
\begin{eqnarray}\label{equ23}
	\sum_{j=1}^{K-1}\left(\Delta^{(j)-+}_{n}C_{n+Q_j}+\Delta^{(j)+-}_{n}C_{n-Q_j}\right)=\left(E-\Delta^{zz}_n\right)C_{n}.\quad
\end{eqnarray}
This equation is reminiscent of the tight-binding problem on a graph. Therefore, the complicated problem of a collection of identical spins interacting pairwise on a simple $1D$ lattice is transformed into a simple problem of a free spinless particle hopping on a complicated graph as also discussed in Ref.\cite{Osborne2006}.
  
We solve \eqref{equ23} analytically by sitting in each sector of the fragmented Hilbert space. We focus exclusively on even $K$ as the case of odd $K$ follows the same logic. We discriminate between the single-state sectors $K_{\downarrow}=0$ and $K_{\downarrow}=K$ where \eqref{equ23} is easily solved as follows:

{\bf Sector $K_{\downarrow}=0$:}   This sector has $\begin{pmatrix} K \\ 0\end{pmatrix}=1$ basis vector with fully aligned spins in the state $\ket{\uparrow}$; the so called {\it pseudo-vacuum state}. Therefore, $\ket{n}\equiv \ket{0}$ and $\Delta^{(j)+-}_{0}=\Delta^{(j)-+}_{0}=0$. The single eigenstate and eigenvalue of the Hamiltonian are respectively  $\ket{\psi^{(0)}_0}=\ket{0}$ and $E_0^{(0)}=\Delta^{zz}_{0}= J_{z}(K-1)/4+hK/2$. This emerges as a consequence of the fact that $s^{\left(j\right)z}_0=1/2$ for all $j$. The superscript on the eigenvalue and the eigenfunction refers to $K_{\downarrow}$.

{\bf Sector $K_{\downarrow}=K$:}  Because this sector has only one state with all spins in the state  $\ket{\downarrow}$, same applies. The unique eigenstate and the eigenenergy are respectively $\ket{\psi^{(K)}_{D-1}}=\ket{D-1}$ and $E_{D-1}^{(K)}=\Delta^{zz}_{D-1}= -J_{z}(K-1)/4-hK/2$ given that $s^{\left(j\right)z}_{D-1}=-1/2$.
 
\begin{figure}[]
	\centering
	\begin{center} 
		\includegraphics[width=7cm, height=6cm]{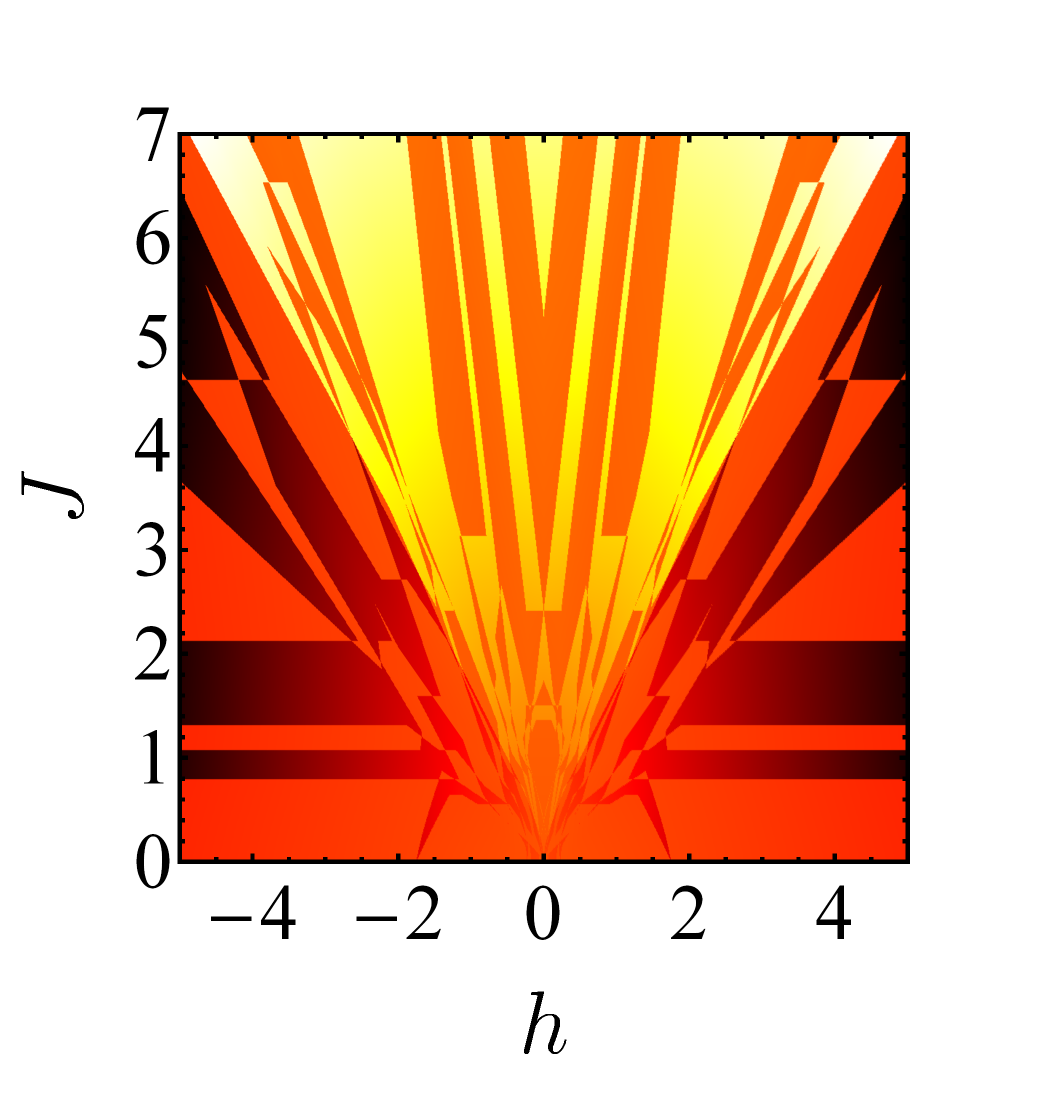}\hspace{-1.8cm}
		\includegraphics[width=2.5cm, height=5.5cm]{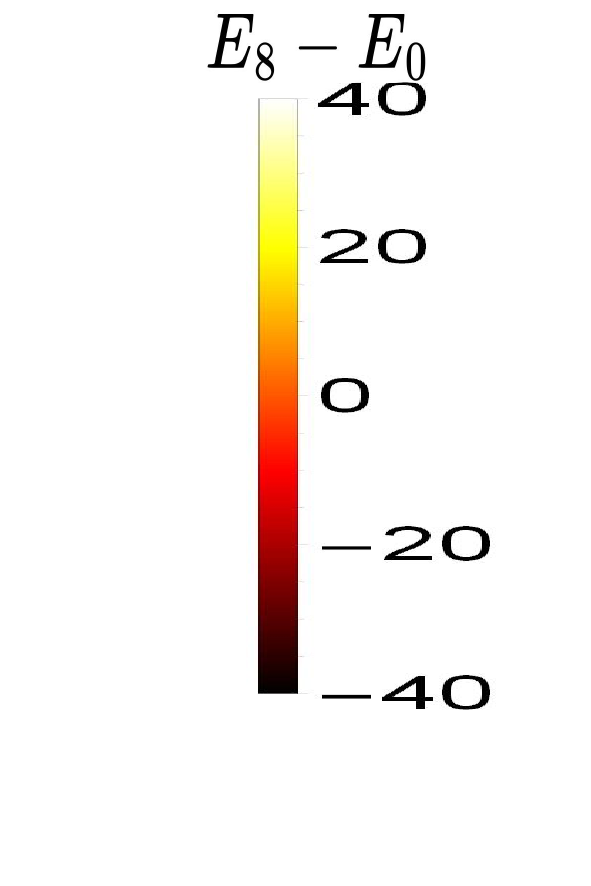}
		\vspace{-0.5cm}
		\caption{Interference pattern in the gap $E_8-E_0$  in the XXZ Heisenberg model for $J_z=0.7$ and $K=8$. The structure of these patterns suggests the existence of level-crossings and/or avoided-level crossing between $E_8$ and $E_0$ by variation of $h$.}
		\label{Figure5} 
	\end{center}
\end{figure}

In the remaining sectors, we set $J_z=0$ for the sake of clarity and simplicity. Interference patterns describing the gap $E_8-E_0$ for $J_z=0.7$ are displayed in Fig.\ref{Figure5} .

{\bf Sector  $K_{\downarrow}=1$:}
All $\binom{K}{1}=K$ basis vectors in this sector have exactly one spin in the down state at site $i$, while all other spins are up.
 They can all be built from $\ket{0}$ as $\{\ket{q_i}\}_{1\leq i\leq K}$ (see previous Section) by flipping one spin. The only anti-parallel bonds are $(i-1,i)$ and $(i,i+1)$ (when they exist); hence only $j=i$ contributes to $\Delta^{(j)+-}_{q_i}$ and only $j=i-1$ contributes to $\Delta^{(j)-+}_{q_i}$ which yields,
\begin{eqnarray}\label{equ24}
	\Delta^{(j)+-}_{q_i} = \frac{J}{2}\delta_{i,j}, \quad \Delta^{(j)-+}_{q_i} = \frac{J}{2}\delta_{i,j+1},
\end{eqnarray}
Assuming a homogeneous magnetic field $h_j=h$ and the convention $s_n^{\left(j\right)z}=1/2$ and $s_n^{\left(j\right)z}=-1/2$ for spins in up and down states respectively, we obtain
\begin{eqnarray}\label{equ21b}
	\Delta^{zz}_{q_i} = h\left(\frac{K}{2}-1\right).
\end{eqnarray}
Considering now Eqs.\eqref{equ24} and \eqref{equ21b} and because $\Delta^{zz}_{q_i}$ is independent on $i$ it enters Eq.\eqref{equ23} as a simple number having only a shift effect. The eigenvalue equation acquires the tight-binding-like form
\begin{eqnarray}\label{equ25}
	\frac{J}{2}\left(C_{q_{j-1}}+C_{q_{j+1}}\right)=\left(E-\Delta^{zz}_{q_i}\right)C_{q_{j}},
\end{eqnarray}
for $1 \leq j\leq K$ with constraint $C_{q_{0}}=C_{q_{K+1}}=0$ by virtue of  \eqref{equ3h}. 
Here, $C_{q_{j}}$ is the probability amplitude for finding the chain in the configuration $\ket{q_j}$. Its governing equation describes an itinerant particle hopping between neighboring sites $j-1$ and $j+1$ with a hopping integral $J/2$. 

Eq.\eqref{equ25} can be reduced to a linear three-term recurrence equation by defining $P_{j}=C_{q_{j}}$ as $P_{j-1}+P_{j+1}=xP_{j}$ with $x=2\left(E-\Delta^{zz}_{q_i}\right)/J$. Boundary conditions become $P_{0}=P_{K+1}=0$. This is a state representation as each state $\ket{q_j}$ is now treated as a point $j$. Upon solving this  equation the eigenfunctions read,
\begin{subeqnarray}
	\slabel{equ26}\ket{\psi^{\left(1\right)}_n}&=&\frac{1}{\mathcal{N}_n}\sum_{j=1}^{K}P_{j}\left(\tilde{E}^{(1)}_n\right)\ket{q_j}, \\ \slabel{equ26a}\mathcal{N}_n&=&\sqrt{\sum_{j=1}^{K}|P_{j}\left(\tilde{E}^{(1)}_n\right)|^2}, 
\end{subeqnarray}
where $\tilde{E}^{(1)}_n = 2\left(E^{(1)}_n-\Delta^{zz}_{q_i}\right)/J$ is a dimensionless energy. The Hamiltonian becomes diagonal in the eigenbasis of the corresponding subspace. The superscript refers as usual to the number of spin in down state. 

At this stage, it is instructive to comment on the inhomogeneous case  $x_j\neq x$. Although simple at first glance does not admit a closed-form solution in terms of elementary functions. This is the consequence of the fact that the recurrence equation $P_{j-1}+P_{j+1}=x_jP_{j}$ can only be solved analytically for restricted cases of sequences $x_j$ (such as constant, periodic, slowly varying, or perturbative deviations from homogeneity). In this piece of work, this case is addressed perturbatively with remarkable agreement with the data from exact diagonalization (see Section \ref{Section4}).

For further purposes, it would be interesting to briefly review how the three-term equation for $P_j$ is solved in the homogeneous case. Because this equation is subject to open boundary conditions, a general solution can be constructed as $P_{j}=\beta^j$ (see Ref.\onlinecite{Yokomizo2019}). Plugging this back into the equation for $P_j$ yields the quadratic equation $\beta^2-x\beta+1=0$ thence the general solution $P_{j}=\phi_1\left(\beta_1\right)^j+\phi_2\left(\beta_2\right)^j$ with $\beta_{1,2}=\left(x\pm\sqrt{x^2-4}\right)/2$. The boundary conditions teach us that $\phi_1=-\phi_2$ and  $\left(\beta_1/\beta_2\right)^{K+1}=1$. We observe that $\beta_1/\beta_2$ is a root of unity. $\beta_1$ and $\beta_2$ must be complex which can only be realized if $x^2-4<0$. If we choose $x=2\cos\theta$ with $0\leq \theta\leq \pi$ then $\beta_1=e^{i\theta}$ and $\beta_2=e^{-i\theta}$. Hence, $\left(\beta_1/\beta_2\right)^{K+1}=e^{2i\pi m}$ with $m=1,\cdots,K$ and whereby $\beta_1/\beta_2=e^{2i\theta_m}$ with $\theta_m=\pi m/(K+1)$. Thus, $\beta^{(m)}_1=e^{i\theta_m}$, $\beta^{(m)}_2=e^{-i\theta_m}$ and  $P^{(m)}_j=A\sin\left(j\theta_m\right)$ where $A$ is a normalization constant evaluated from   \eqref{equ26}. This process yields the eigenvalues\cite{Stolze2018,Benatti2021}
\begin{eqnarray}\label{equ26b}
	E^{\left(1\right)}_{q_m}=\Delta^{zz}_{q_m}+J\cos\left(\frac{\pi m}{K+1}\right),
\end{eqnarray}
and the one-particle eigenstates 
\begin{eqnarray}\label{equ27a}
	\ket{\psi^{\left(1\right)}_{q_m}}=\sum_{j=1}^{K}u_{j,m}\ket{q_j},
\end{eqnarray}
where
\begin{eqnarray}\label{equ27}
	u_{j,m}=\sqrt{\frac{2}{K+1}}\sin\left(\frac{\pi j}{K+1}m\right),
\end{eqnarray}
with $1\leq m\leq K$. Note that $u_{j,m}=u_{m,j}$ form a structured orthogonal symmetric matrix as explained in Ref.\cite{Benatti2021} and $\sum_{k=1}^{K}u_{i,k}u^*_{k,j}=\delta_{i,j}$. The eigenvalues tell us that only the unique spin in down state in the configuration $\ket{q_j}_{1\le j\le K}$ contributes to the energy. The wave function depends only on the total number of spins.

At this stage, we have to indicate that the restriction to open boundary conditions (OBCs) is made  only for analytical convenience and is not related to our Dirac representation. For periodic boundary conditions (PBCs), the representation \eqref{equ3f1}-\eqref{equ3f3} remains unchanged. One only has to include the additional bond connecting the sites $j=K$ and $j=1$. In the sector $K_{\downarrow}=1$, the recurrence equation
$P_{j-1}+P_{j+1}=xP_j$ still holds but the OBCs $P_0=P_{K+1}=0$ are replaced by the periodic boundary condition $P_{j+K}=P_j$. The corresponding eigenmodes are therefore the plane waves $P_j^{(m)}=B e^{ik_m j}$ (with $B$ a normalization constant that can be determined as in the OBC case presented above). Here, $k_m=2\pi m/K$ and $m=0,1,\ldots,K-1$, leading in the homogeneous case $J_z=0$ considered above to
\begin{eqnarray}\label{equ27a}
	E_{m,\mathrm{PBC}}^{(1)} = \Delta_{q_m}^{zz} +
	J\cos\left(\frac{2\pi m}{K}\right),
\end{eqnarray}
with $\Delta^{zz}_{q_m}$ in Eq.\eqref{equ21b}.	
The formalism applies to periodic chains and only the boundary conditions are modified. It worth mentioning that the finite-$K$ results derived above for open chains are exact within the stated assumptions and do not rely on an expansion in the chain length.

{\bf Sector  $K_{\downarrow}=2$:} In the third sector,  all $\begin{pmatrix} K \\ 2\end{pmatrix}=K(K-1)/2$ basis vectors have two spins in down state (two-particle block). The complete set of orthogonal basis vectors in this sector  can be obtained from  $\{\ket{q_{j_1}+q_{j_2}}\}_{1\leq j_1< j_2\leq K}$ corresponding to all distinct ways of arranging two spins in down state in a chain of $K-2$ spins in up state. Hence,
\begin{subeqnarray}\label{equ22e}
\hspace{-0.5cm}	\Delta^{(j)+-}_{q_{j_1}+q_{j_2}} =\frac{J}{2}\left[\delta_{j_1,j}\left(1-\delta_{j_2,j+1}\right)+\delta_{j_2,j}\left(1-\delta_{j_1,j+1}\right)\right],\qquad \\ \Delta^{(j)-+}_{q_{j_1}+q_{j_2}} = \frac{J}{2}\left[\delta_{j_1,j+1}\left(1-\delta_{j_2,j}\right)+\delta_{j_2,j+1}\left(1-\delta_{j_1,j}\right)\right],\qquad
\end{subeqnarray}
and
\begin{eqnarray}\label{equ22ee}
	\Delta^{zz}_{q_{j_1}+q_{j_2}} = -h\left(\frac{K}{2}-2\right).
\end{eqnarray}
It should be noted that under the constraint $i<j$ the term $\delta_{i,j+1}$ contributes only for $i\geq 2$. Indeed, for $i=1$ there is no value of $j\ge1$ such that $i=j+1$. Taking this remark into account Eq.\eqref{equ23} then takes the form
\begin{eqnarray}\label{equ22f}
	\nonumber\frac{J}{2}\left[\left(1-\delta_{j_1-1,j_2}\right)C_{q_{j_1-1}+q_{j_2}} + \left(1-\delta_{j_1+1,j_2}\right)C_{q_{j_1+1}+q_{j_2}} \right] &+&\\\nonumber \frac{J}{2}\left[\left(1-\delta_{j_1,j_2-1}\right)C_{q_{j_1}+q_{j_2-1}} + \left(1-\delta_{j_1,j_2+1}\right)C_{q_{j_1}+q_{j_2+1}} \right] &=&\\\nonumber \left(E-\Delta^{zz}_{q_{j_1}+q_{j_2}}\right)C_{q_{j_1}+q_{j_2}}.\\
\end{eqnarray}
This equation is solved with the conditions $C_{q_{0}+q_{j_2}}=C_{q_{j_1}+q_{0}}=0$ and $C_{q_{K+1}+q_{j_2}}=C_{q_{j_1}+q_{K+1}}=0$ for $j_1<j_2$. As we did for the second sector, we define the auxiliary variable $P_{j_1,j_2} = C_{q_{j_1}+q_{j_2}}$ and obtain the two-dimensional linear recurrence equation
\begin{eqnarray}\label{equ22g}
\nonumber\left[\left(1-\delta_{j_1-1,j_2}\right)P_{j_1-1, j_2} + \left(1-\delta_{j_1+1,j_2}\right)P_{j_1+1,j_2} \right] &+&\\\nonumber \left[\left(1-\delta_{j_1,j_2-1}\right)P_{j_1, j_2-1} + \left(1-\delta_{j_1,j_2+1}\right)P_{j_1,j_2+1} \right]  &=& xP_{j_1,j_2},\\
\end{eqnarray}
with boundary conditions $P_{0, j_2}=P_{j_1,0}=P_{K+1, j_2}=P_{j_1, K+1}=0$ and $x=2\left(E-\Delta^{zz}_{q_{j_1}+q_{j_2}}\right)/J$. Once this equation is solved, the eigenvectors are calculated as
\begin{subeqnarray}\label{equ26g}
	\hspace{-0.5cm}\ket{\psi^{\left(2\right)}_n}&=&\frac{1}{\mathcal{N}_n}\sum_{1\le j_1<j_2\leq K}P_{j_1,j_2}\left(\tilde{E}_n\right)\ket{q_{j_1}+q_{j_2}},\\
	\mathcal{N}_n&=&\sqrt{\sum_{1\le j_1<j_2\leq K}|P_{j_1,j_2}\left(\tilde{E}_n\right)|^2}.
\end{subeqnarray}
 By virtue of the condition $j_1<j_2$ it is legitimate to set $P_{j,j}=0$ (hard-core condition).  For $J_z=0$, this constraint excludes double-occupation of a lattice site allowing the two-particle eigenfunction to be constructed from anti-symmetrized products of the one-particle eigenfunction. Accordingly, the properly constrained two-particle eigenfunction is given by the Slater determinant  as we will see later.
  As yet another consequence of this constraint, all Kronecker delta terms in Eq.\eqref{equ22g} do not contribute. For example $\delta_{j_1-1,j_2}P_{j_1-1, j_2}=\delta_{j_1+1,j_2}P_{j_1+1, j_2}=0$; this is the consequence of the fact that in such terms every attempt of Kronecker delta to contribute is immediately multiplied by $P_{j, j}=0$. These observations greatly simplify Eq.\eqref{equ22g} leading to,
\begin{eqnarray}\label{equ22gg}
	\left[P_{j_1-1, j_2} + P_{j_1+1,j_2} \right]+ \left[P_{j_1, j_2-1} + P_{j_1,j_2+1} \right] = xP_{j_1,j_2}.
\end{eqnarray}
Here, it can be observed that the first brackets gather all terms with the same index $j_2$ while the second one is for those with the same $j_1$. It can then be solved by separating the variable as $P_{j_1, j_2}=X_{j_1}Y_{j_2}$. Inserting this into \eqref{equ22gg} and dividing the resulting equation by $P_{j_1, j_2}$ we find that the consistency of the leading equation is achieved when  $X_{j_1-1}-\eta X_{j_1}+X_{j_1+1}=0$ and $Y_{j_2-1}-\left(x-\eta\right) Y_{j_2}+Y_{j_2+1}=0$ with $\eta$ an arbitrary parameter. These equations are respectively solved with boundary conditions $X_0=X_{K+1}=0$ and $Y_0=Y_{K+1}=0$. We have then transformed \eqref{equ22gg} into two uncoupled linear recurrence equation each describing the sector $K_{\downarrow}=1$. We obtain the eigenvalues ($1\leq m_1<m_2\leq K$),
\begin{eqnarray}\label{equ22gh}
	E^{\left(2\right)}_{q_{m_1}+q_{m_2}}=\Delta^{zz}_{q_{m_1}+q_{m_2}}+J\cos\left(\frac{\pi m_1}{K+1}\right) + J\cos\left(\frac{\pi m_2}{K+1}\right),\quad
\end{eqnarray}
and the eigenstates 
\begin{eqnarray}\label{equ22h}
\ket{\psi^{\left(2\right)}_{m_1,m_2}}=\sum_{1\leq j_1<j_2\leq K}u_{j_1,m_1}u_{j_2,m_2}\ket{q_{j_1}+q_{j_2}}.
\end{eqnarray}
However, \eqref{equ22h} is not the correct expression for eigenstates. Indeed, because local spins are indistinguishable, the wave function should only change by a factor when interchanging particles with spins pointing in the same direction i.e. $\ket{\psi^{\left(2\right)}_{m_1,m_2}}=e^{i\phi}\ket{\psi^{\left(2\right)}_{m_2,m_1}}$.  For states in the sectors $K_{\downarrow}=0$ and $K_{\downarrow}=K$ (fully symmetric states) the phase $\phi$ is chosen as $\phi=0$ while for states in other sectors, the wave function must be anti-symmetric by interchanging particles. The phase is chosen as $\phi=\pi$ and,
\begin{eqnarray}\label{equ22ha}
	\ket{\psi^{\left(2\right)}_{m_1,m_2}}=-\ket{\psi^{\left(2\right)}_{m_2,m_1}}.
\end{eqnarray}
For our construction \eqref{equ22h}, we observe that the two-particle wave function does not fulfill the anti-symmetric requirement \eqref{equ22ha}. This problem is circumvented by re-constructing the wave function as the linear combination of the two solutions in \eqref{equ22ha} i.e.
\begin{eqnarray}\label{equation22hb}
\ket{\psi^{\left(2\right)}_{m_1,m_2}}=\sum_{1\leq j_1<j_2\leq K}\mathsf{D}_{m_1,m_2}^{j_1,j_2}\ket{q_{j_1}+q_{j_2}},
\end{eqnarray}
where
\begin{eqnarray}\label{equ22hc}
	\mathsf{D}_{m_1,m_2}^{j_1,j_2} =
	\left|\begin{array}{cc}
		u_{j_1,m_1} & u_{j_1,m_2}\\ 
		u_{j_2,m_1} & u_{j_2,m_2}
	\end{array}
	\right| = 
	\sum_{i,j=1}^2\epsilon_{i,j}u_{j_1,m_{i}}u_{j_2,m_{j}},
\end{eqnarray}
is nothing but the two-particle Slater determinant\cite{Landau1981} with $1\leq m_1<m_2\leq K$. This is also a consequence of the hard-core condition we mentioned earlier. In these expressions $j_2=j_1+1$ such that for a function $g_{j_1,j_2}$ the double summation in \eqref{equation22hb} can be viewed as $\sum_{1\leq j_1<j_2\leq K}g_{j_1,j_2} = \sum_{j_1=1}^{K}\sum_{j_2=j_1+1}^Kg_{j_1,j_2}$. In Eq.\eqref{equ22hc},
\begin{eqnarray}\label{equ22he}
	\epsilon_{i,j}=\left\{\begin{array}{cc}
		+1, &\quad {\rm if} \quad \left(i,j\right)=\left(1,2\right),\\ 
		-1, &\quad {\rm if} \quad \left(i,j\right)=\left(2,1\right), \\
		0, &\quad {\rm if} \quad i = j
	\end{array}
	\right. 
\end{eqnarray}
is the two-dimensional Levi-Civita symbol. The Slater determinant vanishes by repetition of at least two indices, $\mathsf{D}_{m,m}^{j_1,j_2}=\mathsf{D}_{m_1,m_2}^{j,j}=\mathsf{D}_{m,m}^{j,j}=0$.

It is also instructive to examine how periodic boundary conditions modify the present $K_{\downarrow}=2$ sector. As in the one-spin-down sector, the bulk recurrence equation remains unchanged, while the site indices are now understood modulo $K$, so that sites $K$ and 
$1$ become nearest neighbors. In particular, the configuration $\ket{q_1+q_K}$, which involves the two boundary sites for OBC, becomes a nearest-neighbor configuration on the periodic chain. For $J_z=0$, the two-particle amplitudes can be written in the 
antisymmetrized form
\begin{eqnarray}\label{equ22h1}
	P_{j_1,j_2} = A\left[ e^{i(k_1j_1+k_2j_2)} - e^{i(k_2j_1+k_1j_2)} \right],
	\qquad j_1<j_2 ,
\end{eqnarray}
which automatically satisfies the hard-core condition $P_{j,j}=0$ and $P_{j_1,j_2}=-P_{j_2,j_1}$. The wave-numbers $k_{1,2}$ are determined from boundary conditions and $A$ from the normalization condition. Substituting into the bulk recurrence equation  \eqref{equ22gg} gives
\begin{eqnarray}\label{equ22h2}
	E^{(2)}_{k_1, k_2, \rm PBC} = \Delta_{q_{j_1}+q_{j_2}}^{zz} + J\cos k_1 + J\cos k_2 .
\end{eqnarray}
Given that transporting one coordinate once around the periodic chain
changes its ordering relative to the other coordinate, the antisymmetric amplitude acquires a minus sign. The periodicity condition therefore becomes
\begin{eqnarray}\label{equ22h3}
	e^{ik_\ell K}=-1, \qquad k_\ell=\frac{2\pi}{K} \left(m_\ell+\frac{1}{2}\right),
	\qquad \ell=1,2,
\end{eqnarray}
with distinct quantum numbers $m_1\neq m_2$. Hence, as in the sector $K_{\downarrow}=1$, passing from OBC to PBC requires no modification of the Dirac representation or of the bulk recurrence equations. Only the boundary connectivity and the corresponding momentum quantization are modified.

At this stage, one might be tempted to conjecture about the global structure of the eigensystem. However, in order to ensure a better understanding, it is preferable to first gain further insight into the general structure of the recurrence equation by briefly examining the sector with three spins in the down state.

{\bf Sector  $K_{\downarrow}=3$:}
The sub-Hilbert space corresponds to all possible manners of replacing three spins in a spin chain with $K$ identical spins in up state with three spins in down state. It can quickly be verified that 
\begin{widetext}
	\begin{subeqnarray}\label{equ23a}
		\Delta^{(j)+-}_{q_{j_1}+q_{j_2}+q_{j_3}} &=&\frac{J}{2}\left[\delta_{j_1,j}\left(1-\delta_{j_2,j+1}-\delta_{j_3,j+1}\right)+\delta_{j_2,j}\left(1-\delta_{j_1,j+1}-\delta_{j_3,j+1}\right)+\delta_{j_3,j}\left(1-\delta_{j_1,j+1}-\delta_{j_2,j+1}\right)\right], \\ \Delta^{(j)-+}_{q_{j_1}+q_{j_2}+q_{j_3}} &=&\frac{J}{2}\left[\delta_{j_1,j+1}\left(1-\delta_{j_2,j}-\delta_{j_3,j}\right)+\delta_{j_2,j+1}\left(1-\delta_{j_1,j}-\delta_{j_3,j}\right)+\delta_{j_3,j+1}\left(1-\delta_{j_1,j}-\delta_{j_2,j}\right)\right].
	\end{subeqnarray}
In the same fashion as we did for previous sectors, this leads to the three-dimensional linear recurrence equation
	\begin{eqnarray}\label{equ23b}
		\nonumber\left[\left(1-\delta_{j_1-1,j_2}-\delta_{j_1-1,j_3}\right)P_{j_1-1, j_2, j_3} + \left(1-\delta_{j_1+1,j_2}-\delta_{j_1+1,j_3}\right)P_{j_1+1, j_2, j_3} \right] &+& \\\nonumber \left[\left(1-\delta_{j_2-1,j_1}-\delta_{j_2-1,j_3}\right)P_{j_2-1, j_1, j_3} + \left(1-\delta_{j_2+1,j_1}-\delta_{j_2+1,j_3}\right)P_{j_2+1, j_1, j_3} \right] &+&  \\
		\left[\left(1-\delta_{j_3-1,j_1}-\delta_{j_3-1,j_2}\right)P_{j_3-1, j_1, j_2} + \left(1-\delta_{j_3+1,j_1}-\delta_{j_3+1,j_2}\right)P_{j_3+1, j_1, j_2} \right] &=& xP_{j_1,j_2,j_3}. 
	\end{eqnarray}
\end{widetext}
This equation is solved with boundary conditions $P_{0,j_2,j_3}=P_{j_1,0,j_3}=P_{j_1,j_2,0}=0$, $P_{K+1,j_2,j_3}=P_{j_1,K+1,j_3}=P_{j_1,j_2,K+1}=0$ and $x=2(E-\Delta^{zz}_{q_{j_1}+q_{j_2}+q_{j_3}})/J$. It is important to go deep into this case in order to have a clear comprehension of our ongoing strategy. As in previous cases and by virtue of the assumption $j_1<j_2<j_3$ we can assume that $P_{j_1, j_2, j_3}=0$ (hard-core condition) for every repeated (two or more) indices. For example $P_{j_1, j_1, j_3}=P_{j_1, j_1, j_1}=P_{j_1, j_2, j_2}=0$ and many more. Hence, all Kronecker delta term in \eqref{equ23b} vanish leading us to
\begin{eqnarray}\label{equ23c}
	\nonumber\left[P_{j_1-1, j_2, j_3} + P_{j_1+1, j_2, j_3} \right] + \left[P_{j_1,j_2-1, j_3} + P_{j_1,j_2+1, j_3} \right]\\ + \left[P_{j_1, j_2,j_3-1} + P_{j_1, j_2, j_3+1} \right] = xP_{j_1,j_2,j_3}. 
\end{eqnarray}
By separation of variable as we did so far, one achieves a set of three uncoupled linear-recurrence equations that are solved exactly as for the case of $K_{\downarrow}=1$. This task leads to an eigensystem similar to \eqref{equ22gh} and \eqref{equ22h}. Hence, $\ket{\psi^{\left(3\right)}_{m_1,m_2,m_3}}=\sum_{1\leq j_1<j_2<j_3\leq K}u_{j_1,m_1}u_{j_2,m_2}u_{j_3,m_3}\ket{q_{j_1}+q_{j_2}+q_{j_3}}.$ As already pinpointed, such a wave function does not fulfill the fundamental principle of indistinguishability of particles. Upon overcoming this constraint through the three-particle Slater determinant\cite{Landau1981} we obtain ($1\leq m_1<m_2<m_3\leq K$),
\begin{eqnarray}\label{equ23d}
	\hspace{-1cm}\ket{\psi^{\left(3\right)}_{m_1,m_2,m_3}}=\sum_{1\leq j_1<j_2<j_3\leq K}\mathsf{D}_{m_1,m_2,m_3}^{j_1,j_2,j_3}\ket{q_{j_1}+q_{j_2}+q_{j_3}},
\end{eqnarray}
where
\begin{eqnarray}\label{equ3e}
\nonumber	\mathsf{D}_{m_1,m_2,m_3}^{j_1,j_2,j_3} &=& 
	\left|\begin{array}{ccc}
		u_{j_1,m_1} & u_{j_1,m_2} & u_{j_1,m_3}\\ 
		u_{j_2,m_1} & u_{j_2,m_2} & u_{j_2,m_3}\\
		u_{j_3,m_1} & u_{j_3,m_2} & u_{j_3,m_3}
	\end{array}
	\right| \\
	&=& 
	\sum_{i,j,k=1}^3\epsilon_{i,j,k}u_{j_1,m_{i}}u_{j_2,m_{j}}u_{j_3,m_{k}}.
\end{eqnarray}
Here, $\epsilon_{i,j,k}$ is the three-dimensional Levi-Civita symbol. For a given function $g_{j_1,j_2, j_3}$ the triple summation in \eqref{equ23d} can be replaced by $\sum_{1\leq j_1<j_2<j_3\leq K}g_{j_1,j_2,j_3} = \sum_{j_1=1}^{K}\sum_{j_2=j_1+1}^K\sum_{j_3=j_2+1}^Kg_{j_1,j_2,j_3}$.

We are now in position to extend  our analysis to arbitrary sectors $1\leq K_{\downarrow}\leq K-1$. In order to achieve this goal and for $1\leq j_1<j_2<\cdots <j_{K_{\downarrow}}\leq K$ we have to write our classical variables in each sector. Thus, in general,
\begin{eqnarray}\label{equ23dd}
	\Delta^{(j)+-}_{q_{j_1}+q_{j_2}+q_{j_3}+\cdots+q_{j_{K_{\downarrow}}} }=\frac{J}{2}\sum_{\ell=1}^{K_{\downarrow}}\delta_{j_{\ell},j}\mathcal{X}_{j_\ell+1}, \quad
\end{eqnarray}
\begin{eqnarray}\label{equ23ddd}
	\Delta^{(j)-+}_{q_{j_1}+q_{j_2}+q_{j_3}+\cdots+q_{j_{K_{\downarrow}}} }=\frac{J}{2}\sum_{\ell=1}^{K_{\downarrow}}\delta_{j_{\ell},j+1}\mathcal{X}_{j_\ell-1},
\end{eqnarray}
\begin{eqnarray}\label{ob10b}
	\Delta^{zz}_{q_{j_1}+q_{j_2}+q_{j_3}+\cdots+q_{j_{K_{\downarrow}}}} =h\left(\frac{K}{2}-K_{\downarrow}\right),
\end{eqnarray}
where we have defined
\begin{eqnarray}\label{equ23dd-clean}
	\mathcal{X}_{j_\ell\pm1}=1-\sum_{k\neq\ell}^{K_{\downarrow}}\delta_{j_k,j_\ell\pm1}.
\end{eqnarray}

Although the Kronecker deltas entering $X_{j_\ell\pm1}$ do not vanish identically, every nonzero delta contribution multiplies an amplitude containing at least two identical site indices. Due to the hard-core conditions, these contributions vanish identically.

The bulk recurrence equation \eqref{equ23} with the auxiliary function $P_{j_\ell,\cdots, j_{K_{\downarrow}}}=C_{q_{j_1}+\cdots+ q_{j_\ell}+\cdots+q_{j_{K_{\downarrow}}}}$ takes the generalized form,
\begin{eqnarray}\label{equ23cc}
	\hspace{-0.5cm}\sum_{\ell=1}^{K_{\downarrow}}\left(\mathcal{X}_{j_\ell-1}P_{j_\ell-1,\cdots, j_{K_{\downarrow}}} + \mathcal{X}_{j_\ell+1}P_{j_\ell+1,\cdots, j_{K_{\downarrow}}} \right) = xP_{j_1,\cdots, j_{K_{\downarrow}}}, 
\end{eqnarray}
i.e.
\begin{eqnarray}\label{equ23ee}
	\sum_{\ell=1}^{K_{\downarrow}}\left(P_{j_\ell-1,\cdots, j_{K_{\downarrow}}} + P_{j_\ell+1,\cdots, j_{K_{\downarrow}}} \right) = xP_{j_1,\cdots, j_{K_{\downarrow}}},
\end{eqnarray}
and is associated with the traditional boundary conditions 
$P_{0,j_2,\cdots, j_{K_{\downarrow}}}=P_{j_1,0,\cdots, j_{K_{\downarrow}}}=\cdots=P_{j_1,j_2,\cdots, 0}=0$ and $P_{K+1,j_2,\cdots, j_{K_{\downarrow}}}=P_{j_1,K+1,\cdots, j_{K_{\downarrow}}}=\cdots=P_{j_1,j_2,\cdots, K+1}=0$.
Upon solving Eq.\eqref{equ23ee} by variable separation as promoted so far leads us  independently on the sector to the eigenenergies,
\begin{eqnarray}\label{equ23e}
	E_n = \Delta^{zz}_{n}+ J\sum_{j=1}^{K}s^{(j)+}_{n}\cos\left(\frac{\pi j}{K+1}\right).
\end{eqnarray}
Here, $s^{(j)+}_{n}$ and  $s^{(j)z}_{n}$ are classical spins in \eqref{equ6a} and \eqref{equ6c} respectively. By setting $n=\{q_j\}_{1\leq j\leq K}$ as in the sector $K_{\downarrow}=1$ or $n=\{q_{j_1}+q_{j_2}\}_{1\leq j_1<j_2\leq K}$ as in the sector  $K_{\downarrow}=2$ etc and accounting for \eqref{equ3gh} we retrieve previous results. We observe that the energy linearly changes with $J$.

Our general results for the eigenstates exclude the sectors with extremal value of the total magnetization as they contain fully-symmetric states. They are written here as
\begin{eqnarray}\label{equ22hb}
	\ket{\psi^{\left(K_{\downarrow}\right)}_{\bf i}}=\sum_{\bf j}\mathcal{D}^{\left(K_{\downarrow}\right)}_{{\bf i},{\bf j}}\ket{{\bf n_j}},
\end{eqnarray}
where 
\begin{eqnarray}\label{equ3e-clean}
 \mathcal{D}^{\left(K_{\downarrow}\right)}_{{\bf i},{\bf j}} =  
	\sum_{i_1,i_2,\cdots, i_{K_{\downarrow}}=1}^{K_{\downarrow}}\epsilon_{i_1,i_2,\cdots, i_{K_{\downarrow}}}u_{j_1,m_{i_1}}u_{j_2,m_{i_2}}\cdots u_{j_{K_{\downarrow}},m_{i_{K_{\downarrow}}}},\qquad
\end{eqnarray}
is the multi-particle Slater determinant\cite{Landau1981} with $\mathsf{D}^{\left(K\right)}_{{\bf i},{\bf j}}=1$. Here, ${\bf i}=\{i_1, i_2,\cdots, i_{K_{\downarrow}}\}$, ${\bf j}=\{j_1, j_2,\cdots, j_{K_{\downarrow}}\}$; ${\bf n_k}=q_{k_1}+\cdots+q_{k_{K_{\downarrow}}}$ with ${\bf k}\in ({\bf i}, {\bf j})$; $1\leq i_1<i_2<\cdots<i_{K_{\downarrow}}\leq K$ and $1\leq j_1<j_2<\cdots<j_{K_{\downarrow}}\leq K$ (see also Refs.\cite{Yokomizo2019, Benatti2021}). The multi-dimensional Levi-Civita  $\epsilon_{i_1,i_2,\cdots, i_{K_{\downarrow}}}=0$ if any two indices are the same; $\epsilon_{i_1,i_2,\cdots, i_{K_{\downarrow}}}=+1$ for even permutation of $(1,2,\cdots, K_{\downarrow})$; $\epsilon_{i_1,i_2,\cdots, i_K}=-1$ for odd permutation of $(1,2,\cdots, K_{\downarrow})$. 

\begin{figure}[]
	\centering
	\begin{center} 
		\includegraphics[width=4.5cm, height=4cm]{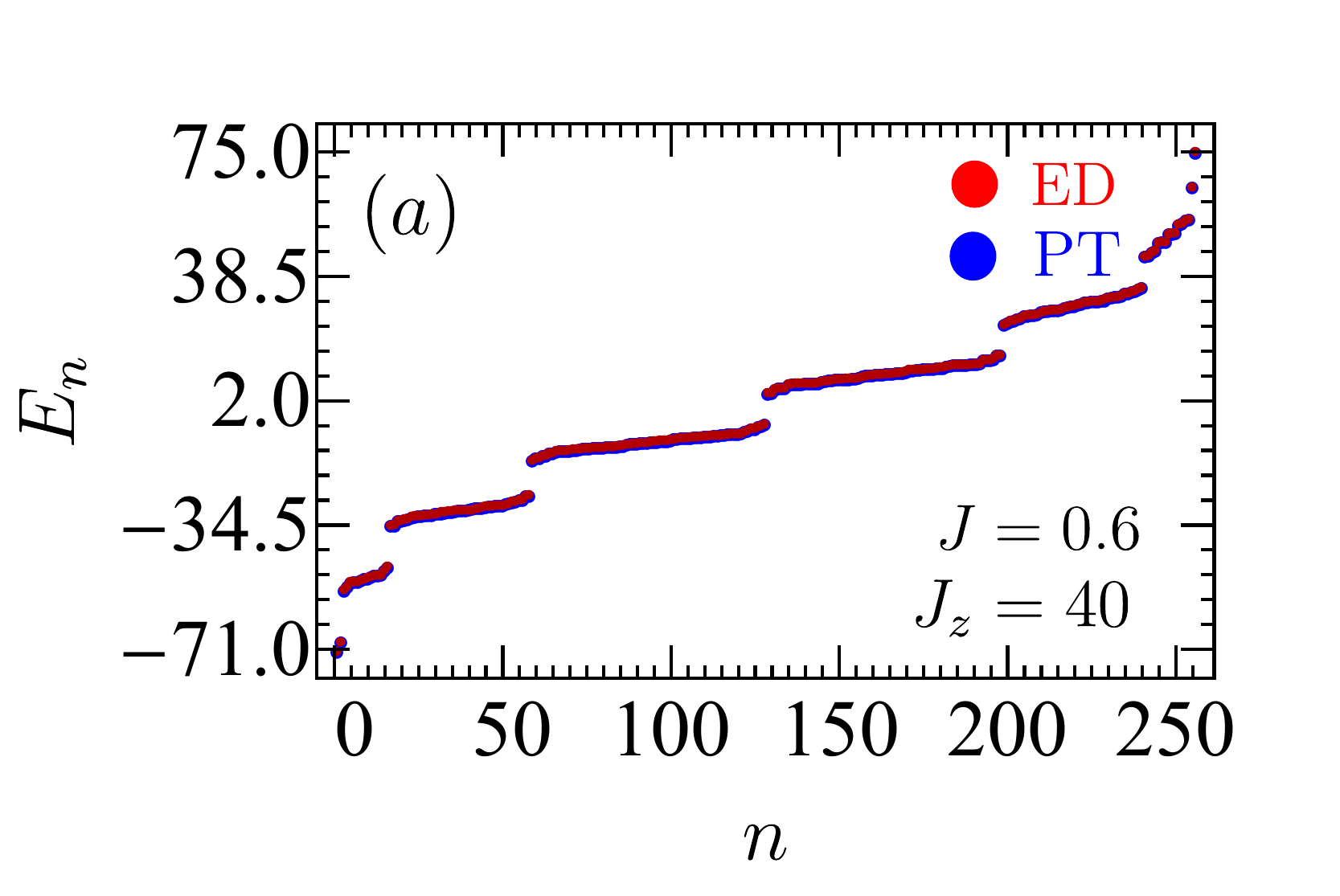}\hspace{-0.5cm}
		\includegraphics[width=4.5cm, height=4cm]{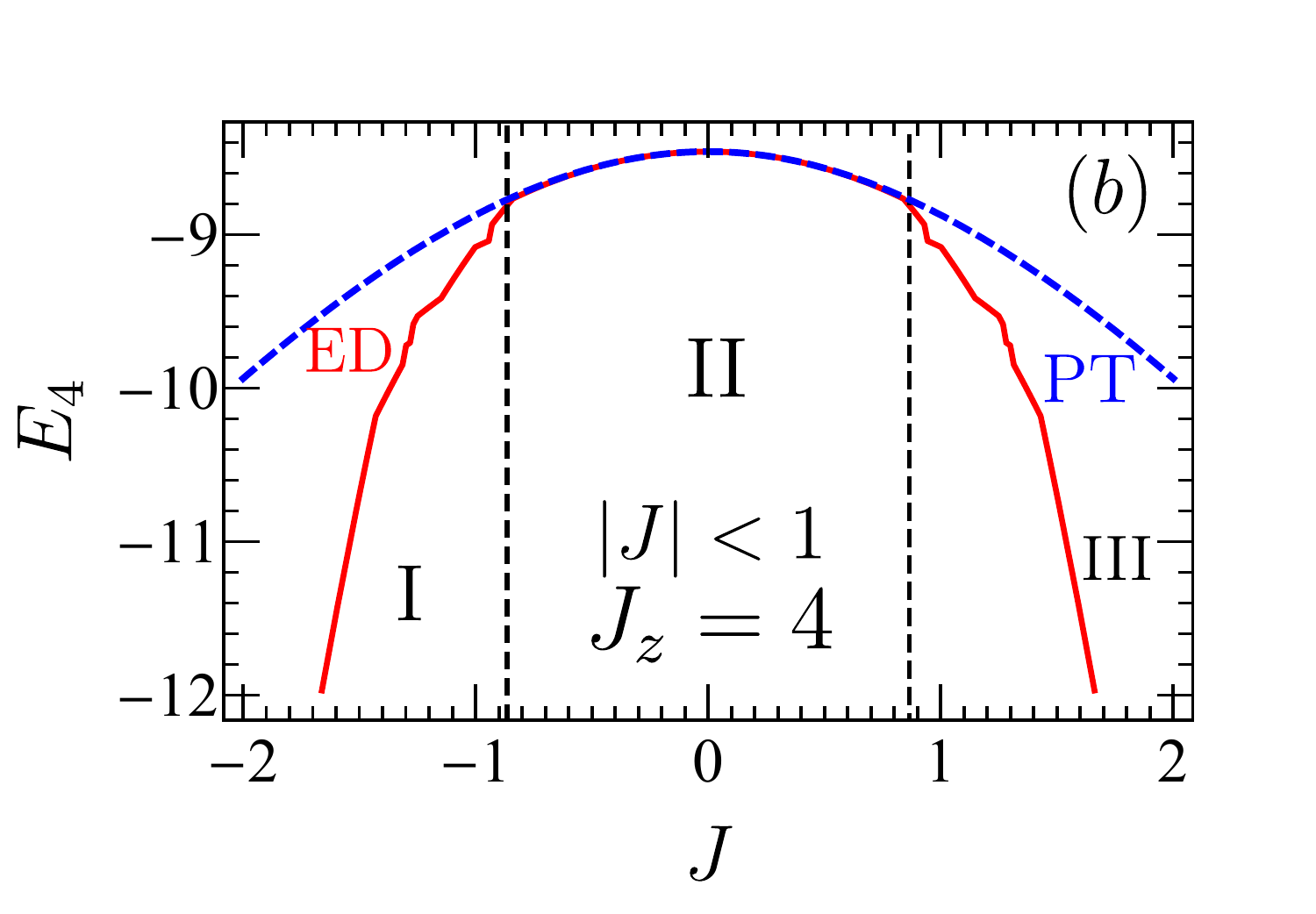}\hspace{-0.1cm}\\\vspace{-0.5cm}
		\includegraphics[width=4.5cm, height=4cm]{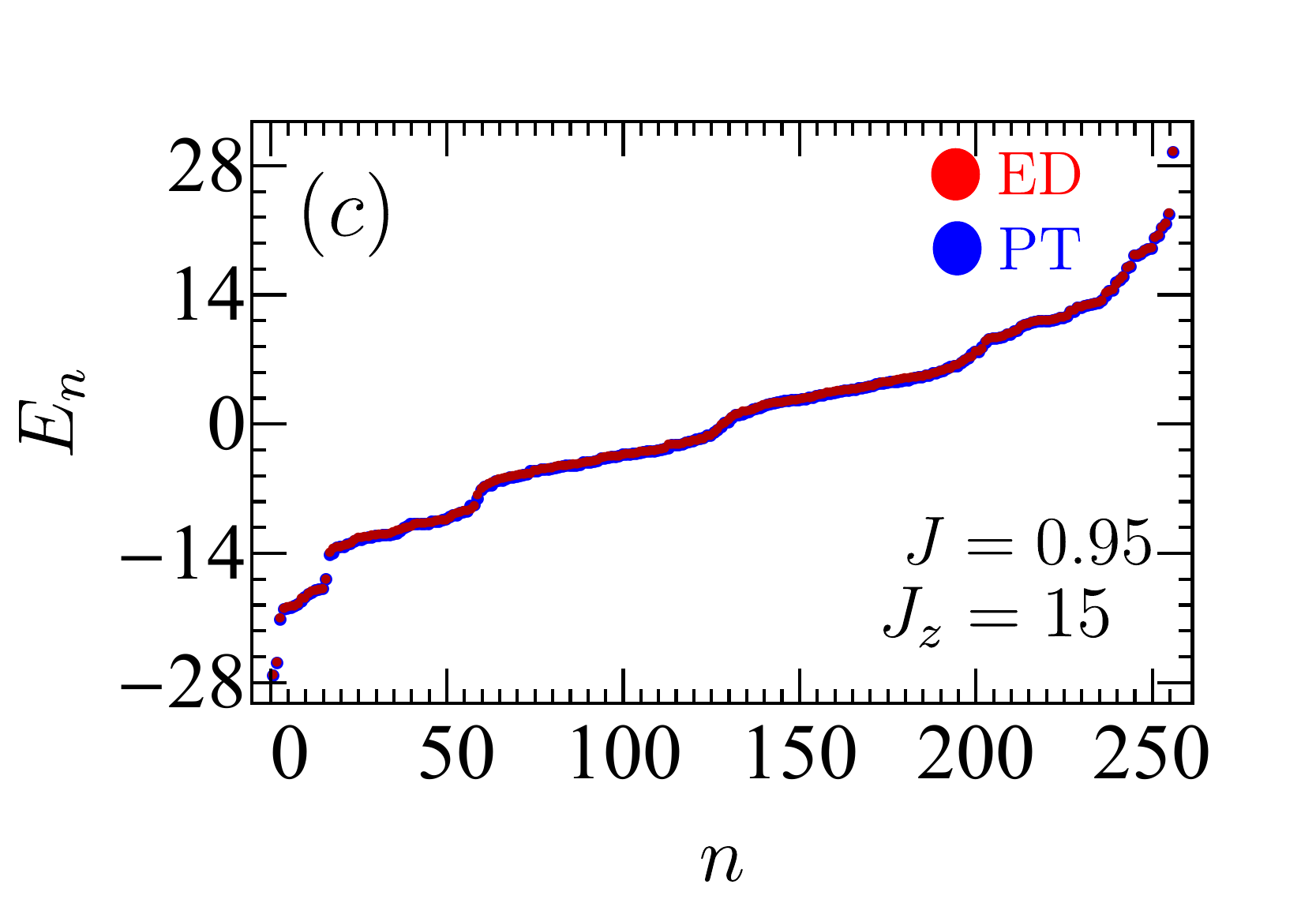}\hspace{-0.5cm}
		\includegraphics[width=4.5cm, height=4cm]{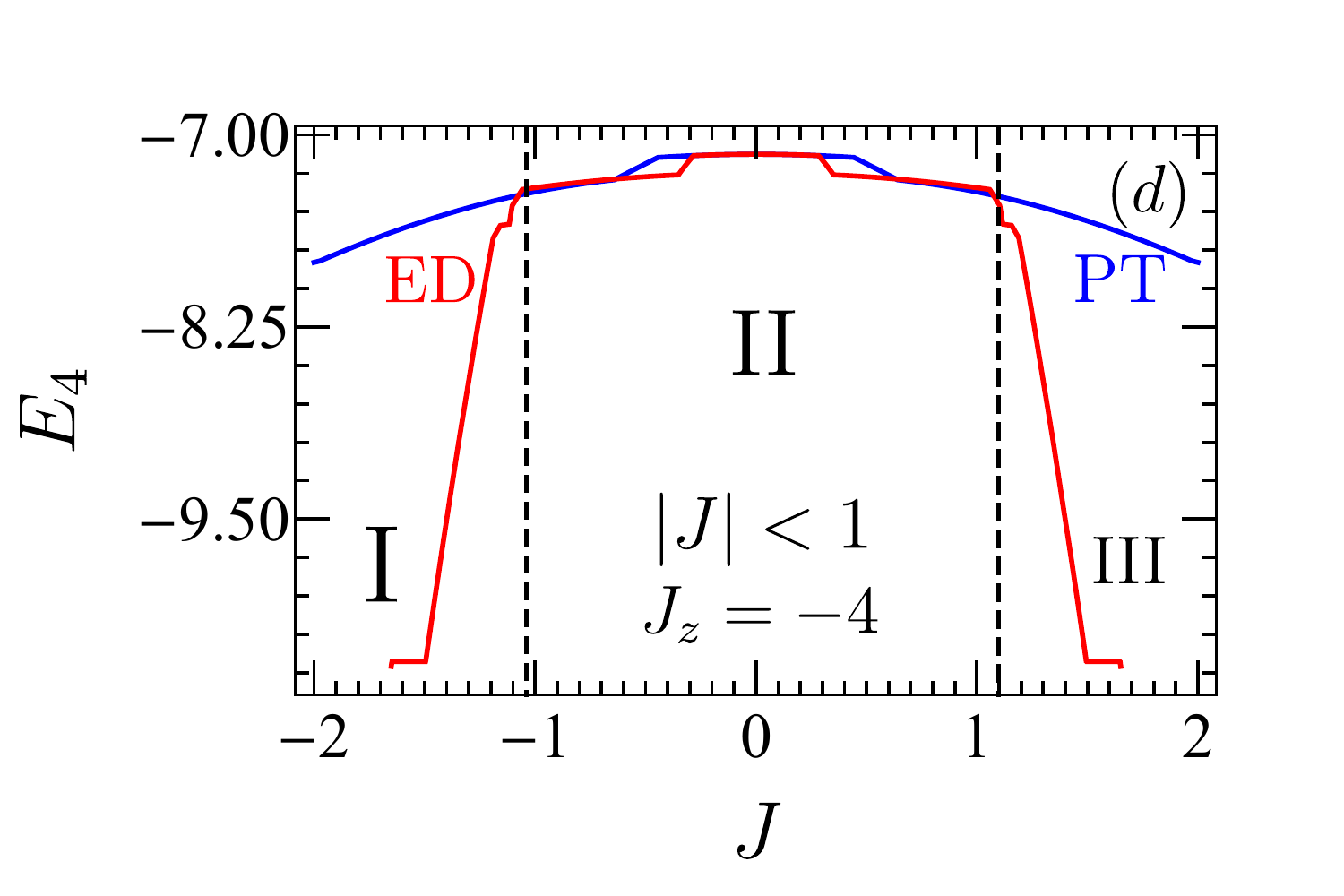}\hspace{-0.1cm}
		\vspace{-0.3cm}
		\caption{{\bf Left-Panel:} Results of exact diagonalization (ED) versus data from perturbation theory (PT) for the random spin-$1/2$ XXZ Heisenberg model with open boundary conditions\cite{Claire2011}. We observe that both data are barely discernible. The values are sorted in ascendant order and this explains why the branches $E_n<0$ and $E_n>0$ as in  Fig.\ref{Figure7} have disappeared.  {\bf Right-Panel:} Range of validity of our PT. We confirm that our PT is valid in the region II ($|J|<1$ and arbitrary $J_z$). It completely fails in the regions I and III. Local magnetic fields (disorder) are withdrawn from a uniform distribution $[-h,h]$. For calculations, we choose $h=3$ and restricted ourselves to small spin system ($K=8$) for the sake of clarity of figures. }
		\label{Figure6} 
	\end{center}
\end{figure}

\subsection{Comparison with previously published exact results}
\label{sec:comparison_exact}

The above obtained analytical results can directly be compared with previously published exact results for the open spin-$1/2$ XXZ chain. This comparison is particularly useful because it provides an independent validation of the recurrence-equation approach derived from our Dirac representation. We consider here the coordinate Bethe-ansatz solution of Kapustin and Skorik\cite{Kapustin1996}, who
studied an open XXZ chain in boundary magnetic fields. Their
Hamiltonian is written in the convention adopted throughout the present work as
\begin{eqnarray}
\nonumber \hat{H}_{\rm KS} = 2\sum_{j=1}^{K-1}
	\left(\hat{S}_j^x\hat{S}_{j+1}^x+\hat{S}_j^y\hat{S}_{j+1}^y+
	\Delta\hat{S}_j^z\hat{S}_{j+1}^z\right)+h_1\hat{S}_1^z+h_2\hat{S}_K^z
	\label{eq:KS_H}.\\
\end{eqnarray}
For vanishing boundary fields, $h_1=h_2=0$, this reduces to the standard open XXZ chain. This Hamiltonian and ours are related through
$J=2$, $J_z=2\Delta$. Consequently, the $J_z=0$ analytical results derived above correspond
to the XX limit $\Delta=0$ of the Bethe-ansatz solution.

Kapustin and Skorik construct the $n$-magnon eigenstates as\cite{Kapustin1996}
\begin{eqnarray}
	|n\rangle=\sum_{x_1<\cdots<x_n}f^{(n)}(x_1,\ldots,x_n)|x_1,\ldots,x_n\rangle ,
\end{eqnarray}
where the coordinate wave function contains all permutations and
sign changes of the magnon momenta $p_j$. The corresponding momenta
are determined by the open-chain Bethe equations. In the XX limit
$\Delta=0$, and in the absence of boundary fields, these equations
reduce to
\begin{eqnarray}
	e^{2i(K+1)p_j}=1,
\end{eqnarray}
and therefore
\begin{eqnarray}
	p_m=\frac{\pi m}{K+1},
	\qquad
	m=1,\ldots,K.
	\label{eq:BA_momenta}
\end{eqnarray}
Eq.(\ref{eq:BA_momenta}) is identical to the
quantization condition obtained directly from the Dirac recurrence
equation, $ \theta_m=\pi m/(K+1)$ which led to Eq.\eqref{equ26b}. For a single spin in the down state, the exact Bethe-ansatz energy in
this limit is
\begin{eqnarray}
E_{\rm BA}^{(1)} = 2\cos p_m = 2\cos\left(\frac{\pi m}{K+1}\right).
	\label{eq:BA_one}
\end{eqnarray}
On the other hand, Eq.\eqref{equ26b} for $h=0$ and $J=2$ yields, $E_{q_m}^{(1)} = J\cos\left(\theta_m\right)$ telling us that $E_{q_m}^{(1)} = E_{\rm BA}^{(1)}$. Thus, the agreement is mode by mode and holds at finite chain length without invoking the thermodynamic limit.

The correspondence extends to the eigenfunctions. The one-particle
wave functions obtained from the recurrence equation are given in Eq.\eqref{equ27}.
The open-chain Bethe wave function contains the two reflected waves with momenta $p_m$ and $-p_m$. Their appropriate linear combination is proportional to
\begin{eqnarray}\label{Eq3.60}
	e^{ip_mj}-e^{-ip_mj} = 2i\sin(p_mj),
\end{eqnarray}
and therefore reproduces Eq.\eqref{equ27}, up to a normalization factor. Hence, both the spectrum and the spatial dependence of the one-magnon eigenstates coincide with the exact known-solution.

The same comparison can be performed in the sector $K_{\downarrow}=2$. For $n=2$, $\Delta=0$, and $h_1=h_2=0$, the Bethe momenta satisfy
\begin{eqnarray}\label{Eq3.61}
	p_\ell=\frac{\pi m_\ell}{K+1},
	\qquad
	1\leq m_1<m_2\leq K,
\end{eqnarray}
with $\ell=1,2$. The Bethe-ansatz energy formula then reads,
\begin{eqnarray}\label{Eq3.62}
E_{\rm BA}^{(2)} = 2\cos\left(\frac{\pi m_1}{K+1}\right) + 2\cos\left(\frac{\pi m_2}{K+1}\right).
\end{eqnarray}
Our Eq.\eqref{equ22gh} evaluated for $h=0$ and $J=2$ gives precisely
$E_{q_{m_1}+q_{m_2}}^{(2)} = E_{\rm BA}^{(2)}$. The equality therefore again holds for every allowed pair $(m_1,m_2)$ at finite $K$.

The agreement also extends to the two-particle wave function. In the
present formulation the hard-core constraint $P_{j,j}=0$ is
implemented by the antisymmetrized combination \eqref{equ22hc}. Using Eq.\eqref{equ27}, this expression becomes
\begin{eqnarray}\label{Eq3.63}
D_{m_1,m_2}^{j_1,j_2} \propto \sin(p_1j_1)\sin(p_2j_2)
	- \sin(p_2j_1)\sin(p_1j_2),\qquad
\end{eqnarray}
which after expressing the sine functions in exponential form, contains precisely the permutations and momentum sign reversals entering the coordinate Bethe-ansatz wave function for an open chain. The two constructions therefore yield the same two-particle eigenspace, up to normalization and overall phase.

These comparisons  explicitly establish that the recurrence equations steming from the Dirac representation reproduce the previously known exact open-chain results in the sectors where a direct analytical comparison can be made. This suggests that the same finite-size spectra are recovered here without resorting to Bethe rapidities or solving the coupled Bethe equations. In the XX limit, they follow directly from the finite-dimensional recurrence relations generated by the lattice-spin Dirac representation.

As a yet another benchmark, we considered the $1D$ transverse-field Ising model exactly solved by Pfeuty\cite{Pfeuty1970}. This comparison is useful because, unlike the XXZ Hamiltonian considered above, this model does not conserve the total magnetization. It therefore stand as another test-bed for the Dirac representation outside the fixed-$K_\downarrow$ decomposition used in the preceding sections. Details of our derivations are deferred in Appendix \ref{AppC}.

\section{Non-integrable version}\label{Section4}
We have so far discussed the integrable version of \eqref{equ18} and got full access to its eigenspectrum in some specific cases. Although this helps gain insight into the physics of spin systems, physics can be radically different when systems are found in more realistic situations. For example spin systems contaminated with sufficiently strong disorder exhibit many-body localization losing their ability to thermalize and explore their entire Hilbert space\cite{Schliemann2021, Siegl2023}. This leads to a new phase of matter. Understanding this phenomenon requires analyzing energy-level spacing or inverse participation ratios (IPR)\cite{Tikhonov2018, Vu2022}. The method presented above fails as a consequence of the fact that the $h_j$ are all distinct and the anisotropy parameter playing a crucial role leads to tight-binding equations that cannot be solved in compact form. It is commonly desired in such cases to use exact-diagonalization (ED) to evaluate the eigenspectrum in order to discriminate between ergodic and MBL phases and to probe precursors of localization and benchmark against ED in spin chains.
As already mentioned the exponential growth of the Hilbert space with the total number of spins, causes finite-size effects dramatically limiting the number of implementable spins. The strongest methods requiring large computer memory and long runtimes can handle only up to $K=24$ spins (to the best of our knowledge) provided one can identify and exploit certain symmetries as we did in Section.\ref{Section3}. Throughout we assume generic continuous disorder, so the unperturbed spectrum is almost surely non-degenerate; near-resonances control the convergence of the expansion.

We propose in this Section to perturbatively calculate the eigenspectrum (up to the fourth order) of the spin-$1/2$ random XXZ Heisenberg model \eqref{equ18} in an open boundary configuration. Analytical expressions are written down and could be used to dig deeper into MBL. Our results are confirmed with data from ED in small chains and could easily be extended to arbitrary $K$ by simply setting the desired value of $K$ into our expressions.  

A perturbation theory including  higher order is elaborated in Appendix \ref{PT}. As traditionally required, the Hamiltonian is split into an unperturbed part $\hat{H}_0$ whose spectrum is well-identified and a perturbed part $\hat{H}_1$ with respect to a perturbative parameter, say $J$ such that $\hat{H}=\hat{H}_0+J\hat{H}_1$. For our case, $\hat{H}_0 =J_z\sum^{K-1}_{j=1}\hat{S}^{z}_{j}\hat{S}^{z}_{j+1}+\sum^K_{j=1}h_{j}\hat{S}^{z}_{j}$ is diagonal with entries $\Delta^{zz}_n$ in Eq.\eqref{equ21} assumed to be  real and non-degenerate while $\hat{H}_1= \sum^{K-1}_{j=1}\left(\hat{S}^{x}_{j}\hat{S}^{x}_{j+1}+\hat{S}^{y}_{j}\hat{S}^{y}_{j+1}\right)$ with the structured symmetric matrix elements $X_{n,m}=\bra{n}\hat{H}_1\ket{m}$ given by,
\begin{eqnarray}\label{ob2}
	X_{n,m} = J^{-1}\sum_{j=1}^{K-1}\left(\Delta_m^{\left(j\right)+-}\delta_{n,m-Q_j} + \Delta_m^{\left(j\right)-+}\delta_{n,m+Q_j}\right).
\end{eqnarray}
The factor $J^{-1}$ is inserted in order to compensate the $J$ in $\Delta_m^{\left(j\right)+-}$ and $\Delta_m^{\left(j\right)-+}$ such that $X_{n,m}$ becomes independent on $J$.
In the Dirac representation, we see that $X_{n,m}$ can be regarded as a generalized hopping model  where $\Delta_m^{\left(j\right)+-}$ and $\Delta_m^{\left(j\right)-+}$ determine the hopping amplitude between neighboring sites separated by $Q_j$ such that $X_{n,n+Q_j}=\Delta_{n+Q_j}^{\left(j\right)+-}=\Delta_{n}^{\left(j\right)-+}$ and $X_{n,n-Q_j}=\Delta_{n-Q_j}^{\left(j\right)-+}=\Delta_{n}^{\left(j\right)+-}$. Note that $X_{n,n}=0$ typically because $Q_j>0$ and there is no $j$ in the interval $1\leq j\leq K-1$ for which $\delta_{n,n+Q_j}=\delta_{n,n-Q_j}=1$.  

For our perturbative analysis, we are interested in the energy and wave function of the $n$th configuration determined up to the fourth order as
\begin{eqnarray}\label{ob3}
	E_n\approx \sum_{k=0}^{3}J^k\mathcal{E}_{n}^{\left(k\right)} +\mathcal{O}\left(J^4\right),
\end{eqnarray}
\begin{eqnarray}\label{ob4}
	\ket{\psi_n}\approx \sum_{k=0}^{2}J^k\ket{\psi_n^{\left(k\right)}}+\mathcal{O}\left(J^3\right).
\end{eqnarray}
Here, $\mathcal{E}^{\left(k\right)}_{n}$ and $\ket{\psi_n^{\left(k\right)}}$ are the $k$-order corrections to the energy and eigenstates respectively. The superscripts should not be mistaken for sectors labels and the subscripts $m$ do not refer to the magnetic quantum number. Foremost, we start by describing the energy. According to the perturbation theory in Appendix \ref{PT} the zero-order contribution to the energy corresponds to the eigenvalue of $\hat{H}_0$  and reads $\mathcal{E}^{\left(0\right)}_{n}=\Delta^{zz}_n$ while the first order $\mathcal{E}^{\left(1\right)}_{n} = X_{n,n} = 0$ for the aforementioned reasons. Other contributions write
\begin{eqnarray}\label{ob5}
	\mathcal{E}^{\left(2\right)}_{n} =\sum_{\substack{m=0 \\ m\neq n}}^{D-1}\frac{X_{n,m}X_{m,n}}{\Lambda_{n,m}},
\end{eqnarray}
and
\begin{eqnarray}\label{ob6}
	\mathcal{E}^{\left(3\right)}_{n} =\sum_{\substack{m_1=0 \\ m_1\neq n}}^{D-1}\sum_{\substack{m_2=0 \\ m_2\neq n,\, m_2\neq m_1}}^{D-1}\frac{X_{n,m_1}X_{m_1,m_2}X_{m_2,n}}{\Lambda_{n,m_1}\Lambda_{n,m_2}},
\end{eqnarray}
where $\Lambda_{n,m}$ denotes the energy difference (gap) $\Lambda_{n,m}=\Delta^{zz}_{n}-\Delta^{zz}_{m}$ between the configurations $\ket{n}$ and $\ket{m}$. It is also the energy accumulated by the system when transitioning from $\ket{n}$ to $\ket{m}$ vice versa. Note that when $J=0$ (Anderson limit) no transition between the eigenstates of $\hat{H}_0$ occurs, $E_n=\Delta_n^{zz}$ as all corrections vanish, the system remains in its original configuration. 

In the expression for the eigenstates, the zero-order correction corresponds to the eigenstate of $\hat{H}_0$ namely $\ket{\psi_n^{\left(0\right)}}=\ket{n}$ whereas the first and second order contributions are respectively given by
\begin{eqnarray}\label{ob7}
	\ket{\psi_n^{\left(1\right)}}=\sum_{\substack{m=0 \\ m\neq n}}^{D-1}\frac{X_{n,m}}{\Lambda_{n,m}}\ket{m},
\end{eqnarray}
and
\begin{eqnarray}\label{ob8}
	\hspace{-0.75cm}	\ket{\psi_n^{\left(2\right)}}=\sum_{\substack{m=0 \\ m\neq n}}^{D-1}\sum_{\substack{m_1=0 \\ m_1\neq n}}^{D-1}\frac{X_{n,m_1}X_{m_1,m}}{\Lambda_{n,m_1}\Lambda_{n,m}}\ket{m}-\frac{1}{2}\left(\sum_{\substack{m=0 \\ m\neq n}}^{D-1}\frac{X_{n,m}X_{m,n}}{\Lambda^2_{n,m}}\right)\ket{n}.
\end{eqnarray}
Albeit the above expressions can readily be implemented in any symbolic calculator, our main goal is to analytically calculate them and break them down into  sectors corresponding to fixed values of the magnetic quantum number.

\subsubsection{\bf Eigenenergies}
We start with the eigenenergies. Thanks to the hopping-shaped form of $X_{n,m}$ in terms of Kronecker Delta symbols plugging this back into Eq.\eqref{ob5} allows us after a few calculations to find that:
\begin{eqnarray}\label{ob9}
	\mathcal{E}^{\left(2\right)}_{n} =J^{-2}\sum_{j=1}^{K-1}\left(\frac{\left(\Delta_n^{\left(j\right)-+}\right)^2}{\Lambda_{n,n+Q_j}}+\frac{\left(\Delta_n^{\left(j\right)+-}\right)^2}{\Lambda_{n,n-Q_j}}\right).
\end{eqnarray}
Similarly, $\mathcal{E}^{\left(3\right)}_{n} = 0$ because every application of 
$\hat{H}_1$ flips a nearest-neighbor pair, so returning to $\ket{n}$  requires an even number of flips (no closed walk of odd length in the configuration graph). As a consequence of these results, $E_n\approx \Delta^{zz}_n + J^2\mathcal{E}^{\left(2\right)}_{n} +\mathcal{O}\left(|J|^4\right)$, the energy holds up to the order $\mathcal{O}\left(|J|^4\right)$. Note that although $J^2\mathcal{E}^{(2)}_n$ is second order in the perturbation $J\hat{H}_1$, the Dirac hopping amplitudes satisfy $\Delta^{(j)\pm\mp}_n\propto J$; using the projector identities $(\Delta_n^{\left(j\right)-+})^2=\frac{J}{2}(\Delta_n^{\left(j\right)-+})$ and $(\Delta_n^{\left(j\right)+-})^2=\frac{J}{2}(\Delta_n^{\left(j\right)+-})$ as elaborated in the Subsection \ref{Section AT} yields an overall correction linear in $J$. Thus,
\begin{eqnarray}\label{cb5}
	E_n\approx \Delta^{zz}_n + \frac{J}{2}\sum_{j=1}^{K-1}\left(\frac{\Delta_n^{\left(j\right)-+}}{\Lambda_{n,n+Q_j}}+\frac{\Delta_n^{\left(j\right)+-}}{\Lambda_{n,n-Q_j}}\right) +\mathcal{O}\left(|J|^4\right).
\end{eqnarray}
This result is numerically verified (small chains, generic disorder realizations) and we found a good agreement with ED for $|J| \lesssim  1$  as depicted in Fig.\ref{Figure6}. We observe that the eigenenergies converge well to the exact diagonalization results. This is a consequence of the fact that $\mathcal{E}_n^{(1)}=\mathcal{E}_n^{(3)}=0$ leading to a high correction. We also observe from Fig.\ref{Figure6}$(a)$ that for large anisotropy ($J_z\gg 1$) when $E_n$ are sorted in ascending order as a function $n$ they depict several clusters where energies are nearly degenerate. This behavior disappears for small and moderate anisotropy as shown in Fig.\ref{Figure6}$(c)$.
 
 Despite the above observations, it remains of paramount importance to break \eqref{cb5} down into sectors (with a focus on even $K$). To this end, we first prove that for $1\leq K_{\downarrow}\leq K-1$ and $1\leq j_1<j_2<\cdots <j_{K_{\downarrow}}\leq K-1$,
 \begin{eqnarray}\label{ob10a}
 	\nonumber\Delta^{zz}_{q_{j_1}+q_{j_2}+\cdots+q_{j_{K_{\downarrow}}}} =\left(\frac{{\bf h}}{2}-\sum_{k=1}^{K_{\downarrow}}h_{j_k}\right)+J_z\left(\frac{K-1}{4}-K_{\downarrow}\right)\\+J_z\sum_{\ell=1}^{K_{\downarrow}}\left(f_{j_\ell}+\sum_{k\neq \ell}^{K_{\downarrow}}\delta_{j_k,j_{\ell}+1}\right),
 \end{eqnarray}
 where $f_{j_\ell} = \left(\delta_{j_\ell,1}-\delta_{j_\ell,K}\right)/2$ is associated with spin chain's gates; $j_\ell=1$ $(f_1=1/2)$ and  $j_\ell=K$ $(f_K=-1/2)$ are respectively the left and right gates. The region $1<j<K$ is the active domain where $f_j=0$.
  Expression \eqref{ob10a} is extended to the last spin i.e. $1\leq j_1<j_2<\cdots <j_{K_{\downarrow}}\leq K$ by replacing $f_{j_\ell}$ by $g_{j_\ell} = \left(\delta_{j_\ell,1}+\delta_{j_\ell,K}\right)/2$. Defining the occupation number $n_i=\sum_{\ell=1}^{K_{\downarrow}}\delta_{i,j_\ell}$ we learn that $\sum_{\ell=1}^{K_{\downarrow}}f_{j_\ell}=\left(n_1-n_K\right)/2$ . Similarly, $\sum_{\ell=1}^{K_{\downarrow}}\sum_{k\neq \ell}^{K_{\downarrow}}\delta_{j_k,j_{\ell}+1}=\sum_{i=1}^{K-1}n_in_{i+1}$ where $n_i=(0,1)$. 
  
   It should be noted that the second order correction \eqref{ob9} has the same structure as \eqref{equ23}. $\mathcal{E}^{\left(2\right)}_{n}$ is therefore expected in each sector to cast the form of the relations for the $P_{j_1,\cdots,K_{\downarrow}}$ as shown below.
 
\begin{figure}[]
	\centering
	\begin{center} 
		\includegraphics[width=7.5cm, height=5cm]{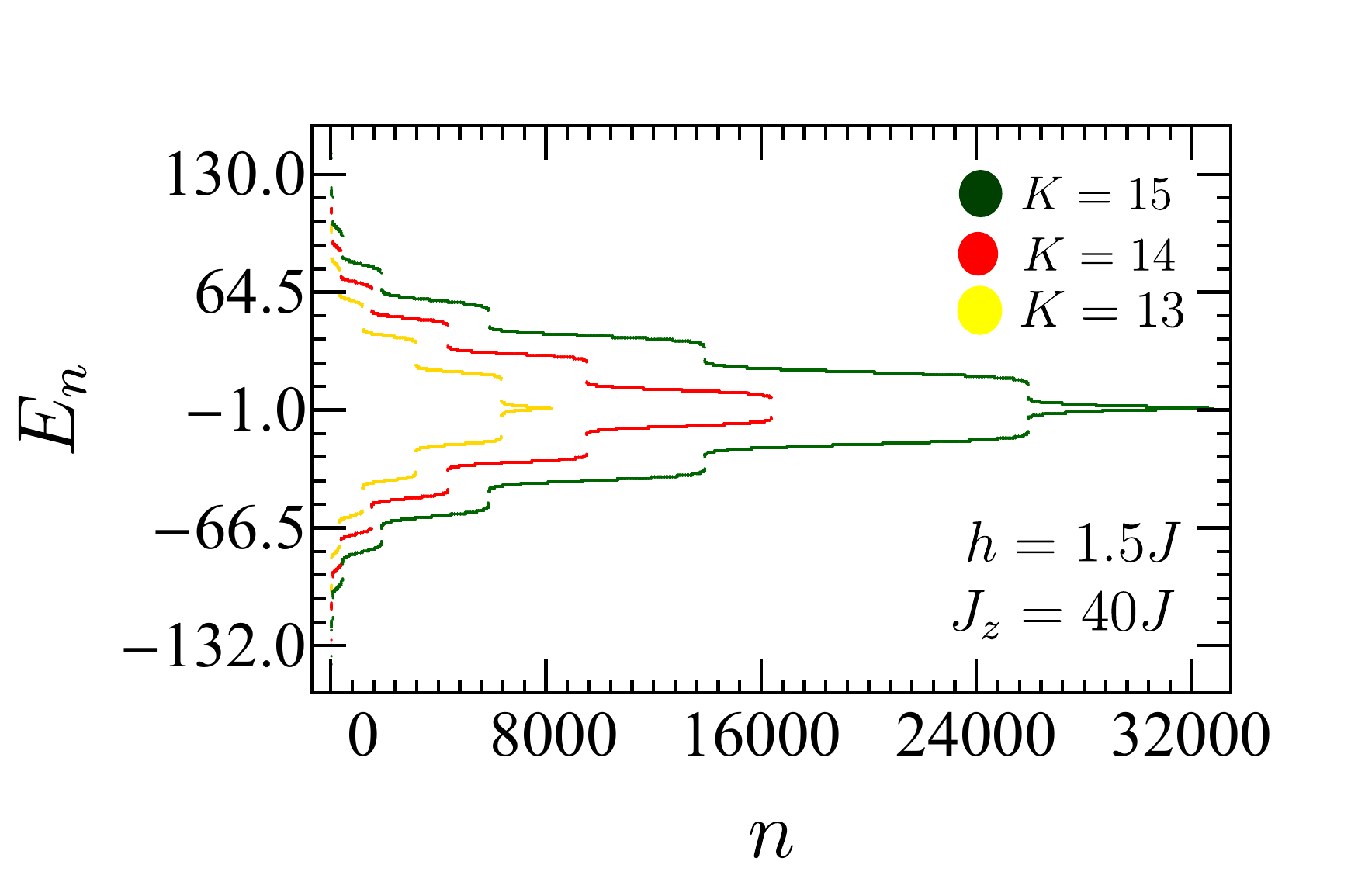}\vspace{-0.4cm}
		\includegraphics[width=7.5cm, height=5cm]{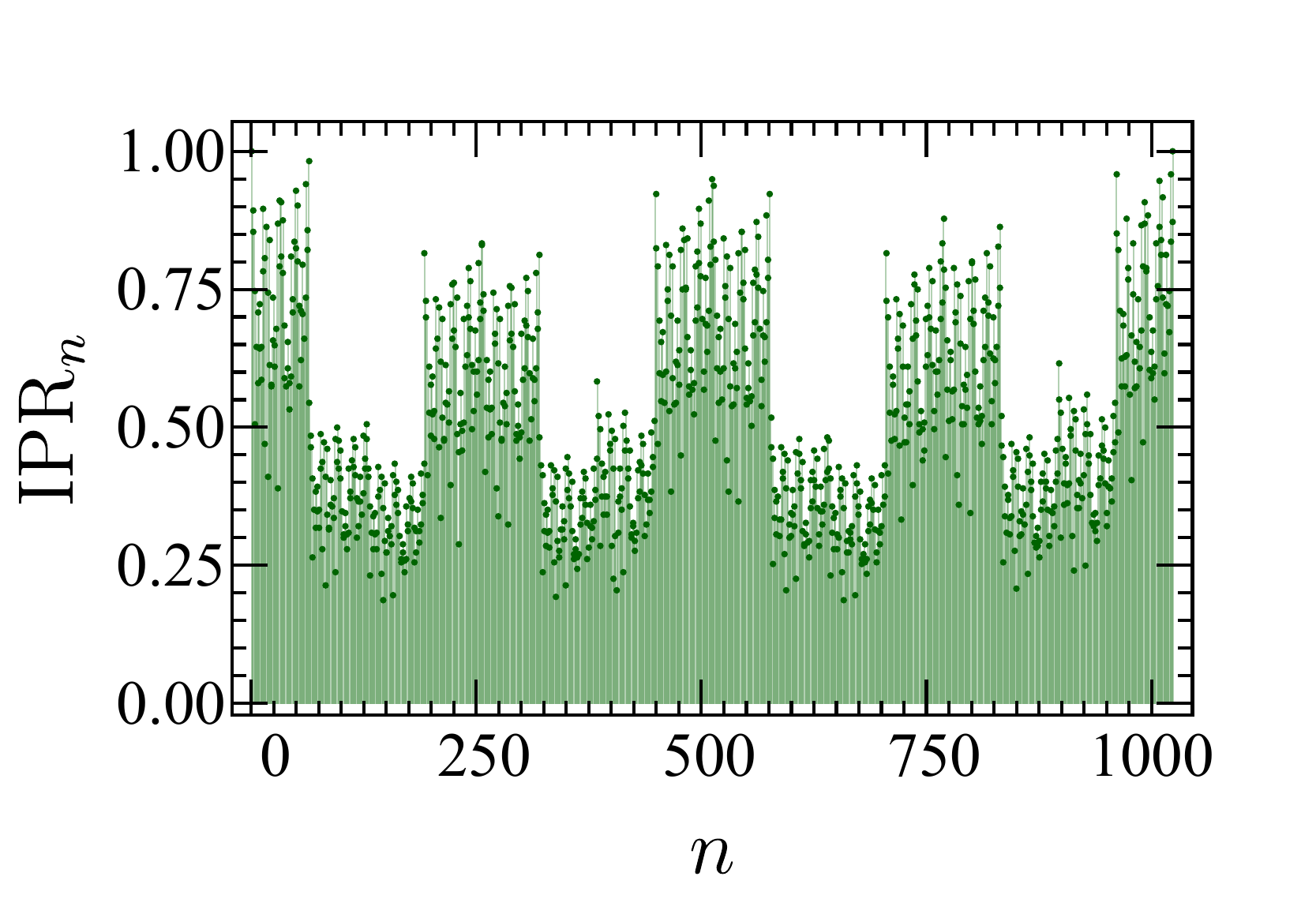}\\
		\vspace{-0.5cm}
		\caption{{\bf Upper panel:} Eigenenergies of the model \eqref{equ18} as a function of $n$ for various values of $K$ and $h\neq 0$ in opposition to Fig.\ref{figure4} where eigenenergies depict vertical curves (for $h=0$) here we have nearly-horizontal lines. For each value of $K$ the figure has two branches consisting of clusters. Each cluster corresponds to a given value of the total magnetization. For even $K$, the branch $E_n>0$ corresponds to the sectors $K/2\leq \mathcal{J}^z\leq 0$ while the branch $E_n<0$ corresponds to the sectors $-K/2\leq \mathcal{J}^z< 0$. We observe that $E_n$ increases with $K$. {\bf Lower panel:} average inverse participation ratio\cite{Misguich2016, Tikhonov2018}, ${\rm IPR}_n=\sum_{m}|\braket{m}{\psi_n}|^4$. Here, $\braket{m}{\psi_n}$ is the projection of $\ket{\psi_n}$ onto the direction of $\ket{m}$ in many-body basis.}
		\label{Figure7} 
	\end{center}
\end{figure}

 {\bf Sector  $K_{\downarrow}=1$:}
  In this sector, classical variables are given by Eq.\eqref{equ24}. The energy of the system reads,
 \begin{eqnarray}\label{ob11a}
 	E^{\left(K_{\downarrow}=1\right)}_{q_j} \approx\Delta^{zz}_{q_{j}}+\left(W_{j,j-1}+W_{j,j+1}\right)+\mathcal{O}\left(J^4\right),
 \end{eqnarray}
 where
  \begin{eqnarray}\label{ob11b}
 	\Delta^{zz}_{q_{j}} =\left(\frac{{\bf h}}{2}-h_{j}\right)+J_z\left(\frac{K-1}{4}-1\right)+J_zg_{j},
 \end{eqnarray}
 and
 \begin{eqnarray}\label{ob11c}
 	W_{i,j} = -\frac{\left(\frac{J}{2}\right)^2}{\left(h_i-h_j\right)-J_z\left(g_i-g_{j}\right)},
 \end{eqnarray}
 is defined with the constraints $W_{i,0}=W_{i,K+1}=W_{i,i}=0$. Eq.\eqref{ob11a} is reminiscent of a $1D$ effective tight-binding picture in configuration space for a free traveling particle. In this context, $\Delta^{zz}_{q_{j}}$ plays the role of the onsite energy while the dimensionless parameter $W_{i,j}$ depicts the interplay between interactions and disorder on one hand, disorder and anisotropy on the other hand and can be interpreted as the transfer rate between the states $\ket{q_{i}}$ and $\ket{q_{j}}$. Population transfer is done by successive jumps between neighboring states. These results could be interesting in deciphering the complex mechanism behind MBL in spin chains. For example, for a sample drawn from a uniform distribution, we have calculated the inverse participation ratio (IPR) in Fig.\ref{Figure7} for $500$ realizations. We see that the wave function tends to concentrate in certain regions more than others indicating the existence of localization.
 

 {\bf Sector  $K_{\downarrow}=2$:} In this subspace, $\Delta^{(j)+-}_{q_{j_1}+q_{j_2}}$ and $\Delta^{(j)-+}_{q_{j_1}+q_{j_2}}$ are given by \eqref{equ22e} while 
 $\Delta^{zz}_{q_{j_1}+q_{j_2}} $ can easily be inferred from  \eqref{ob10a}. As a result of substituting the corresponding expressions into \eqref{cb5} one obtains
  \begin{eqnarray}\label{Ob13}
 \nonumber\left[\left(1-\delta_{j_1-1,j_2}\right)W_{j_1-1, j_2} + \left(1-\delta_{j_1+1,j_2}\right)W_{j_1+1,j_2} \right] &+&\\\nonumber \left[\left(1-\delta_{j_1,j_2-1}\right)W_{j_1, j_2-1} + \left(1-\delta_{j_1,j_2+1}\right)W_{j_1,j_2+1} \right] &\approx&\\ \approx E^{\left(K_{\downarrow}=2\right)}_{q_{j_1}+q_{j_2}}-\Delta^{zz}_{q_{j_1}+q_{j_2}},
\end{eqnarray}
 where
 \begin{eqnarray}\label{ob13a}
 	\nonumber\Delta^{zz}_{q_{j_1}+q_{j_2}} =\left(\frac{{\bf h}}{2}-\sum_{k=1}^{2}h_{j_k}\right)+J_z\left(\frac{K-1}{4}-2\right)\\+J_z\sum_{\ell=1}^{2}g_{j_\ell}+J_z\left(\delta_{j_1,j_{2}+1}+\delta_{j_2,j_{1}+1}\right),
 \end{eqnarray}
 and 
 \begin{subeqnarray}\label{ob14}
 	W_{j_1,j_2\pm 1} =\frac{\left(\frac{J}{2}\right)^2}{\Delta^{zz}_{q_{j_1}+q_{j_2}}-\Delta^{zz}_{q_{j_1}+q_{j_2\pm 1}}},\\
 	W_{j_1\pm 1,j_2} =\frac{\left(\frac{J}{2}\right)^2}{\Delta^{zz}_{q_{j_1}+q_{j_2}}-\Delta^{zz}_{q_{j_1\pm 1}+q_{j_2}}}.
 \end{subeqnarray}
 Here,
\begin{subeqnarray}\label{ob15}
	\nonumber\Delta^{zz}_{q_{j_1}+q_{j_2}}-\Delta^{zz}_{q_{j_1\pm 1}+q_{j_2}}=-\left(h_{j_1}-h_{j_1\pm 1}\right)+J_z\left(g_{j_1}-g_{j_1\pm 1}\right)\\+J_z\left(\delta_{j_1+1,j_2}-\delta_{j_1\pm2,j_2}\right),\quad\\
	\nonumber\Delta^{zz}_{q_{j_1}+q_{j_2}}-\Delta^{zz}_{q_{j_1}+q_{j_2\pm 1}}=-\left(h_{j_2}-h_{j_2\pm 1}\right)+J_z\left(g_{j_2}-g_{j_2\pm 1}\right)\\+J_z\left(\delta_{j_1+1,j_2}-\delta_{j_1,j_2\pm2}\right).\quad
\end{subeqnarray}
 Eq.\eqref{Ob13} is subject to the constraint $W_{0, j_2}=W_{j_1, 0}=W_{j_1+K, j_2}=W_{j_1, j_2+K}=W_{j,j}=0$. It mimics a two-dimensional semi-classical tight-binding model. Population transfer is equally done by successive jumps between neighboring states.
 
The energy in other sectors can be evaluated in a similar fashion. However, as suggested earlier, the perturbative energy being of the form \eqref{equ23}, we have confirmed that in each sector it can be reduced to the tight-binding equation for the  $P_{j_1,\cdots,K_{\downarrow}}$. For instance, \eqref{ob11} in the sector $K_{\downarrow}=1$ mimics \eqref{equ25} while \eqref{Ob13} mimics \eqref{equ22g}. It is then legitimate to claim without further calculations that in the sector $K_{\downarrow}=3$ the energy will be of the form \eqref{equ23b} with the respective constraints. Therefore, in arbitrary sector and without further evidence,
\begin{eqnarray}\label{ob16}
\nonumber	 E^{\left(K_{\downarrow}\right)}_{q_{j_1}+q_{j_2}+\cdots+q_{j_{K_{\downarrow}}}}\approx \Delta^{zz}_{q_{j_1}+q_{j_2}+\cdots+q_{j_{K_{\downarrow}}}}+\hspace{3.2cm}\\+\sum_{\ell=1}^{K_{\downarrow}}\left(\mathcal{X}_{j_\ell-1}W_{j_1,\cdots, j_\ell- 1,\cdots, j_{K_{\downarrow}}} + \mathcal{X}_{j_\ell+1}W_{j_1,\cdots, j_\ell+1,\cdots, j_{K_{\downarrow}}}\right), \qquad
\end{eqnarray}
where
\begin{eqnarray}\label{ob17}
\nonumber	W_{j_1,\cdots, j_\ell\pm 1,\cdots, j_{K_{\downarrow}}} =\frac{\left(\frac{J}{2}\right)^2}{\Delta^{zz}_{q_{j_1}+q_{j_2}+\cdots+q_{j_{K_{\downarrow}}}}-\Delta^{zz}_{q_{j_1}+q_{j_2}\cdots+q_{j_\ell\pm 1}+\cdots+q_{j_{K_{\downarrow}}}}}.\\
\end{eqnarray}
The constraints on the $W_{j_1,\cdots, j_\ell\pm 1,\cdots, j_{K_{\downarrow}}}$ are the same as those on $P_{j_{\ell}},\cdots, j_{K_{\downarrow}}$. The denominators in \eqref{ob17} are calculated from \eqref{ob10a}.

\subsubsection{\bf Eigenstates}
For eigenstates, after calculations one obtains at the first order,
\begin{eqnarray}\label{ob10}
	\hspace{-0.75cm}	\ket{\psi_n^{\left(1\right)}}
	=\sum_{j=1}^{K-1}\left(\frac{\Delta_{n}^{\left(j\right)-+}}{\Lambda_{n,n+Q_j}}\ket{n+Q_j}+\frac{\Delta_{n}^{\left(j\right)+-}}{\Lambda_{n,n-Q_j}}\ket{n-Q_j}\right).
\end{eqnarray}
This relation is also of the form \eqref{equ23}. Thus, without further calculations and for the same reasons as evoked earlier, it can be broken down into sectors exactly as we did for the energies. In the second and third sectors for example:
\begin{eqnarray}\label{ob11}
	\hspace{-0.75cm}	\ket{\psi_{q_j}}
	\approx \ket{q_j}+\left(W_{j,j+1}\ket{q_{j+1}}+W_{j,j-1}\ket{q_{j-1}}\right).
\end{eqnarray}
and
\begin{eqnarray}\label{ob12}
	\nonumber\ket{\psi_{q_{j_1}+q_{j_2}}}\approx\ket{q_{j_1}+q_{j_2}}&+&\Big[\left(1-\delta_{j_1-1,j_2}\right)W_{j_1-1, j_2}\ket{q_{j_1-1}+q_{j_2}} \\\nonumber&+& \left(1-\delta_{j_1+1,j_2}\right)W_{j_1+1, j_2}\ket{q_{j_1+1}+q_{j_2}} \Big]\\\nonumber&+&
	 \Big[\left(1-\delta_{j_1,j_2-1}\right)W_{j_1, j_2-1}\ket{q_{j_1}+q_{j_2-1}}\\\nonumber  &+& \left(1-\delta_{j_1,j_2+1}\right)W_{j_1, j_2+1}\ket{q_{j_1}+q_{j_2+1}} \Big].\\
\end{eqnarray}
The second order correction can be calculated in a similar manner and broken into sectors.
This task is left for future projects.

We learn from \eqref{ob11} that if $h_i$ and $h_j$  become accidentally close while $J_z$ is maintained constant and small compared to the difference $h_i-h_j$, the denominator of $W_{i,j}$ can be very small leading it to acquiring large values while it remains small at other sites. This distorts the wave function causing it to concentrate in certain regions, effectively localizing the states. If the $h_i$ smoothly vary, the eigenstates remain extended but may exhibit weak localization effects.

The above perturbation theory holds for $|J| \lesssim 1$ and arbitrary $J_z$ and local magnetic field. The regime $|J|\gg1$ can be addressed by rotating $\hat{H}_0$ into the eigenbasis of $\hat{H}_1$. In this case $\hat{H}_1$ becomes diagonal with elements the eigenvalues in \eqref{equ23e} and the eigenstates \eqref{equ22hb}. $\hat{H}_0$ now is off-diagonal and the exact above same calculations are repeated by replacing $\ket{n}\to\ket{\psi_n}$ and $\Delta_n^{zz}\to E_n$.

\section{Outlook and perspective: Spin-$1$ Case}\label{Section5}
We have presented in previous Sections the "bra-ket" (Dirac representation) for spin-$1/2$ lattice operators and discussed a few applications. Because the case of spin-1 follows the same algebraic strategy, we present it here for completeness. In order to avoid lenghtly paper, possible applications are deferred to future projects.

In this case, spin projection along the quantization axis takes three values $+1,0,-1$. These local three-level systems (qutrits) constitute the building blocks of ternary quantum computing. They are known to be robust against external unwanted nuisances, encoding more information (already at room temperature) compared to qubits and offering better security during communication\cite{Lanyon2008, Dolde2014}. For these reasons it would be crucial to achieve qutrit without ternary decomposition as we did for qubits. This could improve on encrypting information during communications between quantum devices. 

Foremost, note that qutrits can also be mapped onto their classical analogue $\sigma_{n}^{(j)}=\{0,1,2\}$ (the trits), of the base-$3$. Classical trits cannot be subjected to the principle of quantum superposition nor entanglement in contrast to their quantum analogue. Here, $n$ is the positive integer whose decomposition in base $3$ encapsulates the exact configuration of the projections $+1,0,-1$. It can exactly be written as in Eq.\eqref{equ3} now with $q_j=3^{K-j}$ and $D=3^K$. Without loss of generality, it can be assumed that
$\ket{+1}\to \ket{0}= [1,0,0]^T$, $\ket{0}\to\ket{1}= [0,1,0]^T$ and $\ket{-1}\to\ket{2}= [0,0,1]^T$ yielding,
\begin{eqnarray}\label{equ11}
	\bra{m}S^{x}_j\ket{n}=
	\frac{1}{\sqrt{2}}\left\{\begin{array}{c}
		\delta_{m, \chi_j^{[n,1]}},  \quad {\rm if} \quad \sigma^{(j)}_n=0,\\  
		\delta_{m, \chi_j^{[n,0]}}+\delta_{m, \chi_j^{[n,2]}},  \quad {\rm if} \quad \sigma^{(j)}_n=1,\\  
		\delta_{m, \chi_j^{[n,1]}},  \quad {\rm if} \quad \sigma^{(j)}_n=2,\\  
	\end{array}
	\right. \qquad 
\end{eqnarray}
\begin{eqnarray}\label{equ12}
	\bra{m}S^{y}_j\ket{n}=
	-\frac{i}{\sqrt{2}}\left\{\begin{array}{c}
		-\delta_{m, \chi_j^{[n,1]}},  \quad {\rm if} \quad \sigma^{(j)}_n=0,\\  
		\delta_{m, \chi_j^{[n,0]}}-\delta_{m, \chi_j^{[n,2]}},  \quad {\rm if} \quad \sigma^{(j)}_n=1,\\  
		\delta_{m, \chi_j^{[n,1]}},  \quad {\rm if} \quad \sigma^{(j)}_n=2,\\  
	\end{array}
	\right. \qquad 
\end{eqnarray}
Here, the integer $\chi_j^{[n,a]}$ is the number obtained from the ternary decomposition of $n$ by replacing the $j$th digit by $a$. For the same reasons as for the spin-$1/2$ case, we see that if $\sigma_n^{(j)}=0$ then replacing this digit by $1$ is equivalent to adding $q_j$. Similarly, if $\sigma_n^{(j)}=1$ then replacing this digit by $0$ or $2$ is equivalent to subtracting  $q_j$ or adding $q_j$ respectively. These observations can be translated in a mathematical language as
\begin{eqnarray}\label{equ12a}
	\chi_j^{[n,0]}=
	\left\{\begin{array}{c}
		n,  \quad {\rm if} \quad \sigma^{(j)}_n=0,\\  
		n+q_j,  \quad {\rm if} \quad \sigma^{(j)}_n=1,\\  
		n+2q_j,  \quad {\rm if} \quad \sigma^{(j)}_n=2,\\  
	\end{array}
	\right.
\end{eqnarray}
\begin{eqnarray}\label{equ12b}
	\chi_j^{[n,1]}=
	\left\{\begin{array}{c}
		n+q_j,  \quad {\rm if} \quad \sigma^{(j)}_n=0,\\  
		n,  \quad {\rm if} \quad \sigma^{(j)}_n=1,\\  
		n-q_j,  \quad {\rm if} \quad \sigma^{(j)}_n=2,\\  
	\end{array}
	\right.
\end{eqnarray}
\begin{eqnarray}\label{equ12c}
	\chi_j^{[n,2]}=
	\left\{\begin{array}{c}
		n+2q_j,  \quad {\rm if} \quad \sigma^{(j)}_n=0,\\  
		n-q_j,  \quad {\rm if} \quad \sigma^{(j)}_n=1,\\  
		n,  \quad {\rm if} \quad \sigma^{(j)}_n=2,\\  
	\end{array}
	\right.
\end{eqnarray}
These functions are defined in the same philosophy as the ${\rm\bf flip}()$ function in the spin-$1/2$ case. The promised Dirac notations are constructed from Eqs.\eqref{equ11} and \eqref{equ12} considering Eqs.\eqref{equ12a}-\eqref{equ12c}. After considerable calculations one obtains
\begin{subeqnarray}\label{equ13}
	\slabel{equ13a}	\hat{S}_{j}^{+} &=& \sqrt{2}\sum_{n=0}^{D-1}\Gamma_n^{\left(j\right)+}\ket{n-q_j}\bra{n},\\ 
	\slabel{equ13b}	\hat{S}_{j}^{-} &=& \sqrt{2}\sum_{n=0}^{D-1}\Gamma_n^{\left(j\right)-}\ket{n+q_j}\bra{n}, \\ 
	\slabel{equ13c}	\hat{S}_{j}^{z} &=& \sum_{n=0}^{D-1}\Gamma_n^{\left(j\right)z}\ket{n}\bra{n}.
\end{subeqnarray}
Although a qutrit can exist in three distinct quantum states, we observe that apart from the $\sqrt{2}$ factor  Eqs.\eqref{equ13} have the same structure as Eqs.\eqref{equ3f} where now we have defined the classical trit,
\begin{subeqnarray}\label{equ14}
	\Gamma_n^{\left(j\right)+}=\delta\left(\sigma_n^{(j)},1\right)+\delta\left(\sigma_n^{(j)},2\right), \\ \Gamma_n^{\left(j\right)-}=\delta\left(\sigma_n^{(j)},0\right)+\delta\left(\sigma_n^{(j)},1\right), \\ \Gamma_n^{\left(j\right)z}=\delta\left(\sigma_n^{(j)},0\right)-\delta\left(\sigma_n^{(j)},2\right).
\end{subeqnarray}
Here, we have adopted the notation $\delta\left(n,m\right)\equiv \delta_{n,m}$.
For all possible values of trits, it can be verified that
\begin{eqnarray}\label{SM19}
	\Gamma_n^{\left(j\right)z}=
	\left\{\begin{array}{c}
		+1,  \quad {\rm if} \quad \sigma^{(j)}_n=0,\\  
		0,  \quad {\rm if} \quad \sigma^{(j)}_n=1,\\  
		-1,  \quad {\rm if} \quad \sigma^{(j)}_n=2. 
	\end{array}
	\right.
\end{eqnarray}
For the same reason as in the spin-$1/2$ case, our task at this point is incomplete. Despite the level of complexity has increased with $S$, we have equally disproved the idea that the value of a classical trit at a given location in the ternary representation of a positive integer can only be accessed with aid of a computer. We have conducted a very same experiment as earlier (see Appendix.\ref{AppA}). Deep inspection into the nature of digits in a few numbers brought us to the conclusion that classical trits also enter numbers through specific patterns similar to those for bits (see Fig.\ref{Figure1}). The trits are also constructed in terms of the unit-step function as
\begin{subeqnarray}\label{equ15}
	\slabel{equ15a}\Gamma_n^{\left(j\right)+}&=&\sum_{m=0}^{r_{j-1}-1}\left(u\left(n-(3m+1)q_j\right)-u\left(n-(3m+3)q_j\right)\right),\qquad\\
	\slabel{equ15b}\Gamma_n^{\left(j\right)-}&=&\sum_{m=0}^{r_{j-1}-1}\left(u\left(n-3mq_j\right)-u\left(n-(3m+2)q_j\right)\right),\\
	\nonumber\Gamma_n^{\left(j\right)z}&=&\sum_{m=0}^{r_{j-1}-1}\left(u\left(n-3mq_j\right)-u\left(n-(3m+1)q_j\right)\right)\\\slabel{equ15c}&-&\sum_{m=0}^{r_{j-1}-1}\left(u\left(n-(3m+2)q_j\right)-u\left(n-(3m+3)q_j\right)\right),
\end{subeqnarray}
with $r_j=3^j$. It can be verified that $\Gamma_n^{\left(j\right)-}-\Gamma_n^{\left(j\right)+}=\Gamma_n^{\left(j\right)z}$ and that the following conjugation relations $\Gamma_{n+q_j}^{\left(j\right)+}=\Gamma_n^{\left(j\right)-}$ and $\Gamma_{n-q_j}^{\left(j\right)-}=\Gamma_n^{\left(j\right)+}$ hold. 

Before closing up this section, it is instructive to consider the simplest situation $K=1$. In this case, $D=3$ and $q_1=3^{1-1}=1$. The integer $n\in\{0,1,2\}$ labels directly the three spin projections $+1,0,-1$ according to
$\ket{+1}\equiv\ket{0}$, $\ket{0}\equiv\ket{1}$ and $\ket{-1}\equiv\ket{2}$.
The ternary digit $\sigma_n^{(1)}$ coincides with $n$ itself. From Eqs.~\eqref{equ13}, one obtains
\begin{subequations}
	\begin{align}
		\hat S^{+}
		&= \sqrt{2}\sum_{n=0}^{2}\Gamma_n^{(1)+}\ket{n-1}\bra{n},\\
		\hat S^{-}
		&= \sqrt{2}\sum_{n=0}^{2}\Gamma_n^{(1)-}\ket{n+1}\bra{n},\\
		\hat S^{z}
		&= \sum_{n=0}^{2}\Gamma_n^{(1)z}\ket{n}\bra{n}.
	\end{align}
\end{subequations}
Using Eq.~\eqref{SM19}, one has explicitly $\Gamma_0^{(1)z}=+1$, $\Gamma_1^{(1)z}=0$, and $\Gamma_2^{(1)z}=-1$ so that $\hat S^{z}=\ket{0}\bra{0}-\ket{2}\bra{2}$, which is precisely the diagonal spin-$1$ operator $\mathrm{diag}(1,0,-1)$ in the ordered basis $\{\ket{0},\ket{1},\ket{2}\}$. Similarly, since $\Gamma_1^{(1)+}=\Gamma_2^{(1)+}=1$, 
$\Gamma_0^{(1)-}=\Gamma_1^{(1)-}=1$,  and all other coefficients vanish, one finds
$\hat S^{+}=\sqrt{2}\left(\ket{0}\bra{1}+\ket{1}\bra{2}\right)$, $\hat S^{-}=\sqrt{2}\left(\ket{1}\bra{0}+\ket{2}\bra{1}\right)$,
which are exactly the standard spin-$1$ ladder operators. Therefore, for $K=1$ the present Dirac construction reduces identically to the traditional $3\times 3$ matrix representation of
su(2) in the spin-$1$ irreducible representation. This confirms the internal consistency of our ternary formalism.

 At this point, a key question pops up in mind: can we extend this technique to $S>1$? The answer is affirmative as it is always possible to map quantum spins of size $S>1$ (qudits) onto their classical counterpart. As a consequence, the $d$-decomposition in $d$-trit basis yields exactly the configuration of the $d$-spin. In the general case $q_j=(2S+1)^{K-j}$ and $r_j=(2S+1)^{j}$ (this task is left for future projects).

 It is relevant to emphasize the scope of the present section. In contrast to Sections III and IV where the spin-$1/2$ Dirac representation is applied to specific lattice Hamiltonians,  the purpose of the present section is to establish the extension of the Dirac representation to spin-$1$. Several spin-$1$ chains provide natural benchmarks for future applications of the representation. These include integrable spin-$1$ models admitting Bethe-ansatz solutions\cite{Babujian1982, Alcaraz1988} and the Affleck–Kennedy–Lieb–Tasaki (AKLT) chain, for which an exact valence-bond ground state is known\cite{AKL1987, AKL1988}.
 A detailed application of the present representation to these models
 is left for future works.

\section{Conclusion}\label{Section6}
We have addressed the fundamental problem of constructing suitable representations for lattice spin operators (LSOs) within a fully explicit Dirac (bra–ket) framework. Due to their intrinsic {\it abstract algebra} Dirac representations provide a natural language for manipulating quantum operators and states. We have constructed compact many-body Dirac representations of LSOs for spin-$\tfrac12$ and spin-$1$ systems based on a mapping between quantum spins and classical binary/ternary digit structure. The resulting forms express LSOs through a single summation over the Hilbert space and preserve locality, linearity, and the $su(2)$ algebra. 
To the best of our knowledge, such explicit many-body Dirac representations have not yet been developed beyond the single-spin case.

Our results have potential for improvement not just in physics but broadly in science as they offer the possibility of getting the $j$th bit/trit of an integer of bit/trit -field length $K$ without resorting to a binary/ternary decomposition. The usefulness of our representations are discussed in Physics by examining the spin-$1/2$ XXZ Heisenberg model in an open boundary configuration in the presence of a magnetic field. The prototype model is analyzed both numerically and analytically. We have found that binary representation permits a faster numerical construction of the matrix Hamiltonian by avoiding costly matrix--matrix operations and reducing memory overhead. The global $U(1)$ symmetry is exploited throughout reducing computational complexity due to the exponential growth of the Hilbert space with the total number of spins. Three representative regimes have been considered: (i) absence of local magnetic field $(h_j=0,\, J\neq 0,\, J_z\neq 0)$, (ii) uniform magnetic field $(h_j=h,\, J\neq 0,\, J_z=0)$, and (iii) random magnetic field $(h_j\in[-W,W],\, J\neq 0,\, J_z\neq 0)$ with disorder strength $W$. In case (i), the spin-inversion symmetry in the zero-magnetization sector permit further fragmentation of the sector into two disconnected subspaces, enabling simulations up to $K=18$ spins using a Mathematica implementation with Intel MKL (see \href{https://github.com/Kenmax15}{GitHub repository}). In case (ii), the associated eigenvalue problem is transformed into a recurrence relation, revealing an effective tight-binding evolution on a graph. In case (iii), the full model is discussed. Perturbative expressions for eigenenergies and eigenvectors are derived and show good agreement with exact diagonalization results.
  
Before wrapping up it worth underlying that, a few questions asked in this paper are left here without technical answers and for future projects. For example, while the inverse participation ratio (IPR) has been computed, a detailed analysis of localization properties remains open. The clustering of nearly degenerate energy levels in the presence of disorder and the implications of explicit expressions for $E_{q_j}$ in the context of many-body localization (MBL) also warrant deeper study.  It will be interesting in the future to inspect whether our technique applies to long-range interactions and not only nearest-neighbor interactions. It would also be interesting to investigate the possibility of achieving a Dirac representation for lattice Boson operators.

\appendix

\section*{Acknowledgments}
This paper is part of the project "Building a Comprehensive Quantum Theory for Many-Body Localization in Spin chains".  Fundings for this this paper are provided by the Alexander von Humboldt foundation through the scheme of Georg-Forster fellowship for Postdocs. MBK appreciates the warm hospitality of Prof. John Schliemann at the University of Regensburg where part of this paper was conducted. We also appreciate fruitful discussions with Ateuafack Mathurin and Anicet Kammogne.

\section{Proof of \eqref{equ6a}-\eqref{equ6c}}\label{AppA}
This Appendix is devoted to proofing the formula \eqref{equ6a}-\eqref{equ6c}. This is done in two different ways: (i) an indirect proof by conducting an experiment and (ii) a direct proof based on strict mathematical argument

\subsection{Indirect proof}
 Consider a chain of length $K$ with identical spins of size $S=1/2$. There exist $2^K$ different configurations (possible way of arranging spins in up- and/or down- directions). Mapping the quantum spins onto their classical analogue and sorting the basis vectors $\ket{n}$ in lexicographical order with respect to the site $j$, we construct the diagrams in Fig.\ref{Figure8} for $K=\{3,4,5\}$. Blue and red colors are used to discriminate between possible orientations of spins along the quantization direction. With the aid of such a color palette the diagrams depict a clear pattern. This is clearly seen by standing on each site and looking at the numbers $n$. For instance, at site $j=1$ for $K=5$ (reading direction-- from the left to the right), we observe that for all integer in the interval $0\leq n\leq15$ the first spin is in the state $\sigma_n^{(1)}=0$ whilst for $16\leq n\leq 32$ one always has $\sigma_n^{(1)}=1$. Similarly, at site $j=2$ all integers in the interval $0\leq n\leq7 $ have $\sigma_n^{(2)}=0$ those in $8\leq n\leq15 $ have $\sigma_n^{(2)}=1$ while those in $16\leq n\leq23 $ have $\sigma_n^{(2)}=1$ etc The same observation can be done at every site. 

The very same experiment is repeated for $S=1$ in order to prove \eqref{equ15a}-\eqref{equ15c}. 
\subsection{Spin-1/2}
\begin{figure}[]
	\centering
	\begin{center} 
	    \vspace{2.9cm}\includegraphics[width=8cm, height=14cm]{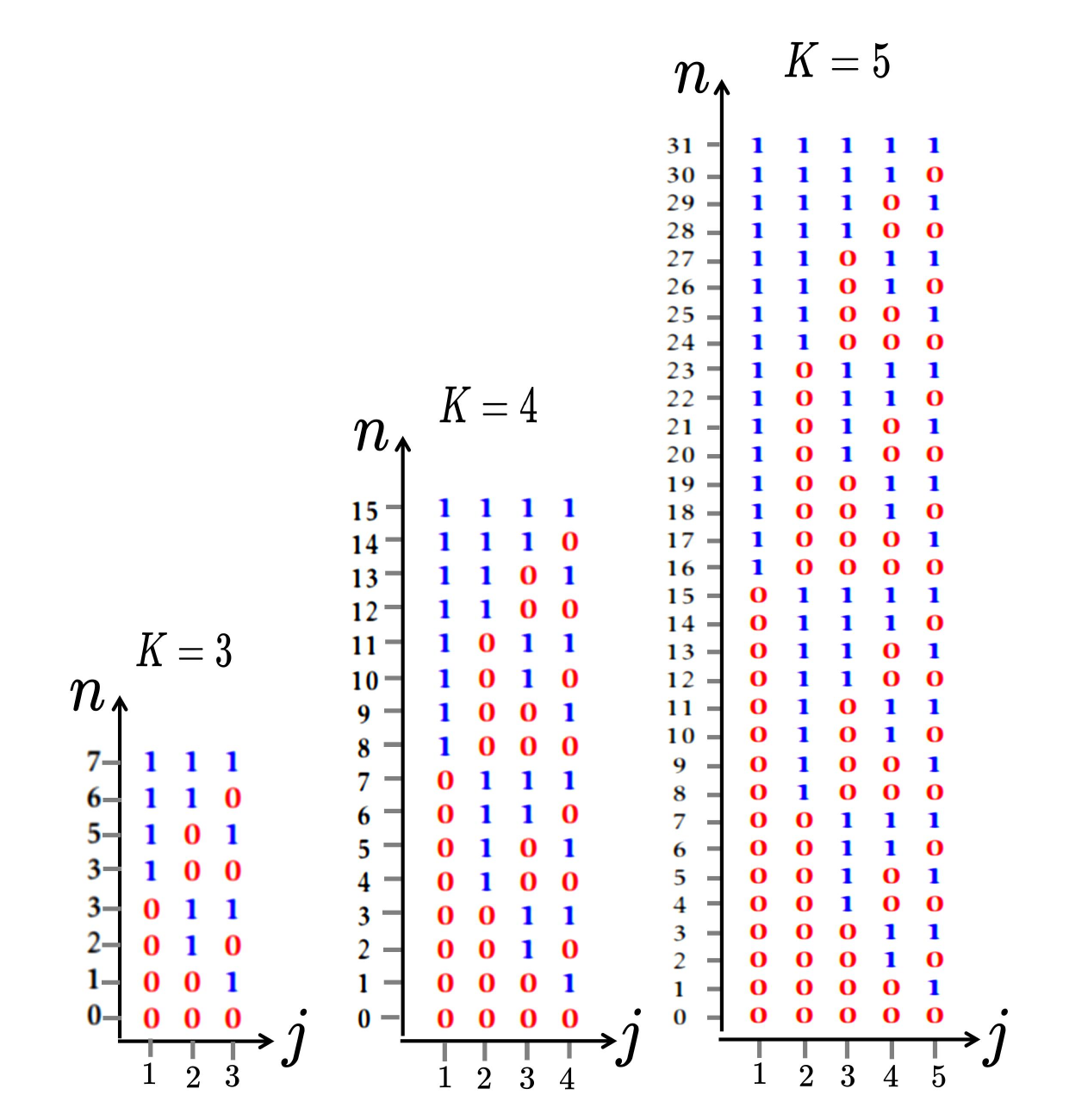}
		\caption{Binary decompositions of integers $0\leq n\le 2^K-1$ for $K=\{3,4,5\}$. Color palette is used to visually distinguish between bits in each representation. We observe that $s^{\left(K\right)+}_n=0$ for even integers $n$ and $s^{\left(K\right)+}_n=1$ for odd integers $n$.}
		\label{Figure8} 
	\end{center}
\end{figure}

The report concerning the above experiment for $K=5$ is presented as follows. Other values of $K$ in the diagram Fig.\ref{Figure8} leading to the same conclusion.

At site $j=1$,
\begin{eqnarray}\label{equationA1}
&\nonumber	0\cdot2^{(5-1)}\le n< 1\cdot2^{(5-1)}, \quad \sigma_n^{(1)}=0,\\
&	1\cdot2^{(5-1)}\le n < 2\cdot2^{(5-1)}, \quad \sigma_n^{(1)}=1.
\end{eqnarray}
At site $j=2$,
\begin{eqnarray}\label{equationA2}
&\nonumber	0\cdot2^{(5-2)}\le n < 1\cdot2^{(5-2)}, \quad \sigma_n^{(2)}=0,\\
&\nonumber	1\cdot2^{(5-2)}\le n < 2\cdot2^{(5-2)}, \quad \sigma_n^{(2)}=1,\\
&\nonumber	2\cdot2^{(5-2)}\le n < 3\cdot2^{(5-2)}, \quad \sigma_n^{(2)}=0,\\
&\nonumber	3\cdot2^{(5-2)}\le n < 4\cdot2^{(5-2)}, \quad \sigma_n^{(2)}=1.
\end{eqnarray}
At site $j=3$,
\begin{eqnarray}\label{equationA3}
&\nonumber	0\cdot2^{(5-3)}\le n < 1\cdot2^{(5-3)}, \quad \sigma_n^{(3)}=0,\\
&\nonumber	1\cdot2^{(5-3)}\le n < 2\cdot2^{(5-3)}, \quad \sigma_n^{(3)}=1,\\
&\nonumber	2\cdot2^{(5-3)}\le n < 3\cdot2^{(5-3)}, \quad \sigma_n^{(3)}=0,\\
&\nonumber	3\cdot2^{(5-3)}\le n < 4\cdot2^{(5-3)}, \quad \sigma_n^{(3)}=1,\\
&	        4\cdot2^{(5-3)}\le n < 5\cdot2^{(5-3)}, \quad \sigma_n^{(3)}=0,\\
&\nonumber	5\cdot2^{(5-3)}\le n < 6\cdot2^{(5-3)}, \quad \sigma_n^{(3)}=1,\\
&\nonumber	6\cdot2^{(5-3)}\le n < 7\cdot2^{(5-3)}, \quad \sigma_n^{(3)}=0,\\
&\nonumber	7\cdot2^{(5-3)}\le n < 8\cdot2^{(5-3)}, \quad \sigma_n^{(3)}=1.
\end{eqnarray}
At site $j=4$,
\begin{eqnarray}\label{equationA4}
&\nonumber	0\cdot2^{(5-4)}\le n < 1\cdot2^{(5-4)}, \quad \sigma_n^{(4)}=0,\\
&\nonumber	1\cdot2^{(5-4)}\le n < 2\cdot2^{(5-4)}, \quad \sigma_n^{(4)}=1,\\
&\nonumber	2\cdot2^{(5-4)}\le n < 3\cdot2^{(5-4)}, \quad \sigma_n^{(4)}=0,\\
&\nonumber	3\cdot2^{(5-4)}\le n < 4\cdot2^{(5-4)}, \quad \sigma_n^{(4)}=1,\\
&	\vdots\\
&\nonumber	13\cdot2^{(5-4)}\le n < 14\cdot2^{(5-4)}, \quad \sigma_n^{(4)}=1,\\
&\nonumber	14\cdot2^{(5-4)}\le n < 15\cdot2^{(5-4)}, \quad \sigma_n^{(4)}=0,\\
&\nonumber	15\cdot2^{(5-4)}\le n < 16\cdot2^{(5-4)}, \quad \sigma_n^{(4)}=1.
\end{eqnarray}
At site $j=5$,
\begin{eqnarray}\label{equationA5}
&\nonumber	0\cdot2^{(5-5)}\le n < 1\cdot2^{(5-5)}, \quad \sigma_n^{(5)}=0,\\
&\nonumber	1\cdot2^{(5-5)}\le n < 2\cdot2^{(5-5)}, \quad \sigma_n^{(5)}=1,\\
&\nonumber	2\cdot2^{(5-5)}\le n < 3\cdot2^{(5-5)}, \quad \sigma_n^{(5)}=0,\\
&	\vdots\\
&\nonumber	29\cdot2^{(5-5)}\le n < 30\cdot2^{(5-5)}, \quad \sigma_n^{(5)}=1,\\
&\nonumber	30\cdot2^{(5-5)}\le n < 31\cdot2^{(5-5)}, \quad \sigma_n^{(5)}=0,\\
&\nonumber	31\cdot2^{(5-5)}\le n < 32\cdot2^{(5-5)}, \quad \sigma_n^{(5)}=1.
\end{eqnarray}

By inspecting the structure of integers, we observe that all integers in an appropriately chosen interval have the same value of the $j$th digit as indicated in Table.\ref{Table1} below. From this table we easily infer \eqref{equ6a}-\eqref{equ6c} and achieve our goal.

In general, a given piece-wise function $f(n)$  defined as
\begin{eqnarray}\label{equA6}
	f(n)=
	\left\{\begin{array}{c}
		n_1, \quad {\rm if} \quad a_1\leq n<a_2\\ 
		n_2, \quad {\rm if} \quad a_2\leq n<a_3\\ 
		n_3, \quad {\rm if} \quad a_3\leq n<a_4\\ 
		\vdots \quad \quad \\
		n_L, \quad {\rm if} \quad a_{L-1}\leq n<a_L
	\end{array}
	\right. 
\end{eqnarray}
can be written in terms of unit-step function $u(x)$ as 
\begin{eqnarray}\label{equA7}
	f(n) = \sum_{j=1}^{L-1}n_j\left[u\left(n-a_j\right)-u\left(n-a_{j+1}\right)\right].
\end{eqnarray}

An interesting fact to be noted is that if we split the set of integers in the interval $0\leq n\leq 2^K-1$ into even and odd integers, perform binary decomposition, replace $0\to\frac{1}{2}$ and $1\to-\frac{1}{2}$, we observe that both results differ only by the least significant digit (LSD). This is illustrated for $K=4$ in the Tables \ref{Table2} and \ref{Table3}. As mentioned in the main text, the LSD is the digit at $j=K$ as we have adopted the reading-coding direction -- from the left to the right.

\begin{table}[]
	\caption{Table of the values of classical spins in the range $0\le n\le 2^K-1$ splitted into $r_j=2^j$  intervals with $q_j=2^{K-j}$.}\label{Table1}
	\centering
	\begin{tabular}{| c | c | c | c |}
		\hline
		\hline
		{\rm Intervals} $\left(S=1/2\right)$  & $s_n^{\left(j\right)+}$ & $s_n^{\left(j\right)-}$ & $s_n^{\left(j\right)z}$ \\ 
		\hline
		\hline
		$0\cdot q_j\le n < 1\cdot q_j$ & 0 & 1 & $\frac{1}{2}$\\ \hline
		$1\cdot q_j\le n < 2\cdot q_j$ & 1 & 0 & $-\frac{1}{2}$\\ \hline
		$2\cdot q_j\le n \le3\cdot q_j$ & 0 & 1 & $\frac{1}{2}$\\ \hline
		\vdots        &  \vdots &  \vdots & \vdots \\ \hline
		$\left(r_j-2\right)\cdot q_j\le n < \left(r_j-1\right)\cdot q_j$ & 0 & 1 & $\frac{1}{2}$\\ \hline
		$\left(r_j-1\right)\cdot q_j\le n < \left(r_j-0\right)\cdot q_j$ & 1 & 0 & $-\frac{1}{2}$\\
		\hline
		\hline
	\end{tabular}
\end{table}

\begin{table}[]
	\caption{Table of the binary decomposition of even integers in bit-field length $K=4$ for numbers in the interval $0\le n\le 2^K-1$}\label{Table2}
	\centering
	\begin{tabular}{| c | c | c | c | c |}
		\hline
		\hline
		even number, $n$  & $j=1$ & $j=2$ & $J=3$  & $J=4$ \\ 
		\hline
		\hline
		$14$ & $-\frac{1}{2}$ & $-\frac{1}{2}$ & $-\frac{1}{2}$ & ${\bf \frac{1}{2} }$ \\ \hline
		$12$ & $-\frac{1}{2}$ & $-\frac{1}{2}$ & $\frac{1}{2}$ & ${\bf \frac{1}{2} }$ \\ \hline
		$10$ & $-\frac{1}{2}$ & $\frac{1}{2}$ & $-\frac{1}{2}$ & ${\bf \frac{1}{2} }$ \\ \hline
		$8$ & $-\frac{1}{2}$ & $\frac{1}{2}$ & $\frac{1}{2}$ & ${\bf \frac{1}{2} }$ \\ \hline
		$6$ & $\frac{1}{2}$ & $-\frac{1}{2}$ & $-\frac{1}{2}$ & ${\bf \frac{1}{2} }$ \\ \hline
		$4$ & $\frac{1}{2}$ & $-\frac{1}{2}$ & $\frac{1}{2}$ & ${\bf \frac{1}{2} }$ \\ \hline
		$2$ & $\frac{1}{2}$ & $\frac{1}{2}$ & $-\frac{1}{2}$ & ${\bf \frac{1}{2} }$ \\ \hline
		$0$ & $\frac{1}{2}$ & $\frac{1}{2}$ & $\frac{1}{2}$ & ${\bf \frac{1}{2} }$\\ \hline 
		\hline
		\hline
	\end{tabular}
\end{table}

\begin{table}[]
	\caption{Same as Table \ref{Table2} for odd integers.}\label{Table3}
	\centering
	\begin{tabular}{| c | c | c | c | c |}
		\hline
		\hline
		odd number, $n$  & $j=1$ & $j=2$ & $J=3$  & $J=4$ \\ 
		\hline
		\hline
		$15$ & $-\frac{1}{2}$ & $-\frac{1}{2}$ & $-\frac{1}{2}$  & ${\bf -\frac{1}{2} }$ \\ \hline
		$13$ & $-\frac{1}{2}$ & $-\frac{1}{2}$ & $\frac{1}{2}$ & ${\bf -\frac{1}{2} }$ \\ \hline
		$11$ & $-\frac{1}{2}$ & $\frac{1}{2}$ & $-\frac{1}{2}$ & ${\bf -\frac{1}{2} }$ \\ \hline
		$9$ & $-\frac{1}{2}$ & $\frac{1}{2}$ & $\frac{1}{2}$ & ${\bf -\frac{1}{2} }$ \\ \hline
		$7$ & $\frac{1}{2}$ & $-\frac{1}{2}$ & $-\frac{1}{2}$ & ${\bf -\frac{1}{2} }$ \\ \hline
		$5$ & $\frac{1}{2}$ & $-\frac{1}{2}$ & $\frac{1}{2}$ & ${\bf -\frac{1}{2} }$ \\ \hline
		$3$ & $\frac{1}{2}$ & $\frac{1}{2}$ & $-\frac{1}{2}$ & ${\bf -\frac{1}{2} }$ \\ \hline
		$1$ & $\frac{1}{2}$ & $\frac{1}{2}$ & $\frac{1}{2}$ & ${\bf -\frac{1}{2} }$ \\ \hline
		\hline
		\hline
	\end{tabular}
\end{table}
We observe that apart from last columns corresponding to least significant digits (LSDs), both Tables are identical. Same observations can be done for $K>4$. This is clearly an indication that if one knows the binary decomposition of an odd/even number, one can infer the binary decomposition of the consecutive even/odd number just by flipping the LSD. For example the two consecutive digits $0$ and $1$ differ only by the last digits etc.

\subsection{Direct proof}

This subsection provides a rigorous proof of the formulas \eqref{equ6a}--\eqref{equ6c} for spin-$\frac12$. Consider again a spin-$\frac12$ chain of length $K$ with each spin mapped onto its classical analogue. As already mentioned in the main text, each integer $n$ admits a unique binary decomposition
\begin{eqnarray}\label{A7}
	n=\left(\sigma_{n_S}^{(1)}\cdots \sigma_{n_S}^{(j)}\cdots \sigma_{n}^{(K)}\right)_{2}, \quad  n=\sum_{j=1}^{K}\left(\sigma_{n_S}^{(j)}\right)q_j,
\end{eqnarray}
where $q_j = 2^{K-j}$. We identified
$s_n^{(j)+} = \sigma_{n}^{(j)}$, 
$s_n^{(j)-} = 1 - \sigma_{n}^{(j)}$, 
$s_n^{(j)z} = \frac12\left(1 - 2 \sigma_{n}^{(j)}\right)$.
Thus proving \eqref{equ6a}--\eqref{equ6c} reduces to determining 
an explicit expression for $\sigma_{n}^{(j)}$.
For the proof, we perform the Euclidean division of $n$ by $q_j$:
\begin{eqnarray}\label{A8}
	n = a q_j + r,
	\qquad
	0 \le r < q_j.
\end{eqnarray}
Then
\begin{eqnarray}\label{A9}
a = \left\lfloor \frac{n}{q_j} \right\rfloor.
\end{eqnarray}
Since $q_j$ isolates the $j$th binary digit, the $j$th digit equals the least significant bit of $a$.
Therefore,
\begin{eqnarray}\label{A10}
\sigma_{n}^{(j)} = a \bmod 2 = \left\lfloor \frac{n}{q_j} \right\rfloor \bmod 2.
\end{eqnarray}
Hence
\begin{eqnarray}\label{A11}
\sigma_{n}^{(j)}=
\begin{cases}
	0 & \text{if } \left\lfloor n/q_j \right\rfloor \text{ is even},\\
	1 & \text{if } \left\lfloor n/q_j \right\rfloor \text{ is odd}.
\end{cases}
\end{eqnarray}
Thus, the integers $0\le n \le 2^K-1$ decompose into 
$r_j = 2^j$ alternating intervals
\[
[0,q_j),\ [q_j,2q_j),\ [2q_j,3q_j),\dots,
[(r_j-1)q_j,r_j q_j),
\]
on each of which $\sigma_{n}^{(j)}$ is constant and alternates between $0$ and $1$. Using the piecewise constant function in Eq.\eqref{equA6} immediately gives Eqs.~\eqref{equ6a}--\eqref{equ6c}.

For $j=K$, one has $q_K=1$, and therefore $\sigma_{n}^{(K)} = n \bmod 2$. Thus consecutive integers differ only in the least significant digit, explaining once again the structure observed in Tables~\ref{Table2} and~\ref{Table3}.
 
\subsection{Spin-1}

\begin{figure}[]
	\centering
	\begin{center} 
		\includegraphics[width=3.4cm, height=12cm]{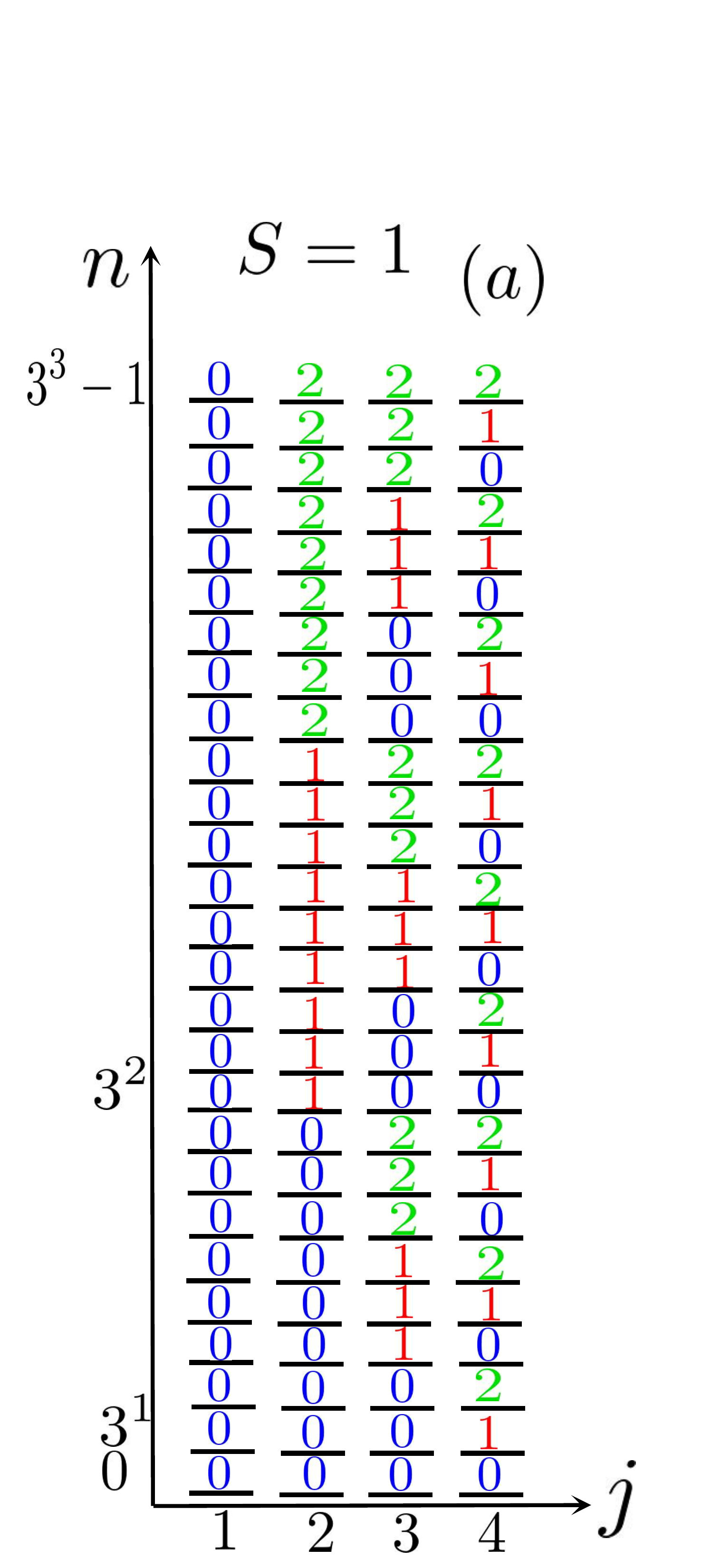}\hspace{-0.7cm}
		\includegraphics[width=3.4cm, height=12cm]{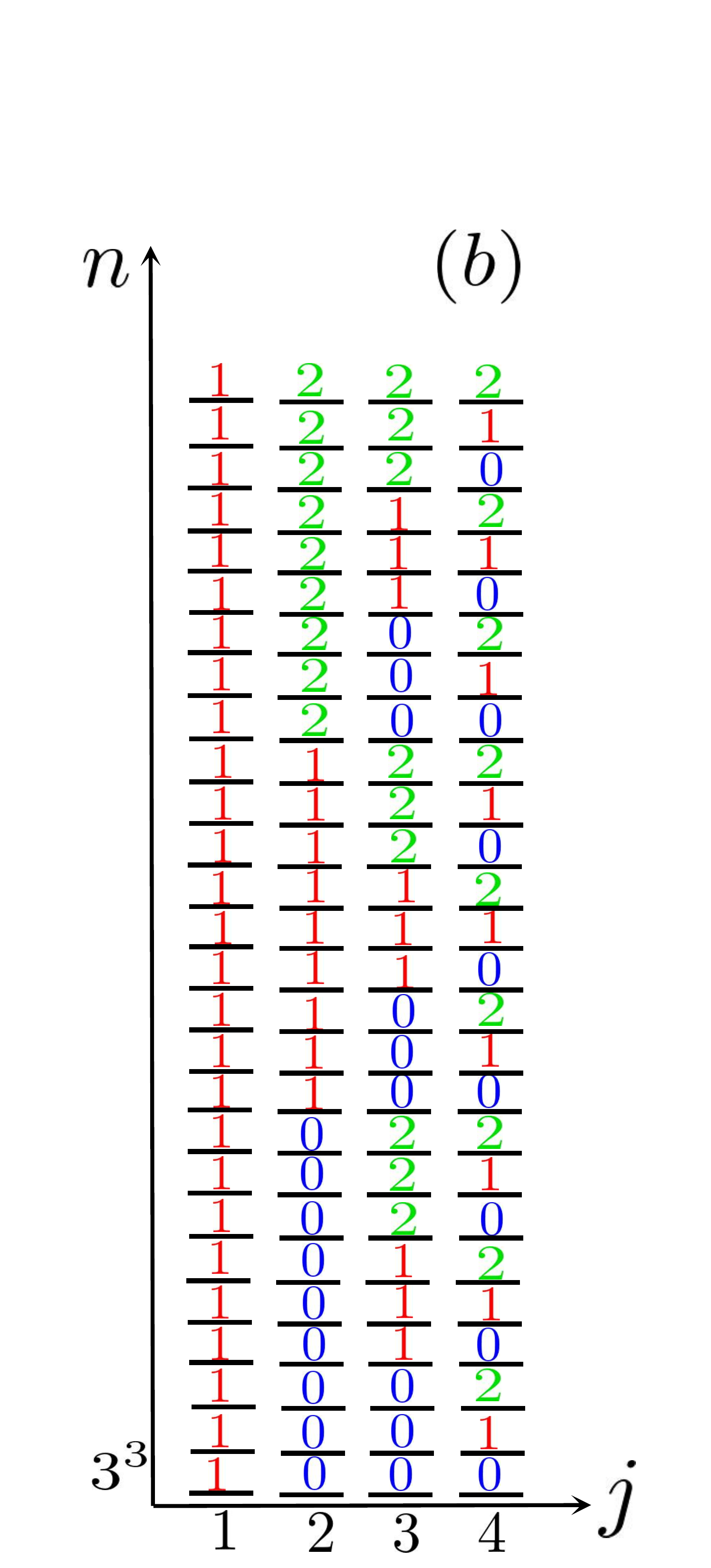}\hspace{-1.cm}
		\includegraphics[width=3.4cm, height=12cm]{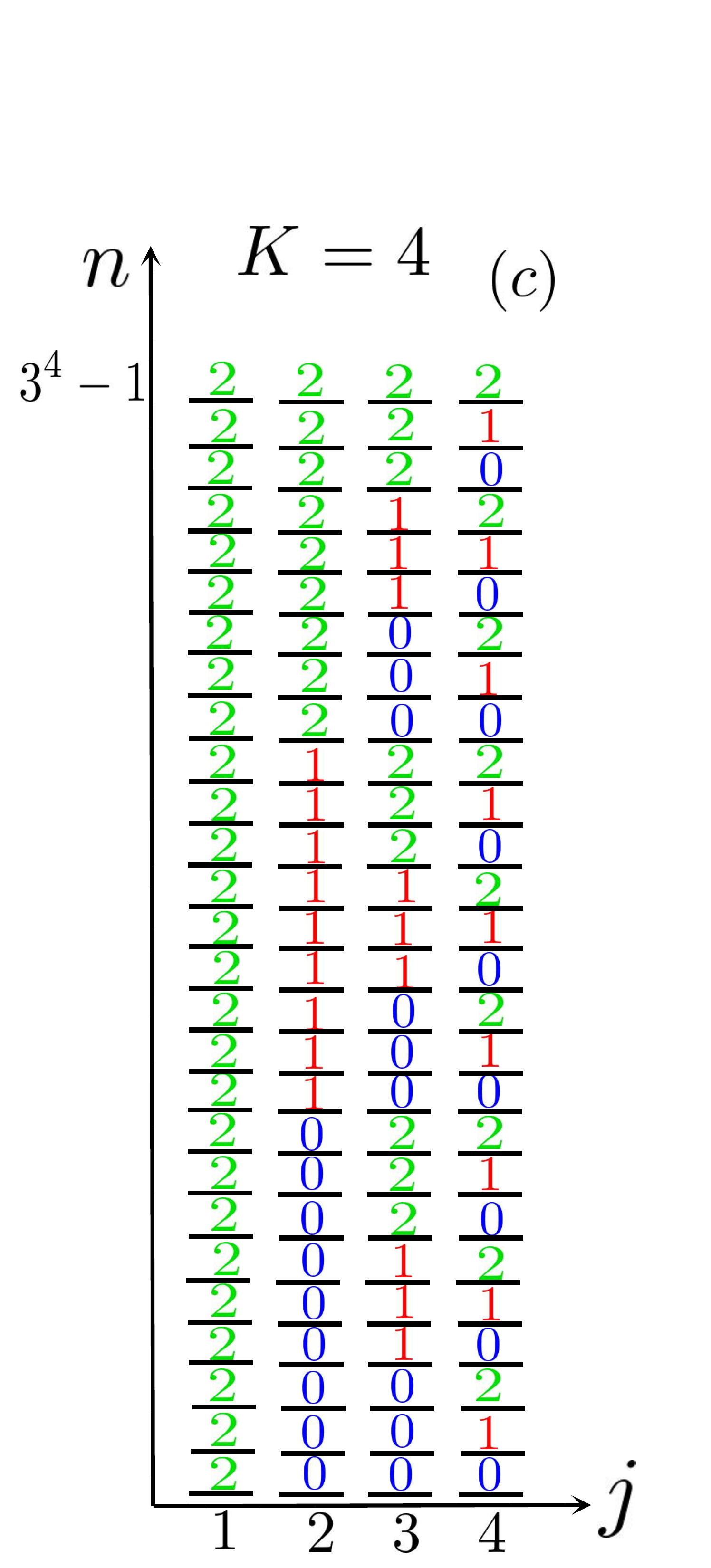}\hspace{-1.cm}
		\caption{Same as \ref{Figure8} for $S=1$.}
		\label{Figure9} 
	\end{center}
\end{figure}

The report concerning the experiment for $S=1$ and $K=4$ is presented as follows:

At site $j=1$,
\begin{eqnarray}\label{equationA6}
	&\nonumber	0\cdot3^{(4-1)}\le n < 1\cdot3^{(4-1)}, \quad \sigma_n^{(1)}=0,\\
	&\nonumber	1\cdot3^{(4-1)}\le n < 2\cdot3^{(4-1)}, \quad \sigma_n^{(1)}=1,\\
	&       	2\cdot3^{(4-1)}\le n < 3\cdot3^{(4-1)}, \quad \sigma_n^{(1)}=2.
\end{eqnarray}
At site $j=2$,
\begin{eqnarray}\label{equationA7}
	&\nonumber	0\cdot3^{(4-2)}\le n < 1\cdot3^{(4-2)}, \quad \sigma_n^{(2)}=0,\\
	&\nonumber	1\cdot3^{(4-2)}\le n < 2\cdot3^{(4-2)}, \quad \sigma_n^{(2)}=1,\\
	&\nonumber	2\cdot3^{(4-2)}\le n < 3\cdot3^{(4-2)}, \quad \sigma_n^{(2)}=2,\\
	&\vdots\\
	&\nonumber	6\cdot3^{(4-2)}\le n < 7\cdot3^{(4-2)}, \quad \sigma_n^{(2)}=0,\\
	&\nonumber	7\cdot3^{(4-2)}\le n < 8\cdot3^{(4-2)}, \quad \sigma_n^{(2)}=1,\\
	&\nonumber	8\cdot3^{(4-2)}\le n < 9\cdot3^{(4-2)}, \quad \sigma_n^{(2)}=2.
\end{eqnarray}
At site $j=3$,
\begin{eqnarray}\label{equationA8}
	&\nonumber	0\cdot3^{(4-3)}\le n < 1\cdot3^{(4-3)}, \quad \sigma_n^{(3)}=0,\\
	&\nonumber	1\cdot3^{(4-3)}\le n < 2\cdot3^{(4-3)}, \quad \sigma_n^{(3)}=1,\\
	&\nonumber	2\cdot3^{(4-3)}\le n < 3\cdot3^{(4-3)}, \quad \sigma_n^{(3)}=2,\\
	&\vdots\\
	&\nonumber	24\cdot3^{(4-3)}\le n < 25\cdot3^{(4-3)}, \quad \sigma_n^{(3)}=0,\\
	&\nonumber	25\cdot3^{(4-3)}\le n < 26\cdot3^{(4-3)}, \quad \sigma_n^{(3)}=1,\\
	&\nonumber	26\cdot3^{(4-3)}\le n < 27\cdot3^{(4-3)}, \quad \sigma_n^{(3)}=2.
\end{eqnarray}
At site $j=4$,
\begin{eqnarray}\label{equationA9}
	&\nonumber	0\cdot3^{(4-4)}\le n < 1\cdot3^{(4-4)}, \quad \sigma_n^{(4)}=0,\\
	&\nonumber	1\cdot3^{(4-4)}\le n < 2\cdot3^{(4-4)}, \quad \sigma_n^{(4)}=1,\\
	&\nonumber	2\cdot3^{(4-4)}\le n < 3\cdot3^{(4-4)}, \quad \sigma_n^{(4)}=2,\\
	&\vdots\\
	&\nonumber	78\cdot3^{(4-4)}\le n < 79\cdot3^{(4-4)}, \quad \sigma_n^{(4)}=0,\\
	&\nonumber	79\cdot3^{(4-4)}\le n < 80\cdot3^{(4-4)}, \quad \sigma_n^{(4)}=1,\\
	&\nonumber	80\cdot3^{(4-4)}\le n < 81\cdot3^{(4-4)}, \quad \sigma_n^{(4)}=2.
\end{eqnarray}

\begin{table}[!ht]
	\caption{Table of the values of classical spins in the range $0\le n\le 3^K-1$ splitted into $r_j=3^j$  intervals with $q_j=3^{K-j}$.}\label{Table4}
	\centering
	\begin{tabular}{| c | c | c | c |}
		\hline
		\hline
		{\rm Intervals} $\left(S=1\right)$  & $\Gamma_n^{\left(j\right)+}$ & $\Gamma_n^{\left(j\right)-}$ & $\Gamma_n^{\left(j\right)z}$ \\ 
		\hline
		\hline
		$0\cdot q_j\le n < 1\cdot q_j$ & 0 & 1 & 1\\ \hline
		$1\cdot q_j\le n < 2\cdot q_j$ & 1 & 1 & 0\\ \hline
		$2\cdot q_j\le n < 3\cdot q_j$ & 1 & 0 & $-1$\\ \hline
		$3\cdot q_j\le n < 4\cdot q_j$ & 0 & 1 & 1\\ \hline
		$4\cdot q_j\le n < 5\cdot q_j$ & 1 & 1 & 0\\ \hline
		$5\cdot q_j\le n < 6\cdot q_j$ & 1 & 0 & $-1$\\ \hline
		\vdots        &  \vdots &  \vdots & \vdots \\ \hline
		$\left(r_j-3\right)\cdot q_j\le n < \left(r_j-2\right)\cdot q_j$ & 0 & 1 & 1\\ \hline
		$\left(r_j-2\right)\cdot q_j\le n < \left(r_j-1\right)\cdot q_j$ & 1 & 1 & 0\\ \hline
		$\left(r_j-1\right)\cdot q_j\le n < \left(r_j-0\right)\cdot q_j$ & 1 & 0 & $-1$\\
		\hline
		\hline
	\end{tabular}
\end{table}

\section{A few properties of the classical spins}\label{AppB}
A few properties of classical spins-$1/2$ are elaborated and presented.

\subsection{Local algebra of classical spins}
\begin{eqnarray}\label{equ9a}
	\left(s_n^{\left(j\right)+}\right)^2=s_n^{\left(j\right)+}, \quad \left(s_n^{\left(j\right)-}\right)^2=s_n^{\left(j\right)-}, \quad 
	\left(s_n^{\left(j\right)z}\right)^2=\frac{1}{4},
\end{eqnarray}
\begin{subeqnarray}\label{equA1a}
s_n^{\left(j\right)+}s_n^{\left(j\right)-}=s_n^{\left(j\right)-}s_n^{\left(j\right)+}=0,
\end{subeqnarray}
\begin{subeqnarray}\label{equA1b}
 s_n^{\left(j\right)+}s_n^{\left(j\right)z}=-\frac{s_n^{\left(j\right)+}}{2},
\end{subeqnarray} 
 \begin{subeqnarray}\label{equA1c}
  s_n^{\left(j\right)-}s_n^{\left(j\right)z}= \frac{s_n^{\left(j\right)-}}{2},
\end{subeqnarray}
and
\begin{eqnarray}\label{equA2}
	s_n^{\left(j\right)+}+s_n^{\left(j\right)-}=1, \quad s_n^{\left(j\right)+}-s_n^{\left(j\right)-}=-2s_n^{\left(j\right)z},
\end{eqnarray}
whereby
\begin{eqnarray}\label{equA3}
	s_n^{\left(j\right)z}=\frac{1}{2}\left(1-2s_n^{\left(j\right)+}\right)=-\frac{1}{2}\left(1-2s_n^{\left(j\right)-}\right),
\end{eqnarray}
\begin{eqnarray}\label{equA3a}
	s_n^{\left(j\right)+}=\frac{1}{2}\left(1-2s_n^{\left(j\right)z}\right), \quad s_n^{\left(j\right)-}=\frac{1}{2}\left(1+2s_n^{\left(j\right)z}\right).
\end{eqnarray}
As a consequence of these first properties $\cos(\alpha s_n^{(j)\pm})=s_n^{(j)\mp}+s_n^{(j)\pm}\cos(\alpha)=1-2s_n^{(j)\pm}\sin^2(\alpha/2)$ and $\sin(\alpha s_n^{(j)\pm})=s_n^{(j)\pm}\sin(\alpha)$. When the angle $\alpha=\pi/2$ then $\cos(\pi s_n^{(j)\pm}/2)=1-2s_n^{(j)\pm}$ and $\sin(\pi s_n^{(j)\pm}/2)=s_n^{(j)\pm}$ thereby $s_n^{\left(j\right)z}=\frac{1}{2}\cos(\pi s_n^{(j)+}/2)$. In the same vein, one easily proves that $\cos(\pi s_n^{(j)z})=0$ and  $\sin(\pi s_n^{(j)z})=2s_n^{(j)z}$.

\subsection{Spin-flip identities}
 From Eqs.\eqref{equ6a}-\eqref{equ6c} we have also verified the relations  
\begin{subeqnarray}\label{equA4}
&\slabel{equA4a}	s_{n+q_j}^{\left(j\right)+} = s_n^{\left(j\right)-}, \quad s_{n-q_j}^{\left(j\right)-} = s_n^{\left(j\right)+}, \\\nonumber\\
&\slabel{equA4c}	s_{D-1-n}^{\left(j\right)+} = s_n^{\left(j\right)-}, \quad s_{D-1-n}^{\left(j\right)-} = s_n^{\left(j\right)+}, \\\nonumber\\
&\slabel{equA4e}	s_{D-1-n-q_j}^{\left(j\right)-} = s_{n+q_j}^{\left(j\right)+}=s_n^{\left(j\right)-}, \quad s_{D-1-n+q_j}^{\left(j\right)+} = s_{n-q_j}^{\left(j\right)-}=s_n^{\left(j\right)+}.
\end{subeqnarray}
Eq.\eqref{equA4a} and \eqref{equA4e} tell us that $s_n^{\left(j\right)\pm}$ has period $2q_j$ by virtue of $s_{n+2q_j}^{\left(j\right)\pm}=s_{n}^{\left(j\right)\pm}$.
Thereby, for $0\le n \le2^K-1$ 
\begin{eqnarray}\label{equA6a}
	s_{n+q_j}^{\left(j\right)z}  = - s_{n}^{\left(j\right)z},\\
	s_{D-1-n}^{\left(j\right)z}  = - s_{n}^{\left(j\right)z},
\end{eqnarray}
and for $n\ge q_j$ 
\begin{eqnarray}\label{equA6b}
	s_{n-q_j}^{\left(j\right)z} = - s_{n}^{\left(j\right)z},
\end{eqnarray}
confirming that the states $\ket{n\pm q_j}$ correspond to $\ket{n}$ with a flipped $j$th spin. Therefore, $s_{n}^{\left(j\right)z}$ has period $2q_j$ since $s_{n+2q_j}^{\left(j\right)z}  = s_{n}^{\left(j\right)z}$ and half-period sign flip (anti-periodic  $q_j$).
\begin{subeqnarray}\label{equA6c}
	s_n^{\left(j\right)+}s_n^{\left(j+1\right)-}s_n^{\left(j\right)z}s_n^{\left(j+1\right)z}=-\frac{1}{4}s_n^{\left(j\right)+}s_n^{\left(j+1\right)-},
\end{subeqnarray}
\begin{subeqnarray}\label{equA6d}
	s_n^{\left(j\right)-}s_n^{\left(j+1\right)+}s_n^{\left(j\right)z}s_n^{\left(j+1\right)z}=-\frac{1}{4}s_n^{\left(j\right)-}s_n^{\left(j+1\right)+}.
\end{subeqnarray}
Eq.\eqref{equ3f} must preserve the $su(2)$ algebra. Demanding that Eq.\eqref{equ3f} preserves the $su(2)$ algebra allows to infer the following properties:
\begin{subeqnarray}\label{equA7}
	s_{n+q_j}^{\left(j\right)+}s_n^{\left(j\right)-}-s_{n+q_j}^{\left(j\right)-}s_n^{\left(j\right)+}=2s_n^{\left(j\right)z},\\\nonumber\\
	s_{n-q_j}^{\left(j\right)z}s_n^{\left(j\right)+}-s_{n}^{\left(j\right)+}s_n^{\left(j\right)z}=s_n^{\left(j\right)+},
	\\\nonumber\\
	s_{n+q_j}^{\left(j\right)z}s_n^{\left(j\right)-}-s_{n}^{\left(j\right)-}s_n^{\left(j\right)z}=-s_n^{\left(j\right)-}.
\end{subeqnarray}
These last relations can be verified with aid of previous relations. It can also be demonstrated that
\begin{subeqnarray}\label{equA8a}
&\slabel{equA8a1}	s_{n-Q_{j}}^{\left(j\right)+}s_n^{\left(j\right)+}=s_{n+Q_{j}}^{\left(j\right)-}s_n^{\left(j\right)-}=0, \\\nonumber\\ &\slabel{equA8a2} s_{n+Q_{j}}^{\left(j\right)+}s_n^{\left(j+1\right)-}=s_{n}^{\left(j\right)+}s_n^{\left(j+1\right)-}, \\\nonumber\\
&\slabel{equA8a3}	s_{n-Q_{j}}^{\left(j\right)-}s_n^{\left(j+1\right)+}=s_{n}^{\left(j\right)-}s_n^{\left(j+1\right)+}.
\end{subeqnarray}

\subsection{Reconstruction of identities}
The following identities can easily be verified:
\begin{eqnarray}\label{equA9a}
	\hspace{-0.75cm} \sum_{j=1}^{K}\sum_{m=0}^{r_{j-1}-1}\left[u\left(n-2mq_j\right)-u\left(n-(2m+1)q_j\right)\right]q_j = n,  
\end{eqnarray}
For any integer $0\leq n\leq 2^K-1$ and from Eq.\eqref{equA2} and \eqref{equ6a}-\eqref{equ6c},
\begin{eqnarray}\label{equA9c}
	\sum_{m=0}^{r_{j-1}-1}\left(u\left(n-2mq_j\right)-u\left(n-(2m+2)q_j\right)\right)=1,
\end{eqnarray}
as a consequence, summing \eqref{equA9c} over $j$ yields
\begin{eqnarray}\label{equA9b}
	\hspace{-0.75cm} \sum_{j=1}^{K}\sum_{m=0}^{r_{j-1}-1}\left(u\left(n-2mq_j\right)-u\left(n-(2m+2)q_j\right)\right)=K,
\end{eqnarray}
and multiplying \eqref{equA9c} by $q_j$ and summing over $j$ leads to 
\begin{eqnarray}\label{equA9d}
	\hspace{-0.75cm} \sum_{j=1}^{K}\sum_{m=0}^{r_{j-1}-1}\left[u\left(n-2mq_j\right)-u\left(n-(2m+2)q_j\right)\right]q_j=2^K-1,\quad
\end{eqnarray}
since $\sum_{j=1}^{K}q_j=2^K-1$.
\begin{eqnarray}\label{equA9e}
	\hspace{-0.75cm} \sum_{j=1}^{K}\sum_{\ell=0}^{r_j-1}\left(-1\right)^\ell\left[u\left(n-\ell q_j\right)-u\left(n-(\ell+1)q_j\right)\right]=2m^z_n,
\end{eqnarray}
where $m^z_n$ is the $n$th magnetization quantum number.  

\section{Transverse-field Ising model in the Dirac representation}\label{AppC}

In this Appendix, we illustrate yet another application of the Dirac representation to a $1D$  Ising model with a transverse field. The exact solution to this model is well established\cite{Pfeuty1970}. Our goal is therefore not to provide a new solution but essentially to show how the known solution can directly be derived from a Dirac-basis Schr\"odinger equation.

\subsection{Dirac-basis Schr\"odinger equation}

We consider the following open $1D$ transverse-field Ising chain\cite{Pfeuty1970}
\begin{eqnarray}\label{EqC1}
	\hat{H} = h\sum_{j=1}^{K}\hat{S}_j^z + J\sum_{j=1}^{K-1}\hat{S}_j^x\hat{S}_{j+1}^x, 
\end{eqnarray}
This model is particularly instructive typically because in contrast to the XXZ model discussed in the main text, the total magnetization is not conserved,
\begin{eqnarray}
	\left[\hat{H}, \hat{\mathcal{J}}^z\right]\neq0.
\end{eqnarray}
 As a consequence, the fixed-$K_\downarrow$ decomposition used for the XXZ chain is
 no longer applicable.  We cannot solve the corresponding recursive equation in each sector associated with a magnetic quantum number. Thus, the change in the block structures is a property of the symmetry of the Hamiltonian and not a limitation inherent to the Dirac representation \eqref{equ3f}.
 
The prototype Hamiltonian \eqref{EqC1} nevertheless has a symmetry. Let $K_\downarrow$ denotes the spin occupation number. In contrast to the longitudinal XXZ model, 
$\hat{S}_j^x\hat{S}_{j+1}^x = \frac{1}{4}\left(\hat{S}_j^+\hat{S}_{j+1}^+ + \hat{S}_j^+\hat{S}_{j+1}^- + \hat{S}_j^-\hat{S}_{j+1}^+ + \hat{S}_j^-\hat{S}_{j+1}^-\right)$. The first term annihilates two neighboring occupations whereas the last term creates them. Let $\hat{\Pi}=\left(-1\right)^{K_\downarrow}$ be a parity operator; its action changes a state just by a sign $\hat{\Pi}\ket{q_{j_1}+\cdots +q_{j_p}}=\left(-1\right)^{p}\ket{q_{j_1}+\cdots +q_{j_p}}$. Thus, $\hat{\Pi}\ket{\psi_{\rm even}}=+\ket{\psi_{\rm even}}$, $\hat{\Pi}\ket{\psi_{\rm odd}}=-\ket{\psi_{\rm odd}}$ and 
\begin{eqnarray}
	\left[\hat{H}, \hat{\Pi}\right] = 0.
\end{eqnarray}
The parity of $K_\downarrow$ is conserved meanwhile $K_\downarrow$ itself is not. The Hilbert space can then be subdivided into even- and odd- subspaces $\mathcal{H}=\mathcal{H}_{\rm even}\oplus \mathcal{H}_{\rm odd}$ with $\mathcal{H}_{\rm even}=\mathcal{H}_{0}\oplus\mathcal{H}_{2}\oplus\mathcal{H}_{4}\oplus\cdots $ while $\mathcal{H}_{\rm odd}=\mathcal{H}_{1}\oplus\mathcal{H}_{3}\oplus\mathcal{H}_{5}\oplus\cdots $ where $\mathcal{H}_{p}$ represents a $p$-state occupation subspace. Although $K_\downarrow$ is not a conserved quantum number, sectors with the same parity are coupled among themselves whereas sectors of opposite parity remain completely decoupled.

We consider for our demonstration only the even subspace and show that the $2p$-spin probability amplitudes in each subspace can be organized in terms of the two-spin amplitudes. The corresponding two-spin equation takes the form of an algebraic matrix Riccati equation\cite{Poznyak2008} which can subsequently
be written as a linear eigenvalue problem and solved exactly.

In order to apply our Dirac representation, no modification of the representation is required to construct the Hamiltonian. One simply has to substitute the corresponding Dirac forms of $\hat{S}_j^z$ and $\hat{S}_j^x$ in the model \eqref{EqC1}. The eigenvalue equation is transformed into a bulk recurrence equation by constructing the wave-function as in Eq.\eqref{equ22},
\begin{eqnarray}\label{EqC2}
\nonumber\sum_{j=1}^{K-1}\Big[\Delta_{n}^{\left(j\right)++}C_{n-Q_j^+} &+& \Delta_{n}^{\left(j\right)-+}C_{n+Q_j^-}
 \\\hspace{-3cm}+\Delta_{n}^{\left(j\right)+-}C_{n-Q_j^-} &+& \Delta_{n}^{\left(j\right)--}C_{n+Q_j^+}\Big] = (2E-\Delta_n^z)C_n .
\end{eqnarray}
Here,
\begin{eqnarray}\label{EqC3}
\Delta_{n}^{\left(j\right)++} = \frac{J}{2}s_n^{\left(j\right)+}s_n^{\left(j+1\right)+},\quad \Delta_{n}^{\left(j\right)--}= \frac{J}{2}s_n^{\left(j\right)-}s_n^{\left(j+1\right)-},
\end{eqnarray}
and
\begin{eqnarray}\label{EqC4}
\Delta_n^{zz} = 2h\sum_{j=1}^{K}s_n^{\left(j\right)z}, 
\end{eqnarray}
with $\Delta_{n}^{\left(j\right)+-}$ and $\Delta_{n}^{\left(j\right)-+}$ given in \eqref{equ20}.  Following the notations in the main text, we defined $Q_j^+=q_j+q_{j+1}$, $Q_j^-=q_j-q_{j+1}$.  Eq. \eqref{EqC2} is our central equation.

\subsection{The eigenspectrum}

As we restrict ourselves to the even sectors, we can write
\begin{eqnarray}\label{EqC5}
	\begin{split}
		\ket{\psi_{\rm even}}={}& C_0\ket{0} + \sum_{j_1<j_2} C_{j_1,j_2}\ket{
		q_{j_1}+q_{j_2}}
		\\&+ \sum_{j_1<j_2<j_3<j_4} C_{j_1,j_2,j_3,j_4}\ket{q_{j_1}+q_{j_2}+q_{j_3}+q_{j_4}}
		+\cdots ,
	\end{split}
\end{eqnarray}
where for simplicity, $C_{j_1,\ldots, j_{2p}} \equiv C_{q_{j_1}+\cdots+q_{j_{2p}}}$ and can be interpreted as the probability amplitude to observe $2p$ spins pointing in the down direction. For instance, $C_0$ is the probability amplitude to find the system in configuration $\ket{0}$ where no  spin points in the down direction. Similarly, $C_{j_1,j_2}$ is the one to observe two spins in positions $j_1$ and $j_2$ pointing in the down direction. We refer to each fixed-occupation contribution to the wave-function $\ket{\psi_{\rm even}}$ as an occupation sector and use the occupation number $K_{\downarrow}$ to distinguish between them as in the previous case.

Assuming $C_0\neq0$, it is convenient to introduce
\begin{eqnarray}\label{EqC6}
P_{j_1,\cdots, j_{2p}} = \frac{C_{j_1,\cdots, j_{2p}}}{C_0},
\qquad
P_{\varnothing}=1.
\end{eqnarray}
As already mentioned in the main text, it has two important properties: it is anti-symmetric i.e. its sign changes by permutation of two particle and vanish when two indices are identical. In a two-particle sector for instance $P_{j_1, j_2}=-P_{j_2, j_1}$ and $P_{j, j}=0$.

{\bf Sector: $K_{\downarrow}=0$}. This is the zero-occupation sector corresponding to $n=0$. In this sector, $\Delta_{0}^{\left(j\right)++} = \Delta_{0}^{\left(j\right)+-} = \Delta_{0}^{\left(j\right)-+} = 0$, and $\Delta_{0}^{\left(j\right)--} = J/2.$
Equation (\ref{EqC2}) readily gives
\begin{eqnarray}\label{EqC7}
2E-\Delta_0^z = \frac{J}{2}\sum_{j=1}^{K-1}P_{j,j+1}.
\end{eqnarray}
The energy depends only on the {\it neighboring} two-spin occupation amplitudes $P_{j,j+1}$. This relation plays a key role in our demonstration. It enters the Dirac-basis Schr\"odinger equation independently on $K_{\downarrow}$. We also show that the whole wave-function depends only on $P_{j_a,j_b}$ for $j_a<j_b$. Knowing the eigenspectrum of the two-spin sector fully solves our problem. 

{\bf Sector: $K_{\downarrow}=2$}. This is the two-particle sector where 
$n=q_{j_1}+q_{j_2}$ with $j_1<j_2$.  Our classical variables acquire the forms,
 \begin{subeqnarray}\label{EqC8}
 	\hspace{-0.5cm}	
 	\slabel{EqC8a}\Delta^{(j)++}_{q_{j_1}+q_{j_2}} &=&\frac{J}{2}\left[\left(\delta_{j_1,j}+\delta_{j_2,j}\right)\left(\delta_{j_1,j+1}+\delta_{j_2,j+1}\right)\right],\qquad \\ 
 	\slabel{EqC8b}\Delta^{(j)--}_{q_{j_1}+q_{j_2}} &=& \frac{J}{2}\left[\left(1-\delta_{j_1,j}-\delta_{j_2,j}\right)\left(1-\delta_{j_1,j+1}-\delta_{j_2,j+1}\right)\right].\qquad
 \end{subeqnarray}
 The terms $\Delta^{(j)+-}_{q_{j_1}+q_{j_2}}$ and $\Delta^{(j)-+}_{q_{j_1}+q_{j_2}}$ are given in Eq.\eqref{equ22e}. So, we do not need to calculate the second and third terms in Eq.\eqref{EqC2}. We only need to calculate the first and last terms. 
The first term reads,
\begin{eqnarray}\label{EqC9}
\sum_{j=1}^{K-1}\Delta_{q_{j_1}+q_{j_2}}^{\left(j\right)++}C_{q_{j_1}+q_{j_2}-q_j-q_{j+1}}=\frac{J}{2}\delta_{j_1+1,j_2}C_0,
\end{eqnarray}
while the last term is more subtle and requires a special care. The factors in \eqref{EqC8b} vanish in any of the following situations $j=j_1$ or $j=j_2$ or $j=j_1-1$ or $j=j_2-1$. Thus,
\begin{eqnarray}\label{EqC10}
	\sum_{j=1}^{K-1}\Delta_{q_{j_1}+q_{j_2}}^{\left(j\right)--}C_{q_{j_1}+q_{j_2}+q_j+q_{j+1}} = \frac{J}{2} \sum_{\substack{j=1\\j\notin\Omega_2}}^{K-1}C_{q_{j_1}+q_{j_2}+q_j+q_{j+1}},
\end{eqnarray}
where $\Omega_2=\{j_1,j_2,j_1-1,j_2-1\}$. The two-occupation equation is achieved by plugging Eqs.\eqref{EqC9} and \eqref{EqC10} in Eq.\eqref{EqC2} while considering \eqref{EqC7}. Dividing the resulting equation by $C_0$ and finally taking into account the definition \eqref{EqC6} we achieve, 
\begin{eqnarray}\label{EqC11}
\nonumber\delta_{j_2,j_1+1} + \left(\mathcal L^{(2)}P\right)_{j_1,j_2}
+ \sum_{\substack{j=1\\j\notin\Omega_2}}^{K-1}P_{j_1, j_2, j, j+1} = \left(\lambda_2 + \sum_{j=1}^{K-1}P_{j,j+1}\right)P_{j_1,j_2},\\
\end{eqnarray}
where
\begin{eqnarray}\label{EqC12}
\left(\mathcal L^{(2)}P\right)_{i,j} = P_{i-1,j}+P_{i+1,j}+P_{i,j-1}+P_{i,j+1},
\end{eqnarray}
describes the hopping term (see Eq.\eqref{equ22gg}); 
$
\lambda_2 = 2(\Delta_0^{zz}-\Delta_{2}^{zz})/J
$
where $\Delta_0^{zz}=hK$ and $\Delta_{2}^{zz}\equiv \Delta_{q_{j_1}+q_{j_2}}^{zz}=h(K-4)$ is, in our case, independent on $j_1$ and $j_2$. They can be extracted from Eq.\eqref{ob10a} in the main text by setting $J_z=0$ and $h_j=h$ (homogeneous case).

We see that Eq.\eqref{EqC11} involves a four-spin amplitudes $P_{j_1, j_2, j,j+1}$ confirming the pattern $0\longleftrightarrow2\longleftrightarrow4$ predicted by the symmetry of the Hamiltonian. We can now show how they can be
organized in terms of $P_{i,j}$. 

Since the transverse field in the Ising model changes the spin occupation number by a pair, the two-occupation amplitude $P_{i,j}$ stands for the elementary building block of the even sector. For four-occupied sites with $j_1<j_2<j_3<j_4$, there are three distinct ways of partitioning the four sites into two non-overlapping pairs avoiding double-occupation. Anti-symmetrization of these pairings gives,
\begin{eqnarray}\label{EqC14}
P_{j_1, j_2, j_3, j_4} = P_{j_1, j_2}P_{j_3, j_4} - P_{j_1, j_3}P_{j_2, j_4} + P_{j_1, j_4}P_{j_2, j_3}.
\end{eqnarray}
This is known in the Physics of fermionic pairing problem and in Mathematics as the Pfaffian operator\cite{Kasteleyn1961, Lieb1968}. For an anti-symmetric $4\times 4$ matrix we have,
\begin{eqnarray}\label{EqC15}
	\hspace{-1cm}
\operatorname{Pf}
	\begin{pmatrix}
		0 & P_{12} & P_{13} & P_{14}\\
		-P_{12} & 0 & P_{23} & P_{24}\\
		-P_{13} & -P_{23} & 0 &P_{34}\\
		-P_{14} & -P_{24} & -P_{34} & 0
	\end{pmatrix}
	= P_{12}P_{34} - P_{13}P_{24} + P_{14}P_{23}.
\end{eqnarray}
and can be denoted in ou case as
\begin{eqnarray}\label{EqC16}
	P_{j_1,j_2,j_3,j_4} = \operatorname{Pf}\left[P_{j_a, j_b}\right]_{a,b=1}^{4}.
\end{eqnarray}
We now substitute Eq.(\ref{EqC14}) into the last term in the left hand side of 
Eq.(\ref{EqC11}). For arbitrary $j$, one has
$P_{j_1,j_2,j,j+1} = P_{j_1,j_2}P_{j,j+1} - P_{j_1,j}P_{j_2,j+1} + P_{j_1,j+1}P_{j_2,j}$. This relation confirms that $P_{j_1,j_2,j,j+1}$ automatically  vanishes when any newly created pair overlaps an already occupied site; in other words when two indices are identical (hard-core condition). The restricted sum in Eq.(\ref{EqC11}) may therefore be replaced by the full sum,
\begin{eqnarray}\label{EqC17}
\nonumber	\sum_{\substack{j=1\\j\notin\Omega_2}}^{K-1}P_{j_1, j_2, j, j+1}
		&=& P_{j_1, j_2} \sum_{j=1}^{K-1}P_{j,j+1}
		\\&-& \sum_{j=1}^{K-1}P_{j_1, j}P_{j_2,j+1} + \sum_{j=1}^{K-1}P_{j_1,j+1}P_{j_2, j}.
\end{eqnarray}
The first term on the right-hand side cancels exactly the term
$P_{j_1, j_2}\sum_jP_{j,j+1}$ appearing on the right-hand side of Eq.\eqref{EqC11}. We are left
with
\begin{eqnarray}\label{EqC18}
\nonumber \lambda_2 P_{i, j} = \delta_{j,i+1} + \left(\mathcal L^{(2)}P\right)_{i,j} - \sum_{k=1}^{K-1}P_{i, k}P_{j,k+1} + \sum_{k=1}^{K-1}P_{i,k+1}P_{j, k}.\\
\end{eqnarray}
Equation (\ref{EqC18}) is remarkable because the four-spin amplitudes have disappeared. The same Pfaffian recursion applies successively to the $4$, $6$, $\ldots$ occupation sectors.
Thus the complete even-sector hierarchy closes in terms of the two-spin amplitudes $P_{i,j}$.

The structure of Eq.\eqref{EqC18} becomes particularly simple in matrix notation.
Define the symmetric nearest-neighbor matrix $T_{i, j} = \delta_{i+1,j} + \delta_{i-1,j}$
and the antisymmetric nearest-neighbor matrix $ A_{i,j} = \delta_{j, i+1} -\delta_{j, i-1}$. 
Since $P^T=-P$, the hopping terms can be written as $ (TP+PT)_{i, j} = P_{i-1, j} + P_{i+1, j} + P_{i, j-1} + P_{i, j+1}$ while the nonlinear contribution is
\begin{eqnarray}\label{EqC19}
(PAP)_{i,j} = -\sum_kP_{i,k}P_{j,k+1} + \sum_kP_{i,k+1}P_{j,k}.
\end{eqnarray}
Equation (\ref{EqC18}) therefore becomes
\begin{eqnarray}\label{EqC20}
A+TP+PT+PAP-\lambda_2P=0.
\end{eqnarray}
Introducing $ B=T-c\hat{I}$ where $\hat{I}$ is an identity matrix and $c=\lambda_2/2$, we obtain the algebraic Riccati equation\cite{Poznyak2008}
\begin{eqnarray}\label{EqC21}
		A+BP+PB+PAP=0.
\end{eqnarray}
It is worth noting that the Riccati equation has not been introduced as an independent ansatz. It follows directly from the Dirac-basis Schr\"odinger equation after the Pfaffian structure of the even-sector amplitudes has been taken into account.

Let us briefly describe how Eq.\eqref{EqC21} can be linearized and the energies obtained. We introduce the $2K\times2K$ symmetric block matrix
\begin{eqnarray}\label{EqC23}
\mathcal{M}
		=
		\begin{pmatrix}
			-B&-A\\
			A&B
		\end{pmatrix},
\end{eqnarray}
where $A^T=-A$ and $B^T=B$. Let $X$ and $Y$ be two $K\times K$ matrices satisfying
\begin{eqnarray}\label{EqC24}
	\mathcal{M}
	\begin{pmatrix}
		X\\Y
	\end{pmatrix}
	=
	\begin{pmatrix}
		X\\Y
	\end{pmatrix}R,
\end{eqnarray}
where $R$ is a $K\times K$ matrix. This equation is equivalent to $-BX-AY=XR$ (i) and $AX+BY=YR$ (ii). If $X$ is invertible, we define
\begin{eqnarray}
		P=YX^{-1}.
\end{eqnarray}
Using $Y=PX$, equation (i)  readily gives $R=X^{-1}(-B-AP)X$.
Substituting this expression into equation (ii) leads precisely to Eq.\eqref{EqC21} ascertaining a linearization of our algebraic matrix Riccati equation.

Let us show how this relates to the energy. To this end let us introduce the eigenvalues equation,
\begin{eqnarray}\label{EqC26}
	\mathcal{M} 
	\begin{pmatrix}
		\hat{x}\\\hat{y}
	\end{pmatrix} =
	\mu \begin{pmatrix}
		\hat{x}\\\hat{y}
	\end{pmatrix},
\end{eqnarray}
where $\hat{x}$ and $\hat{y}$ are two block vectors and $\mu$ the eigenvalue of $\mathcal{M}$. 
Taking the trace of the matrix $R$ yields,
\begin{eqnarray}\label{EqC27}
	{\rm Tr}R = -{\rm Tr}B - {\rm Tr}\left(AP\right).
\end{eqnarray}
If we choose the block-matrix vector $\left(X \hspace{0.2cm} Y\right)^\mathcal{T}$ ( where $\mathcal{T}$ denotes the transpose operation) such that their columns are the $K$ independent eigenvectors of $\mathcal{M}$ i.e. $\hat{x}_i=\sum_{j}X_{i,j}\ket{j}$ and $\hat{y}_i=\sum_{j}Y_{i,j}\ket{j}$ then  $\mathcal{M}$ is diagonal in this basis and $R={\rm diag}\left(\mu_1,\cdots, \mu_K\right)$ i.e. a diagonal matrix with the eigenvalues of $\mathcal{M}$ on the main diagonal. Thus,
\begin{eqnarray}\label{EqC28}
	{\rm Tr}R = \sum_k \mu_k.
\end{eqnarray}
On the other hand 
\begin{eqnarray}\label{EqC29}
\nonumber	{\rm Tr}B = - \frac{\lambda_2}{2}{\rm Tr}\hat{I} = -\frac{4hK}{J}=-\frac{4\Delta^{zz}_{0}}{J},
\end{eqnarray}
since ${\rm Tr}T=0$. Also,
\begin{eqnarray}\label{EqC30}
\nonumber	{\rm Tr}\left(AP\right)&=&\sum_{i,j}A_{i,j}P_{j,i}=\sum_{i,j}\left(\delta_{j,i+1}P_{j,i} -\delta_{j,i-1}P_{j,i}\right)
	\\&=&\sum_{i}P_{i+1,i} - \sum_{j}P_{j,j+1} = -2\sum_{j}P_{j,j+1},
\end{eqnarray}
since $P_{j,j+1}=-P_{j+1, j}$. Returning to \eqref{EqC27} equipped with Eqs.\eqref{EqC28}-\eqref{EqC30} we obtain,
\begin{eqnarray}\label{EqC31}
	\sum_{j}P_{j,j+1} = \frac{1}{2}\sum_k \mu_k  -\frac{2\Delta^{zz}_{0}}{J}.
\end{eqnarray}
The desired energy is inferred from Eq.\eqref{EqC7} as
\begin{eqnarray}\label{EqC32}
	E = \frac{J}{8}\sum_k \mu_k.
\end{eqnarray}
The energy now depends solely on the eigenvalues of $\mathcal{M}$. We will show that the wave-functions are fully determined by the eigenfunctions of $\mathcal{M}$. The remaining task is to solve Eq.\eqref{EqC26}. We then introduce the auxiliary vectors $\hat{r}=\hat{x}+\hat{y}$ and $\hat{s}=\hat{x}-\hat{y}$.
The two block equations immediately give
$\mu \hat{r}=(A-B)\hat{s}$ and $\mu \hat{s}=-(A+B)\hat{r}$.
Eliminating $\hat{s}$ in the equation for $\hat{r}$ yields $\mu^2\hat{r}=-(A-B)(A+B)\hat{r}=\left(B^2-A^2+[A,B]\right)\hat{r}$. In the bulk $\left(1<j<K\right)$, $[A,B]=0$. Indeed, the operators $A$ and $T$ can be rewritten as $A=R-L$ and $T=R+L$ where $L$ and $R$ are two shift operators defined such that $\left(Lv\right)_j=v_{j-1}$ and  $\left(Rv\right)_j=v_{j+1}$. In the bulk, $RL=LR=\hat{I}$. Since $B=T-c\hat{I}$, one has ${A}B-B{A}=[A,B]=2RL-2LR=0$.  This reduces the equation for $\hat{r}$ to a nearest-neighbor difference equation,
\begin{eqnarray}\label{EqC33}
	\mu^2\hat{r}_j = (c^2+4)\hat{r}_j - 2c(\hat{r}_{j-1}+\hat{r}_{j+1}),
\end{eqnarray}
with boundary conditions  $\hat{r}_{K+1}=0$ and $\hat{r}_0=\lambda \hat{r}_1$ with $\lambda=2/c=J/2h$. The first boundary condition allows to choose a solution $\hat{r}_j=F \sin\left[\left(K+1-j\right)k\right]$ (where $F$ is a normalization constant) leading to\cite{Elliot1961, Pfeuty1970},
\begin{eqnarray}\label{EqC34}
 \mu^2_k = c^2+4-4c\cos k.
\end{eqnarray}
After expressing $c$ and $\mu$ in terms of the conventional dimensionless parameters in the Ising model, Eq.\eqref{EqC33} reproduces the standard transverse-field Ising dispersion relation\cite{Elliot1961, Pfeuty1970}. The sign difference of the cosine resulting from the convention used in the Hamiltonian in Eq.\eqref{EqC1} can equivalently be removed by shifting the momentum as $k\rightarrow\pi-k$.

The second boundary equations determine the allowed momenta. One recovers the standard open-chain quantization condition\cite{Elliot1961, Pfeuty1970}
\begin{eqnarray}\label{EqC35}
 \sin[(K+1)k] = \lambda \sin(Kk).
\end{eqnarray}

{\bf Sector: $K_{\downarrow}=2p$}. The generalization to an arbitrary even sector  is straightforward thanks to the Pfaffians and by using the very same arguments as for $K_{\downarrow}=2$. No additional ingredients or arguments are necessary. Setting $n=\sum_{\ell=1}^{2p}q_{j_\ell}$ our central equation \eqref{EqC2} is transformed with 
\begin{eqnarray}\label{EqC36}
	\Delta^{(j)++}_{q_{j_1}+q_{j_2}+\cdots+q_{j_{K_{\downarrow}}} }=\frac{J}{2}\sum_{\ell=1}^{K_{\downarrow}}\sum_{k\neq\ell}^{K_{\downarrow}}\delta_{j_{\ell},j}\delta_{j_k, j_{\ell}+1}, \quad
\end{eqnarray}
\begin{eqnarray}\label{EqC37}
	\Delta^{(j)--}_{q_{j_1}+q_{j_2}+\cdots+q_{j_{K_{\downarrow}}} }=\frac{J}{2}\left(1-\sum_{\ell=1}^{K_{\downarrow}}\delta_{j_{\ell},j}\right)\left(1-\sum_{\ell=1}^{K_{\downarrow}}\delta_{j_{\ell},j+1}\right),\quad
\end{eqnarray}
and
\begin{eqnarray}\label{EqC38}
	\Delta^{zz}_{q_{j_1}+q_{j_2}+\cdots+q_{j_{K_{\downarrow}}}} =2h\left(\frac{K}{2}-K_{\downarrow}\right).
\end{eqnarray}
So far, the origin of the Pfaffian structure is understood directly from the occupation representation. For $2p$ occupied sites the same construction consists of summing over all $(2p-1)!!$ perfect pairings with the sign associated with the corresponding permutation. Let
\begin{eqnarray}\label{EqC39}
	I=(j_1,\ldots,j_{2p}),
	\qquad
	j_1<\cdots<j_{2p},
\end{eqnarray}
and assume that the even-sector amplitudes are written as
\begin{eqnarray}\label{EqC40}
P_I = \operatorname{Pf}
		\left[P_{j_a, j_b}\right]_{a,b=1}^{2p},
\end{eqnarray}
We show that the complete hierarchy of the $2p$-spin probability amplitudes in the even-sector Schr\"odinger equations reduces to the equation for the two-spin amplitudes. The standard Pfaffian expansion implies the identity
\begin{eqnarray}\label{EqC41}
pP_I = \sum_{1\leq a<b\leq 2p} (-1)^{a+b+1} P_{j_a j_b}P_{I\setminus\{j_a,j_b\}}.
\end{eqnarray}
Here, the notation $P_{I\setminus\{j_a,j_b\}}$ should be understood that the indices $j_a$ and $j_b$ are removed from the list $I$. The pair-annihilation contribution in Eq. \eqref{EqC2} can be written as
\begin{eqnarray}\label{EqC42}
	\mathcal{A}_I = \frac{2}{J}\sum_{j=1}^{K-1}\Delta_{q_{j_1}+q_{j_2}+\cdots+q_{j_{K_{\downarrow}}}}^{\left(j\right)++}\frac{C_{q_{j_1}+q_{j_2}+\cdots+q_{j_{K_{\downarrow}}}-q_j-q_{j+1}}}{C_0}.
\end{eqnarray}
Considering Eq.\eqref{EqC36}, we achieve
\begin{eqnarray}\label{EqC43}
\mathcal A_I = \sum_{a<b} (-1)^{a+b+1}A_{j_a, j_b}P_{I\setminus\{j_a,j_b\}},
\end{eqnarray}
Similarly, the hopping contribution satisfies
\begin{eqnarray}\label{EqC44}
\mathcal L^{(2p)}P_I = \sum_{a<b}(-1)^{a+b+1}\left(\mathcal L^{(2)}P\right)_{j_a j_b}
		P_{I\setminus\{j_a,j_b\}},
\end{eqnarray}
with $\left(\mathcal L^{(2)}P\right)_{i,j}$ given in E.\eqref{EqC12}
Terms leading to coincident indices vanish automatically because
of the antisymmetry of the Pfaffian.

Finally, expansion of the Pfaffian after creation of a pair on the
bond $(k,k+1)$ yields the last term 
\begin{eqnarray}\label{EqC45}
\nonumber	\mathcal{B}_I = \frac{2}{J}\sum_{j=1}^{K-1}\Delta_{q_{j_1}+\cdots+q_{j_{K_{\downarrow}}}}^{\left(j\right)--}\frac{C_{q_{j_1}+\cdots+q_{j_{K_{\downarrow}}}+q_j+q_{j+1}}}{C_0}=\sum_{\substack{j=1\\j\notin\Omega_{2p}}}^{K-1}P_{I, j, j+1}.\\
\end{eqnarray}
where $\Omega_{2p}=\{j_a,j_{a}-1\}_{a=1}^{2p}$ and
\begin{eqnarray}\label{EqC46}
\nonumber	\sum_k P_{I,k,k+1} = P_I\sum_kP_{k,k+1} + \sum_{a<b}(-1)^{a+b+1}(PAP)_{j_a,j_b}P_{I\setminus\{j_a,j_b\}}\\
\end{eqnarray}
with $(PAP)_{i,j}$ given in Eq.\eqref{EqC19}. If the newly created pair overlaps an already occupied site, the corresponding Pfaffian vanishes. The restricted sum over free bonds can therefore be replaced by the full sum in
Eq.\eqref{EqC46}.

Combining Eqs.\eqref{EqC42}, \eqref{EqC44}, and \eqref{EqC45}, and using the zero-occupation equation \eqref{EqC7}, the Schr\"odinger equation in the $2p$-occupation sector   acquires the form  $\mathcal{A}_I + \mathcal{L}^{\left(2p\right)}P + \mathcal{B}_I = \left(2/J\right)\left(2E-\Delta^{zz}_{2p}\right)P_I$ that is
\begin{eqnarray}\label{EqC47}
\lambda_{2p}P_I = \sum_{a<b}(-1)^{a+b+1}\left[A+\mathcal L^{(2)}P+PAP\right]_{j_a,j_b}
		P_{I\setminus\{j_a,j_b\}}.\qquad
\end{eqnarray}
If the two-spin amplitudes satisfy $A+\mathcal L^{(2)}P+PAP=\lambda_2P$ (see Eq.\eqref{EqC20}) Eq.\eqref{EqC47} becomes
\begin{eqnarray}\label{EqC48}
\lambda_{2p}P_I = \lambda_2\sum_{a<b}(-1)^{a+b+1}P_{j_a,j_b}P_{I\setminus\{j_a,j_b\}}.
\end{eqnarray}
Using the Pfaffian identity \eqref{EqC41}, we finally obtain
\begin{eqnarray}\label{EqC49}
	\lambda_{2p}=p\lambda_2.
\end{eqnarray}
This is precisely consistent with
\begin{eqnarray}
	\Delta_0^z-\Delta_{2p}^z=4hp,
\end{eqnarray}
which independently gives
\begin{eqnarray}
	\lambda_{2p} = \frac{2(\Delta_0^z-\Delta_{2p}^z)}{J} = p\lambda_2.
\end{eqnarray}
Hence, once the antisymmetric two-spin matrix $P$ satisfies Eq.\eqref{EqC20}, the Pfaffian amplitudes Eq.\eqref{EqC40} satisfy all even-sector Schr\"odinger
equations and this fully determines the wave-function \eqref{EqC5}.

The last question is how to determine $X$ and $Y$ in order to have $P$ at hand. Indeed, $\hat{x}$ and $\hat{y}$ are the columns of $X$ and $Y$ respectively. We have established that $\hat{r}_j(k)= \sin\left[\left(K+1-j\right)k\right]$ whereby $\hat{s}_j(k)= \left(c\sin\left[\left(K+1-j\right)k\right] - 2\sin\left[\left(K-j\right)k\right]\right)/\mu\left(k\right)$. The components of the eigenfunctions are
\begin{eqnarray}
	\nonumber\hat{x}_j\left(k\right)= \frac{1}{2}\left[\left(1 + \frac{c}{\mu(k)}\right)\sin\left[\left(K+1-j\right)k\right] - \frac{2}{\mu(k)}\sin\left[\left(K-j\right)k\right]\right].\\
\end{eqnarray}
\begin{eqnarray}
	\nonumber\hat{y}_j\left(k\right)= \frac{1}{2}\left[\left(1 - \frac{c}{\mu(k)}\right)\sin\left[\left(K+1-j\right)k\right] + \frac{2}{\mu(k)}\sin\left[\left(K-j\right)k\right]\right].\\
\end{eqnarray}
$P=Y.X^{-1}$ can now be completely constructed. Let $k_1,\cdots k_K$ be the $K$ allowed solutions of Eq.\eqref{EqC35}. For each $k_\ell$ we choose a branch $\mu_\ell = \pm \sqrt{c^2 + 4 -4c\cos\left(k_\ell\right)}$ yielding the elements $X_{j,\ell} = \hat{x}_j\left(k_\ell\right)$ and $Y_{j,\ell} = \hat{y}_j\left(k_\ell\right)$. Note that the allowed momenta may include complex solutions, for which the expressions above are understood by analytic continuation. The branch should also be chosen such that $X$ is invertible.  The normalization of each eigenvector is not important. Multiplying the column by a non-zero constant does not affect $P$. Therefore $C_I = C_0 \operatorname{Pf}\left(P_I\right)$. A Mathematica program that verifies the calculations in this Section is added to our \href{https://github.com/Kenmax15}{GitHub repository}.

\section{Perturbation theory}\label{PT}
No binary theory is used in this Appendix. According to the traditional perturbation theory, the eigenvalues and eigenstates of a partitioned Hamiltonian $\hat{H}=\hat{\Lambda}+\lambda\hat{\Delta}$ with $\hat{\Lambda}$  a diagonal matrix in the original Hilbert space and $\hat{\Delta}$ an off-diagonal matrix obeying $\hat{H}\ket{\psi_n}=E_n\ket{\psi_n}$ can  be constructed as (see Ref.\cite{Kenmoe2022})
\begin{eqnarray}\label{PT1}
	E_n=\sum_{k=0}^{\mathcal{K}}\lambda^k\mathcal{E}_n^{\left(k\right)}=\epsilon_{n}+\sum_{k=1}^{\mathcal{K}}\lambda^k\mathcal{E}_n^{\left(k\right)},
\end{eqnarray}
and
\begin{eqnarray}\label{PT2}
	\ket{\psi_n}=\sum_{k=0}^{\mathcal{K}}\lambda^k\ket{\psi_n^{\left(k\right)}}=\ket{n}+\sum_{k=1}^{\mathcal{K}}\lambda^k\ket{\psi_n^{\left(k\right)}},
\end{eqnarray}
where $\hat{\Lambda}\ket{n}=\epsilon_{n}\ket{n}$ with $\mathcal{K}$ the order of the correction. We assume a non-degenerate for $\hat{\Lambda}$ i.e. $\epsilon_{n}\neq \epsilon_{m}$ for $n\neq m$. Here, $\mathcal{E}_n^{\left(k\right)}$ and $\ket{\psi_n^{\left(k\right)}}$ are respectively the $k$th order corrections to the energy and wave function.  Plugging Eqs.\eqref{PT1} and \eqref{PT2} into the eigenvalue equation,
\begin{eqnarray}\label{PT3}
	\sum_{k=0}^{\mathcal{K}}\lambda^k\left(\hat{\Lambda}+\lambda\hat{\Delta}\right)\ket{\psi_n^{\left(k\right)}}=\left(\sum_{k=0}^{\mathcal{K}}\lambda^k\mathcal{E}_n^{\left(k\right)}\right)\left(\sum_{k=0}^{\mathcal{K}}\lambda^k\ket{\psi_n^{\left(k\right)}}\right).
\end{eqnarray}
The product of power series in the above equation is reduced to a single power series with aid of the Cauchy product formula. Comparing the left and right hand sides of the resulting equation leads for $k=0$ to $\hat{\Lambda}\ket{\psi_n^{\left(0\right)}}=\mathcal{E}_n^{\left(0\right)}\ket{\psi_n^{\left(0\right)}}$ and, 
\begin{eqnarray}\label{PT4}
	\left(\hat{\Lambda}-\mathcal{E}_n^{\left(0\right)}\hat{I}\right)\ket{\psi_n^{\left(k\right)}}=\sum_{q=1}^{k}\mathcal{E}_n^{\left(q\right)}\ket{\psi_n^{\left(k-q\right)}}-\hat{\Delta} \ket{\psi_n^{\left(k-1\right)}}, \quad k\ge 1.  \qquad
\end{eqnarray}
We immediately verify from here that $\mathcal{E}_n^{\left(0\right)}=\epsilon_{n}$ and  $\ket{\psi_n^{\left(0\right)}}=\ket{n}$. Eq.\eqref{PT4} is further transformed by expanding the correction $\ket{\psi_n^{\left(k\right)}}$ into the eigenbasis of $\hat{\Lambda}$ as
\begin{eqnarray}\label{PT5}
	\ket{\psi_n^{\left(k\right)}}=\sum_{j=1}^{D}A_{j,n}^{\left(k\right)}\ket{j}.
\end{eqnarray}
We observe that the coefficients  $A_{j,n}^{\left(k\right)}$ of the expansion fully determine the wave function. We show below that they also determine the eigenenergies. Before coming to their evaluation, let us establish a few properties. We note from $\ket{\psi_n^{\left(0\right)}}=\ket{n}$ that $A_{j,n}^{\left(0\right)}=\delta_{j,n}$ (which we label as condition (i)). For normalized wave functions, the requirement $\braket{\psi_n}{\psi_n}=1$ also reads $\sum_{k,k'=0}^{\mathcal{K}}\lambda^{k+k'}\braket{\psi_n^{\left(k\right)}}{\psi_n^{\left(k'\right)}}=1$. However, this requirement is met for $k=k'=0$, therefore, $\sum_{k+k'\ge 1}^{\mathcal{K}}\lambda^{k+k'}\braket{\psi_n^{\left(k\right)}}{\psi_n^{\left(k'\right)}}=0$. Given that the scalar products $\braket{\psi_n^{\left(k\right)}}{\psi_n^{\left(k'\right)}}$ are defined to all be positive, we infer that $\braket{\psi_n^{\left(k\right)}}{\psi_n^{\left(k'\right)}}=0$ for $k+k'\ge 1$. Eq.\eqref{PT5} teaches us that $\sum_{j=1}^{D}A_{j,n}^{\left(k'\right)}A_{j,n}^{\left(k\right)}=0$ for $k+k'\ge 1$. By setting $k'=0$ in the previous relation, we obtain $\sum_{j=1}^{D}\delta_{j,n}A_{j,n}^{\left(k\right)}=0$ i.e. $A_{n,n}^{\left(k\right)}=0$  (condition (ii)) which we combine with condition (i) to obtain the important relation $A_{n,n}^{\left(k\right)}=\delta_{k,0}$ (condition (iii)).

We can now construct the equation for the coefficients $A_{j,n}^{\left(k\right)}$. This is done by substituting \eqref{PT5} into \eqref{PT4} and sandwiching the resulting equation from the left with $\bra{i}$. This recipe leads us to 
\begin{eqnarray}\label{PT6}
	A_{i,n}^{\left(k\right)}\left(\mathcal{E}_i^{\left(0\right)}-\mathcal{E}_n^{\left(0\right)}\right)=\sum_{q=1}^{k}\mathcal{E}_n^{\left(q\right)}A_{i,n}^{\left(k-q\right)}-\sum_{j=1}^{D}A_{j,n}^{\left(k-1\right)}\hat{\Delta}_{i,j}, 
\end{eqnarray}
where $\hat{\Delta}_{i,j}=\bra{i}\hat{\Delta}\ket{j}$. As mentioned earlier, the coefficients $A_{j,n}^{\left(k\right)}$ in this equation fully determine the eigenvalues as well. Indeed, setting $i=n$ and considering condition (iii) elaborated above readily yields,
\begin{eqnarray}\label{PT7}
	\mathcal{E}_n^{\left(k\right)}=\sum_{j=1}^{D}A_{j,n}^{\left(k-1\right)}\hat{\Delta}_{n,j}=\left(\hat{\Delta} A^{\left(k-1\right)}\right)_{n,n}.
\end{eqnarray}
Plugging this back into \eqref{PT6} followed by a few manipulations and finally changing $i\to j$ leads us to 
\begin{eqnarray}\label{PT8}
	A_{j,n}^{\left(k\right)}=\vartheta_{j,n}\left(\sum_{q=1}^{k}\left(\hat{\Delta} A^{\left(q-1\right)}\right)_{n,n}A_{j,n}^{\left(k-q\right)}-\left(\hat{\Delta} A^{\left(k-1\right)}_{j,n}\right)\right), 
\end{eqnarray}
where we have defined the matrix $\vartheta$ with matrix elements
\begin{eqnarray}\label{PT9}
	\vartheta_{j,n}=\left(\mathcal{E}_j^{\left(0\right)}-\mathcal{E}_n^{\left(0\right)}\right)^{-1},\quad  \vartheta_{j,j}=0.
\end{eqnarray}
Let us denote as $A^{\left(k\right)}$ a matrix whose entries are the desired coefficients. Further inspection of Eq.\eqref{PT8} reveals that the  former  obeys the fixed-point matrix equation,
\begin{eqnarray}\label{PT10}
	A^{\left(k\right)}=\vartheta\circ\left(\sum_{q=1}^{k}A^{\left(k-q\right)}\hat{\mathcal{D}}\left(\hat{\Delta} A^{\left(q-1\right)}\right)-\left(\hat{\Delta} A^{\left(k-1\right)}\right)\right).
\end{eqnarray}
Here, the symbol $\circ$ denotes the elements-wise product operator (Hadamard product\cite{Hadamard1}). We have defined the operator $\hat{\mathcal{D}}\left(X\right)$ which project onto the diagonal part of $X$ i.e. the $(\ell, n)$ element resulting from its action on a matrix $X$ reads $\hat{\mathcal{D}}\left(X\right)_{\ell, n}=\delta_{\ell, n}X_{n,n}$.  The eigenvalues and eigenstates are obtained from Eq.\eqref{PT10} as
\begin{eqnarray}\label{PT11}
\nonumber	E_n=\epsilon_{n}+\sum_{j=1}^{D}\hat{\Delta}_{n,j}\left(\sum_{k=1}^{\mathcal{K}}\lambda^k A_{j,n}^{\left(k\right)}\right)=\epsilon_{n}+\sum_{k=1}^{\mathcal{K}}\lambda^k\left(\hat{\Delta} A^{\left(k-1\right)}\right)_{n,n},\\
\end{eqnarray}
and
\begin{eqnarray}\label{PT12}
	\ket{\psi_n}=\ket{n}+\sum_{j=1}^{D}\left(\sum_{k=1}^{\mathcal{K}}\lambda^k A_{j,n}^{\left(k\right)}\right)\ket{j}.
\end{eqnarray}
For example, 
\begin{eqnarray}\label{PT16}
	A^{\left(1\right)}=-\left(\vartheta\circ\hat{\Delta}\right),
\end{eqnarray}
\begin{eqnarray}\label{PT17}
	A^{\left(2\right)}=-\vartheta\circ\left(\hat{\mathcal{D}}\left(\hat{\Delta}\left(\vartheta\circ\hat{\Delta}\right)\right)-\hat{\Delta}\left(\vartheta\circ\hat{\Delta}\right)\right).
\end{eqnarray}
These formulae are used to build the perturbation theory for random magnetic fields in the main text.
\bibliography{Mybib1} 
	
\end{document}